\documentclass[useAMS,usenatbib,usegraphicx, letterpaper]{mn2e} 
\usepackage{amssymb} 
\usepackage{times} 
\usepackage{aas_macros} 
\usepackage[left=2.2cm,right=2.2cm,bottom=2.5cm,top=2.5cm]{geometry}
\usepackage[usenames]{color} 
\usepackage{color}

\title[The single-sided pulsator CO~Cam]{The single-sided pulsator CO~Camelopardalis} 

\author[D.W. Kurtz et al.]{D. W. Kurtz$^{1,2}$\thanks{E-mail: kurtzdw@gmail.com}, G. Handler$^{3}$, S. A. Rappaport$^4$, H. Saio$^5$, J. Fuller$^{6}$, T. Jacobs$^7$, \newauthor A. Schmitt$^8$, D. Jones$^{9,10}$, A. Vanderburg$^{11}$, D. LaCourse$^{12}$,  L. Nelson$^{13}$, \newauthor F. Kahraman Ali\c{c}avu\c{s}$^{3,14}$ and M. Giarrusso$^{15}$
\\
$^{1}$Centre for Space Research, Physics Department, North West University, Mahikeng 2745, South Africa\\
$^{2}$Jeremiah Horrocks Institute, University of Central Lancashire, Preston PR1 2HE, UK\\
$^{3}$Nicolaus Copernicus Astronomical Center, Polish Academy of Sciences, ul. Bartycka 18, 00-716, Warszawa, Poland\\
$^4$Department of Physics, and Kavli Institute for Astrophysics and Space Research, M.I.T., Cambridge, MA 02139, USA\\
$^5$Astronomical Institute, Graduate School of Science, Tohoku University, Sendai 980-8578, Japan\\
$^{6}$Division of Physics, Mathematics and Astronomy, California Institute of Technology, Pasadena, CA 91125, USA\\
$^7$Amateur Astronomer, 12812 SE 69th Place, Bellevue, WA 98006, USA\\
$^8$Citizen Scientist, 616 W. 53rd. St., Apt. 101, Minneapolis, MN 55419, USA\\
$^9$Instituto de Astrof\'isica de Canarias, E-38205 La Laguna, Tenerife, Spain \\
$^{10}$Departamento de Astrof\'isica, Universidad de La Laguna, E-38206 La Laguna, Tenerife, Spain\\
$^{11}$ Department of Astronomy, The University of Texas at Austin, 2515 Speedway, Stop C1400, Austin, TX 78712, USA\\
$^{12}$Amateur Astronomer, 7507 52nd Place NE, Marysville, WA 98270, USA \\
$^{13}$ Department of Physics and Astronomy, Bishop's University, 2600 College St., Sherbrooke, QC J1M 1Z7, Canada \\
$^{14}$ \c{C}anakkale Onsekiz Mart University, Faculty of Sciences and Arts, Physics Department, 17100,
\c{C}anakkale, Turkey\\
$^{15}$ INFN, Laboratori Nazionali del Sud, Via S. Sofia 62, I-95123 Catania, Italy
}

\date{Accepted XXX. Received YYY; in original form ZZZ}

\pubyear{2019}

\begin{document}
\label{firstpage}
\pagerange{\pageref{firstpage}--\pageref{lastpage}}
\maketitle

\begin{abstract} 
CO~Cam (TIC 160268882) is the second ``single-sided pulsator'' to be discovered.  These are stars where one hemisphere pulsates with a significantly higher amplitude than the other side of the star.  CO~Cam is a binary star comprised of an Am $\delta$~Sct primary star with $T_{\rm eff} = 7070 \pm 150$\,K, and a spectroscopically undetected G main-sequence secondary star. The dominant pulsating side of the primary star is centred on the L$_1$ point.  We have modelled the spectral energy distribution combined with radial velocities, and independently the {\em TESS} light curve combined with radial velocities. Both of these give excellent agreement and robust system parameters for both stars. The $\delta$~Sct star is an oblique pulsator with at least four low radial overtone (probably) f~modes with the pulsation axis coinciding with the tidal axis of the star, the line of apsides. Preliminary theoretical modelling indicates that the modes must produce much larger flux perturbations near the L$_1$ point, although this is difficult to understand because the pulsating star does not come near to filling its Roche lobe. More detailed models of distorted pulsating stars should be developed. These newly discovered single-sided pulsators offer new opportunities for astrophysical inference from stars that are oblique pulsators in close binary stars. 
\end{abstract} 


\section{Introduction}

\label{sec:intro}

Stars are not perfect spheres. Even KIC\,11145123 \citep{2016SciA....2E1777G,2019ApJ...871..135H,2014MNRAS.444..102K}, which was widely reported to be the ``roundest object ever observed in nature,''\footnote{see, e.g., phys.org/news/2016-11-distant-star-roundest-nature.html} has an equatorial radius $3 \pm 1$\,km larger than its polar radius. It is an oblate spheroid due to its slow ($P_{\rm rot} \sim 100$\,d) rotation, and we expect all rotating stars to be oblate spheroids in the absence of magnetic fields or tides, the other strong global forces that distort stars from sphericity.

The fundamental data of asteroseismology are mode frequencies with mode identifications \citep{2010aste.book.....A}, where for convenience the modes are described by spherical harmonics, even though those are technically only correct for perfect spheres. An exception to this simple description is for very rapidly rotating stars that are highly oblate; these have novel modes known as ``island'' and ``whispering gallery'' modes (see, e.g., \citealt{2019MNRAS.483L..28M,2009A&A...506..189R,2009ASPC..416..395R}; \citealt[chapter 3.8.7]{2010aste.book.....A}). 

Rotation breaks the spherical symmetry in a star, and it was long a standard assumption that the pulsation axis and the rotation axis in a pulsating star coincide. That assumption is generally good, as in the case of KIC\,11145123, and it is a fundamental assumption of the asteroseismic method for determining a star's rotational inclination and studying spin-orbit alignment, or misalignment, in exoplanet systems (see, e.g., \citealt{2018MNRAS.479..391K}). But there are other forces that can distort a star from spherical symmetry: particularly magnetic fields and tides. 

\citet{1982MNRAS.200..807K} discovered high radial overtone p-mode pulsations in strongly magnetic Ap stars, which he called rapidly oscillating Ap stars, a name later contracted to roAp stars. Many of those stars show equally split frequency multiplets that previously would have been interpreted as rotationally split multiplets, where the splitting only differs slightly from the rotation frequency of the star by a factor of $(1- C_{n\ell})$, where $C_{n\ell}$ is the well-known ``Ledoux constant" (\citealt{1951ApJ...114..373L}; see also, \citealt{2010aste.book.....A}, chapter 1). 

The Ap stars have strong, global magnetic fields with strengths of about a kG to a few 10s of kG. Because of atomic diffusion in the presence of strong, stable magnetic fields, these stars have long-lived spots that generate stable rotational light curves, hence the rotational frequencies can be precisely determined. \citet{1982MNRAS.200..807K} found that the frequency multiplets in the roAp stars are split by  the rotation frequency to such precision that the Ledoux constant would have to be improbably small if those multiplets were rotational, and he showed by arguments of phase matching between the pulsation amplitude and rotational (spot) phase that the splitting is exactly equal to the rotational splitting. This led to the oblique pulsator model, where the pulsation axis was shown not to be the rotation axis, but was instead assumed to be the magnetic axis. Later theoretical work \citep{1985PASJ...37..245S, 1993PASJ...45..617S, 1995PASJ...47..219T, 2002A&A...391..235B,2011A&A...536A..73B} showed that both the centrifugal and Lorentz forces govern the pulsation axis, which is inclined to the rotation axis, but does not have to coincide exactly with the magnetic axis. 

The oblique pulsator model was subsequently discussed in the context of magnetic $\beta$\,Cephei stars \citep{2000ApJ...531L.143S}, the Blazhko Effect in RR\,Lyrae stars \citep{2000ASPC..203..299S}, white dwarf stars \citep{2010ApJ...716...84M}, and even in pulsars \citep{2004ApJ...609..340C}, where magnetic field strengths of TeraGauss to PetaGauss would suggest that oblique pulsation is obligatory. \citet{2016ApJ...824...14C} discuss strong magnetic fields in the cores of red giants in the context of the observed suppression of dipole modes in many red giants, and they discuss more generally asteroseismic implications of strong magnetic fields in white dwarf stars, neutron stars and possibly subdwarf B stars, although not in the context of oblique pulsation. \citet{2018CoSka..48...73W} also discuss oblique pulsation in a  general context of asteroseismology of magnetic stars. 

With all of this in mind, it has seemed evident since the discovery of the roAp stars that in close binary stars, strong tidal distortion might force the pulsation axis to coincide with the tidal axis. In binaries that are in contact or only slightly detached, it seems plausible that the pulsation axis would be the axis of greatest distortion of the stars from spherical symmetry. Hence we have been searching for stars with their pulsation axis aligned with the tidal axis for over 40 years, since the discovery of oblique pulsation. 

\citet{2019A&A...630A.106G} conducted the first systematic survey of pulsating stars in eclipsing binaries in the {\it Kepler} eclipsing binary catalogue, which includes stars with orbital periods as short as 0.05\,d.  They found 13 systems with periods less than 0.5\,d that show $\gamma$~Dor or $\delta$~Sct pulsations, although they could not rule out false positives, e.g., potential third components, or background contamination. Interestingly, they could not come to a conclusion about whether being in a close binary suppresses pulsation. The statistics are not yet good enough to tell whether the incidence of pulsation in close binary stars differs from that of (seemingly) single stars or those in wide binaries. 

There is no other systematic study of $\delta$~Sct pulsation in near contact binary stars. There are many individual studies of such pulsation in widely separated binaries, including eclipsing binary stars (see, e.g., \citealt{2015EPJWC.10104003M}), and a catalogue of 199  binary systems with $\delta$~Sct components was provided by \citet{2017MNRAS.465.1181L}, with orbital periods as short as 0.7\,d. \citet{2018MNRAS.474.4322M} even developed a novel technique that studied the orbits of 341 new binaries using the $\delta$~Sct pulsations as frequency standards, but, because of requirements of the technique, those stars all have orbital periods greater than 20\,d, so are not close to contact. 

Other studies have probed tidally induced pulsations in ``Heartbeat Stars'' (see \citealt{2019arXiv191108687G} and references therein), but even in those extremely eccentric systems with periastron distances of only a few stellar radii, it has appeared, or at least been assumed, that both the g~modes and p~modes have pulsations axes aligned with the rotation/orbital axis. We ourselves have (unsystematically) also noticed in our examination of both {\it Kepler} and {\em TESS} data that close binary stars with components in the $\delta$~Sct instability strip often show no pulsations at all, but some do. \citet{2019A&A...630A.106G}
have made a good start in examining this. Now the statistics of pulsations in close binary stars needs to be studied even more systematically, in comparison with the statistics for pulsations in wide binaries and single stars. The growing {\em TESS} data set plus the {\it Kepler} data -- both for all binaries and for single stars -- allow for that.

A question then is: are there close binary stars with $\delta$~Sct components that have the pulsation axis aligned with the tidal axis? That is the strongest distortion of the pulsating star from spherical symmetry, so such stars should exist and, because the tidal axis is inclined $90^\circ$ to the orbital/rotation axis, they will be oblique pulsators. The answer to this question is yes, these stars have now been discovered. \citet{2020NatAs.tmp...45H} presented the first such star, HD\,74423 (TIC 355151781). Remarkably, not only does this star pulsate with an axis aligned with the tidal axis, it pulsates primarily on one side, hence has been called a ``single-sided pulsator''. 

Thus HD\,74423 was the first oblique pulsator to be found in a close binary, and its pulsation axis is the tidal axis. What we have been expecting to find for 40 years has now been observed. In this paper we present the second discovery of a ``single-sided pulsator'', the known close binary star CO~Cam, whose radial velocity variations have been studied for over a century. Where HD\,74423 has a single pulsation frequency, making the detection and interpretation of oblique pulsation through the frequency multiplet pattern straightforward, CO~Cam is multi-periodic, hence more complex to decipher.  We have successfully done so, and present the analysis showing this in Section\,\ref{sec:cocam_frequency}.

CO~Cam and HD\,74423 are the first two members of the new class of single-sided pulsators -- stars that are oblique pulsators where we can study the interaction of tidal distortion and pulsation. This will enhance our understanding of pulsation in binary stars in general, and it will provide observational constraints on the stellar structure of tidally distorted stars, particularly in the outer acoustic cavity. 

\section{CO~Cam}
\label{sec:cocam}

CO~Cam (HR\,4646; HD\,106112; TIC\,160268882) is a bright, $V = 5.14$\,mag, northern binary star in the constellation Camelopardalis. It was earlier known as 4~Draconis until the constellation boundaries were formalised by the IAU at its General Assembly in Leiden in 1928 and this star was definitively placed in Camelopardalis.  CO~Cam is a close binary with an orbital period of $1.2709927 \pm 0.0000007$\,d (this work). In Table \ref{tbl:mags} we summarise some information about the system's magnitudes, distance, proper motions and other fundamental parameters. 

We note that Gaia parallaxes of binary stars measure not just the distance to the stars, but also their motion about each other on the plane of the sky. The binary semi-major axis determined in this work (and given in Section 6.2 below) gives a maximum separation of the two components on the sky of 910\,$\umu$as. We find a mass ratio below of 0.58, hence the primary pulsating star changes position on the sky over an orbit by  334\,$\umu$as. This is measurable by Gaia. If we take that change of position of the primary to be the uncertainty to the parallax (cf. Table~1), we find that the distance is still known to a precision of about 1~per~cent. Averaged over the many measurements of Gaia, the uncertainty is more likely larger than that quoted in Table~1, but this does not impact significantly on the results in the paper.

Diagnostics for the Gaia astrometric solution indeed indicates that the orbital motion of CO~Cam contributes some scatter, but likely not enough to significantly decrease the quality of the fit. The Gaia archive reports excess astrometric noise of 930\,$\umu$as about the best fit, nearly three times larger than the expected orbital motion calculated above. Most of this excess noise is likely instrumental; bright stars such as CO~Cam tend to show similar levels of excess astrometric noise. The Renormalized Unit Weight Error (RUWE, \citealt{ruwe}) statistic calculated by the Gaia team compares the measured excess scatter about the best-fit astrometric solution for a given star to other stars with the same brightness and colours, where a value close to 1 indicates a typical level of scatter, and a value larger than about 1.3 indicates some other source of astrometric scatter may be present (usually a close binary companion).  The astrometric solution for CO~Cam has RUWE = 1.30, indicating marginally elevated astrometric noise given the star's magnitude and colours. The Gaia astrometric solution for CO Cam therefore is only slightly affected by the star's binary motion.

\begin{table}
\centering
\caption{Properties of the CO~Cam System}
\begin{tabular}{lc}
\hline
\hline
Parameter & Value   \\
\hline
RA (J2000) (h m s)& 12:12:11.96  \\  
Dec (J2000) ($^\circ \ ^\prime \ ^{\prime\prime}$) &  77:36:58.78 \\ 
$T$$^a$ & $4.857 \pm 0.007$ \\
$V^a$ & $5.130 \pm 0.023$ \\
$K^b$ & $4.371 \pm 0.029$ \\
W1$^c$ & $4.374 \pm 0.235$ \\
W4$^c$ & $4.341 \pm 0.028$  \\
$T_{\rm eff}$ (K)$^d$ & $7075 \pm 100$  \\
$R$ (${\rm R}_\odot$)$^e$ & $1.89 \pm 0.2 $  \\
$L$ (${\rm L}_\odot$)$^e$ & $7.96 \pm 0.05 $  \\
Orbital Period (d)$^d$ & $1.2709927 \pm 0.0000007$ \\
$K_1$ (km~s$^{-1}$)$^f$ & $72.3 \pm 0.3$\\
$\gamma$ (km~s$^{-1}$)$^f$ & $0.7 \pm 0.3$ \\
Distance (pc)$^e$ & $ 33.56 \pm 0.17$  \\   
$\mu_\alpha$ (mas ~${\rm yr}^{-1}$)$^e$ & $+10.2 \pm 0.3$   \\ 
$\mu_\delta$ (mas ~${\rm yr}^{-1}$)$^e$ &  $+19.6 \pm 0.3$   \\ 
\hline
\label{tbl:mags}  
\end{tabular}

{\bf Notes.}  (a) ExoFOP (exofop.ipac.caltech.edu/tess/index.php).  (b) 2MASS catalog \citep{2006AJ....131.1163S}.  (c) WISE point source catalog \citep{2013yCat.2328....0C}.  (d) This work. (e) Gaia DR2 \citep{2018A&A...616A...2L}. (f) \citet{2017AN....338..671B}. 
\end{table}

One of the spectral classifications of CO~Cam is kA6hF0mF0(III) \citep{2003AJ....126.2048G}, meaning that it is a marginal metallic-lined A-F star, and suggesting that it is a giant near the TAMS. \citet{1961ApJS....6...37A} gave a spectral type of kA5hF2mF5(IV), hence Abt saw the spectrum as being a classical Am star a bit cooler than F0, but also only slightly evolved. Its Gaia DR2 distance \citep{2016A&A...595A...2G,2018A&A...616A...1G,2018A&A...616A...9L} is $33.56 \pm 0.17$\,pc. Neglecting reddening for such a close star gives an absolute visual magnitude of $M_V = 2.51$\,mag. 

The Str\"omgren colours, $b-y = 0.198$, $m_1 = 0.221$ and  $c_1 = 0.710$ give $\delta m_1 = 0.029$ and  $\delta c_1 = 0.034$ \citep{1979AJ.....84.1858C}; the $\delta m_1$ index is consistent with an Am star with slightly enhanced metallicity, and the $\delta c_1$ index suggests a star closer to the main sequence than the spectral classifications suggest. However, the atmospheric structure -- particularly the temperature gradient -- and the abundance anomalies in Am stars can distort estimates of luminosity both from spectral classification and the $\delta c_1$ index. 

Those same standard relations \citep{1979AJ.....84.1858C} give an absolute magnitude of $M_V = 2.62$\,mag, which is in agreement with the spectral classification of the star being evolved off the main-sequence. Calibrations of Str\"omgren photometry by \citet{1979AJ.....84.1858C}, coupled to the grids of \citet{1985MNRAS.217..305M}, give an estimate of $T_{\rm eff} = 7300$\,K (consistent with the $T_{\rm eff}$ we derive in this work -- see Table\,\ref{tbl:mags} and Section\,\ref{sec:systemparms}), and $\log g = 4.0$, consistent with the luminosity class, IV, of \citet{1961ApJS....6...37A}. All estimates from the Str\"omgren photometry presume that the secondary is too faint to perturb the photometric indices, and do not take interstellar reddening into account as no H$\beta$ value is available.  

The orbit of CO~Cam was first discussed by \citet{1916ApJ....43..320L} who found an orbital period of $P_{\rm orb} = 1.27100 \pm 0.00002$\,d, $K = 65$\,km~s$^{-1}$ and a barycentric velocity of $\gamma = 1.5$\,km~s$^{-1}$. He stated that ``...the spectrum contains numerous good lines'', with which we agree, and his result has stood the test of time of more than a century. Based on observations obtained less than two years later \citet{1925PulOB..10..19W} reported radial velocity variations of the star and derived an orbital period of 1.271\,d.

\citet{1961ApJS....6...37A}, in his famous paper where he showed that almost all Am stars are in short period binaries, gave orbital parameters for CO~Cam of $P_{\rm orb} = 1.2709934 \pm 0.0000007$\,d, $K = 69.8$\,km~s$^{-1}$, $\gamma = -2.2$\,km~s$^{-1}$, $e = 0$ and a mass function of $f(M) = 0.0449$\,M$_\odot$. Abt's observations were made at McDonald Observatory in 1959; he combined them with those taken by Lee in 1913 -- 1916 at Yerkes Observatory to derive the orbital period.  
\citet{1992A&AS...93..545M} collected radial velocities from several literature sources ranging in time from \citet{1916ApJ....43..320L} to \citet{1961ApJS....6...37A} to derive an improved orbital period of $1.2709943 \pm 0.0000008$\,d$^{-1}$. 

\begin{figure}
\centering
\includegraphics[width=1.0\linewidth,angle=0]{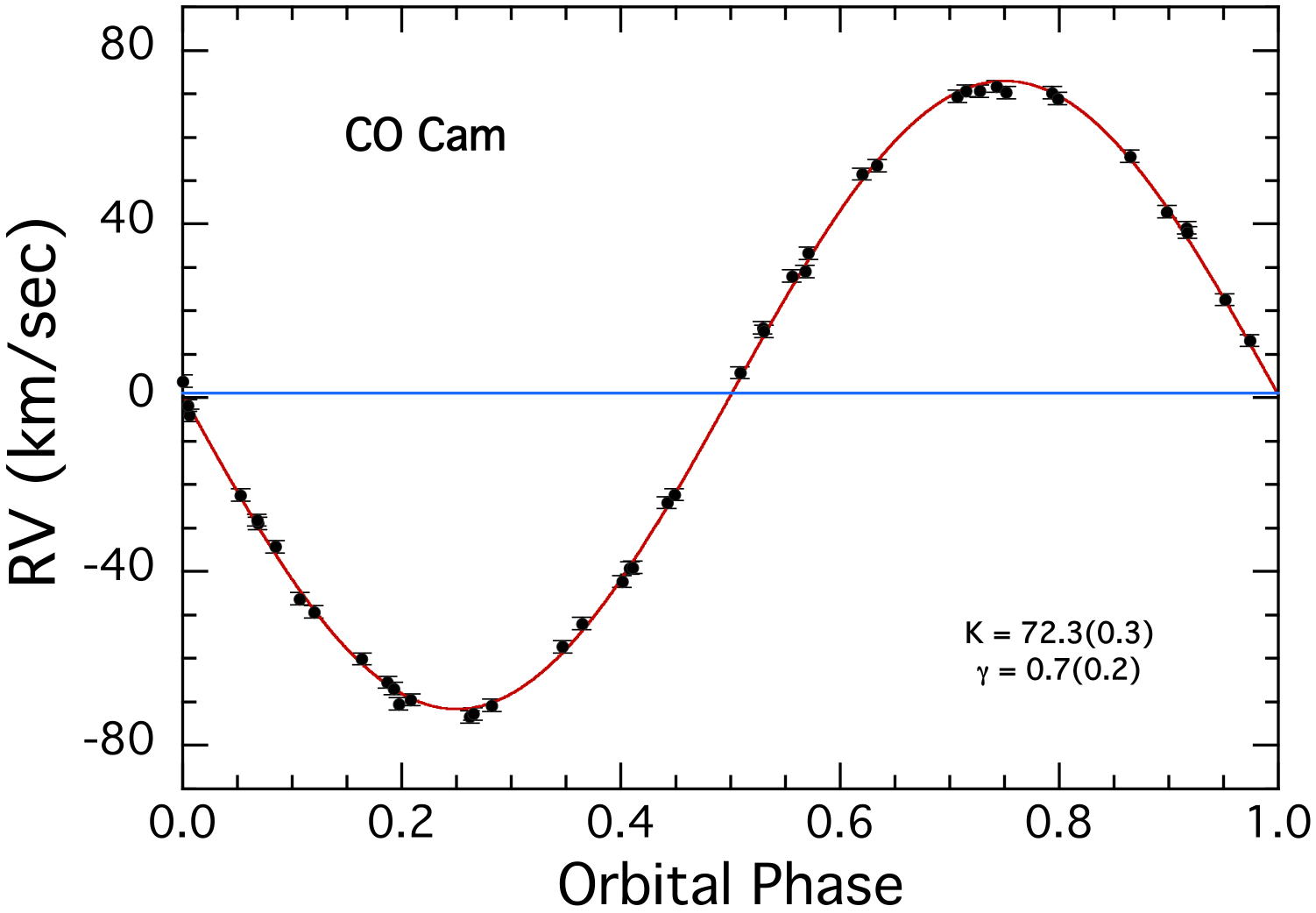}	
\caption{Radial velocity curve for CO~Cam based on the data of \citet{2017AN....338..671B}. The data were rephased with respect the TESS epoch of photometric superior conjunction of the primary star, and refitted with a circular orbit. The $K$ and $\gamma$ velocities and their uncertainties remain unchanged from those given by \citet{2017AN....338..671B}.}
\label{fig:RV}
\end{figure}

The most recent radial velocity (RV) data on CO~Cam were given by \citet{2017AN....338..671B}.  We used the data from their Table A1 to reproduce the RV curve which we show here in Fig.\,\ref{fig:RV}; our version of the RV curve is phased to the {\em TESS} epoch of superior conjunction of the primary. This is the time when the ${\rm L}_1$ side of the primary is facing the line of sight, and it coincides with pulsation amplitude maximum, as we show below in Section~4. We find the best values for $K$ and $\gamma$ from the \citet{2017AN....338..671B} RV data are unchanged from their values, $72.3 \pm 0.3$ and $0.7 \pm 0.2$\,km~s$^{-1}$, respectively. 

\begin{figure}
\centering
\includegraphics[width=1.0\linewidth,angle=0]{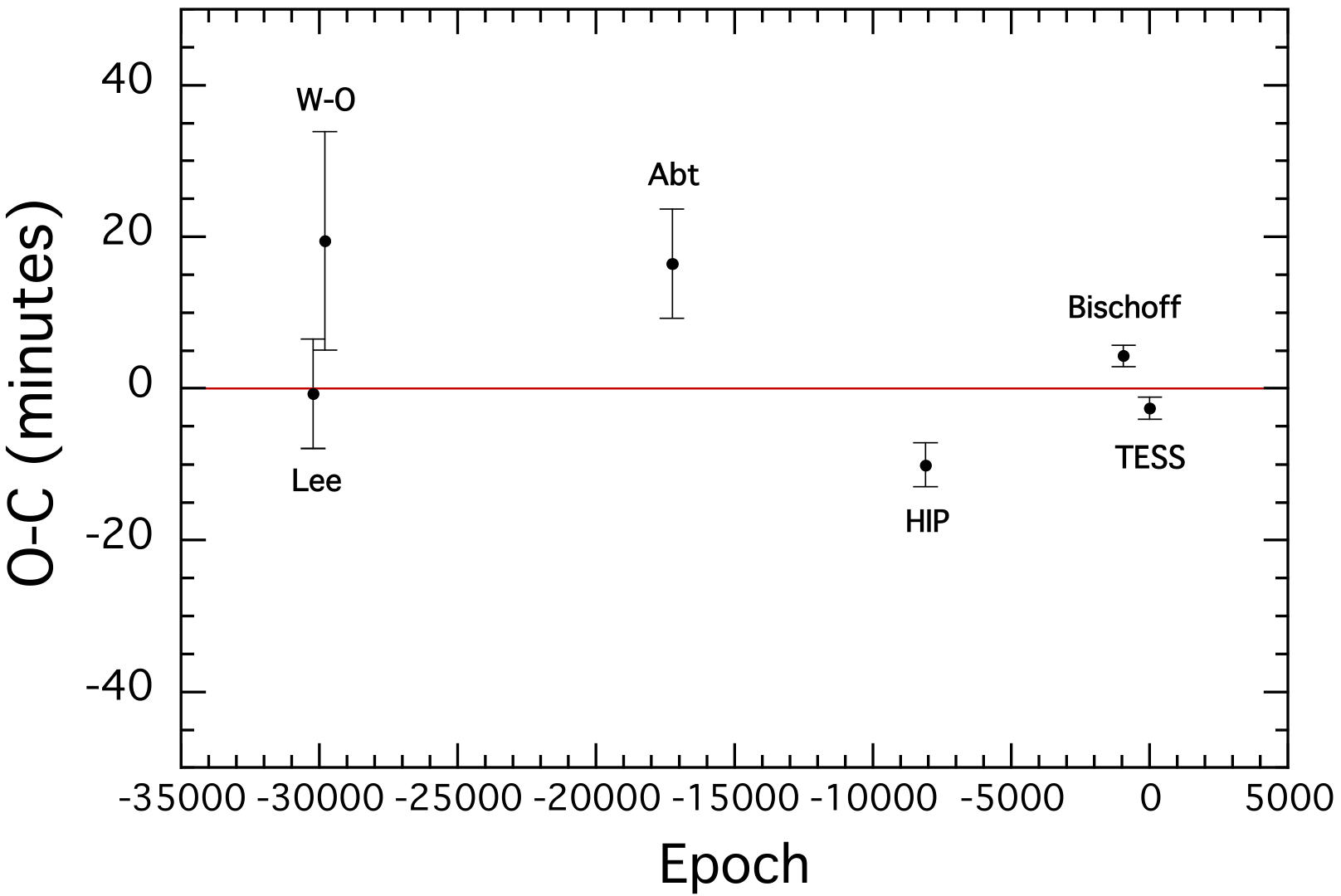}	
\caption{The $O-C$ diagram for CO~Cam based on the data points given in Table \ref{tbl:orbit}. The reference epoch and period were found to be $t_0 = {\rm BJD}~2458693.3449 \pm 0.0035$ and $P = 1.2709927 \pm 0.0000007$\,d. References to the labeled points can be found in Table \ref{tbl:orbit}.  The red line is a linear $\chi^2$ fit to the $O-C$ curve.}
\label{fig:OmC}
\end{figure}

 In Table~\ref{tbl:orbit} and Fig.\,\ref{fig:OmC} we summarise all the available orbital phase information for the past century. The references for these data are given in the notes to Table \ref{tbl:orbit}.  Fig.\,\ref{fig:OmC} is presented in the form of an $O-C$ diagram. While there are some few-$\sigma$ discrepancies\footnote{Some of the small discrepancies in the $O-C$ curve may come from the lack of barycentric corrections, and from the unknown reference time for the time stamps and unknown integration times in the earliest historical data.}, for the most part the orbital period is consistent with being constant at $P = 1.2709927$ days over an interval of 107\,yr.  This is the value for $P$ that we adopt for this work.

\begin{table}
\centering
\caption{Historical Binary Phase Information}
\begin{tabular}{cccc}
\hline
\hline
Epoch of Phase Zero$^a$ & $O-C^b$ & Method & Ref.$^c$ \\
\hline
2420285.2148 & $-0.6 \pm 7.2$ min & RV & 1 \\  
2420822.8587 & $+19.5 \pm 14.4$ min & RV & 2 \\
2436758.5636 &  $+16.5 \pm 7.2$  min & RV & 3 \\  
2448399.5677 & $-10.1 \pm 2.9$ min & Photom. & 4 \\ 
2457499.8857 & $+4.3 \pm 1.4$ min  & RV & 5 \\ 
2458693.3431 & $-2.6 \pm 1.4$ min & Photom. & 6 \\ 
\hline
\label{tbl:orbit}  
\end{tabular}

{\bf Notes.}  (a) Time of superior conjunction of the primary star. (b) Referenced to an orbital period of 1.2709927 d with an epoch zero time of BJD 2458693.3449 during the {\em TESS} observations. (c) References: (1) \citet{1916ApJ....43..320L}; (2) \citet{1925PulOB..10..19W}; (3) \citet{1961ApJS....6...37A}, after taking into account that Lee et al.~used an earlier definition of ``astronomers' GMT''; (4) \citet{1997ESASP1200.....E}; (5) \citet{2017AN....338..671B}; (6) Current work.
\end{table}

\section{CO~Cam: Discovery of pulsations and frequency analysis}
\label{sec:cocam_frequency}

CO~Cam was observed by {\em TESS} in sector 14 in 2-min cadence.   A number of the authors (TJ, DL, SR and AV) have participated in a continuing program of visual inspection of {\em Kepler}, {\em K2} and {\em TESS} light curves as a supplement to the usual computer searches for periodic signals.  The main goal of this program is to find unusual objects or phenomena that might be missed by the more conventional searches. Some of the interesting objects that have been found as part of this program are listed in \citet{2019MNRAS.488.2455R}. The visual presentation of the light curves is greatly facilitated by the software {\tt LcTools} developed by one of us (AS; \citealt{2019arXiv191008034S}).  As part of this program of visual inspection, we first noticed the unusual modulation of the pulsation amplitude in CO~Cam with orbital phase.  This generated sufficient interest to motivate the follow-up studies that are presented in this work.

The data are available in both SAP (simple aperture photometry) and PDCSAP (presearch-data conditioning SAP); we found the SAP data to have better low frequency characteristics for studying the orbital variations, and to be similar at the higher pulsation frequencies, hence we have performed our frequency analysis using the SAP data. The astrophysical results are not significantly affected by the choice of data set. The data have a time span of 26.85\,d with a centre point in time of $t_0 = {\rm BJD}~2458696.77442$, and comprise 18581 data points following the removal of 61 outliers after inspection by eye. 

\begin{figure}
\centering
\includegraphics[width=1.0\linewidth,angle=0]{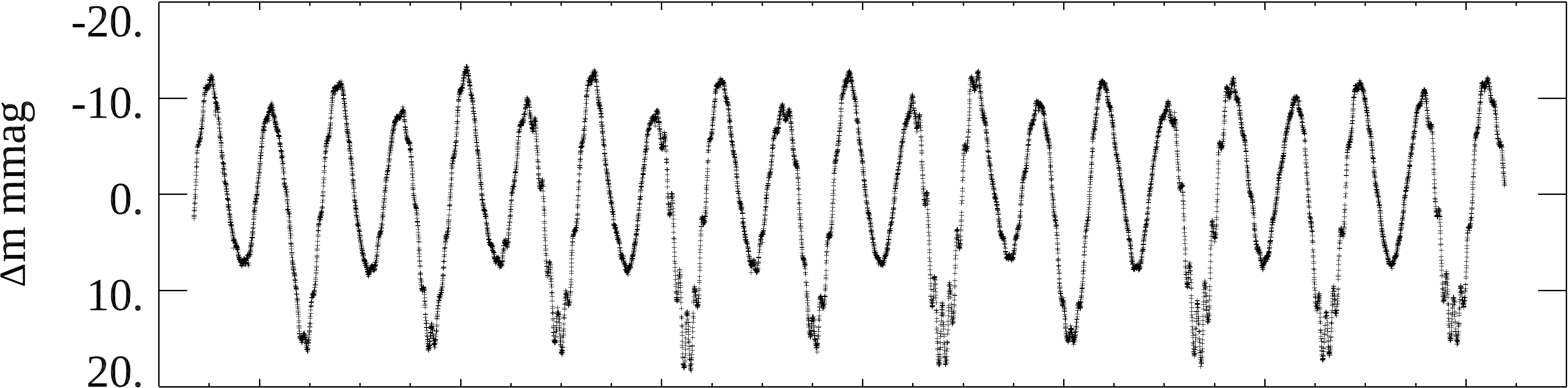} \vglue0.1cm
\includegraphics[width=1.0\linewidth,angle=0]{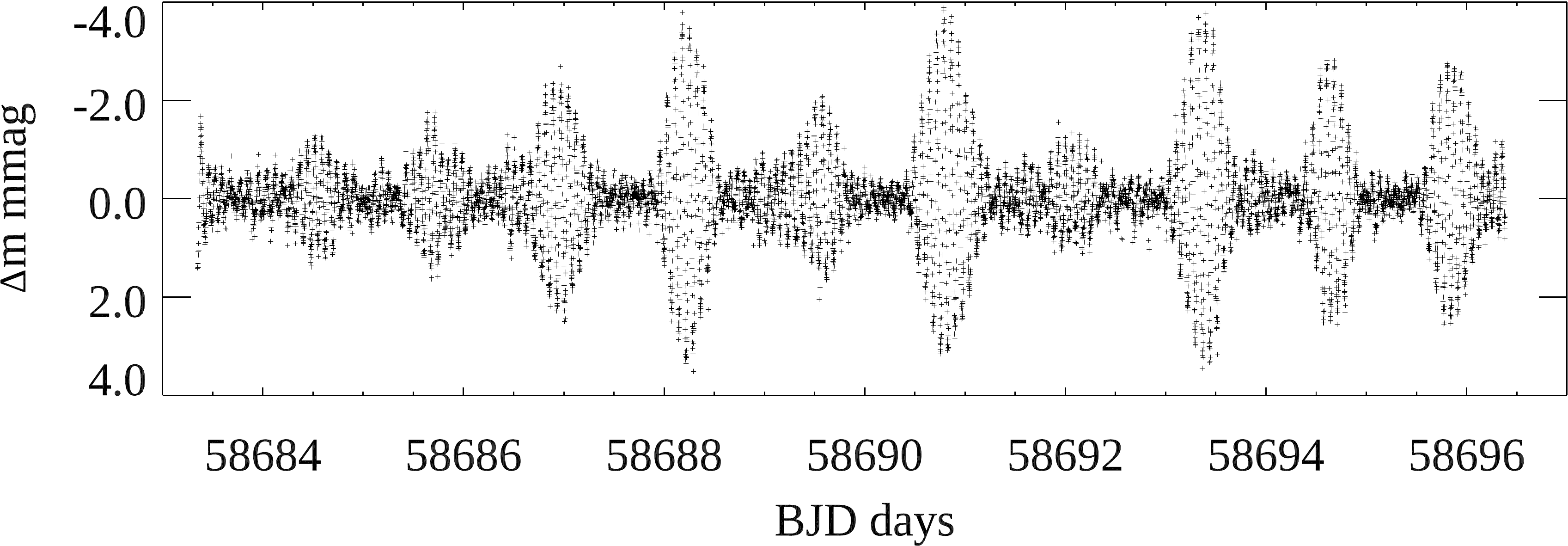}	
\caption{Top: A section of the initial light curve showing the clear ellipsoidal variations. The section of the light curve not shown is similar. Bottom: The same section of the light curve after pre-whitening the orbital variations, g~mode frequencies and low frequency artefacts. The amplitude modulation of the pulsational variations with the orbital period is striking and clear, as is the beating among the four principal mode frequencies. }
\label{fig:lc1}
\end{figure}  

The top panel of Fig.\,\ref{fig:lc1} shows a section of the initial light curve (the full light curve is too compressed to see the details at publication scale) where the orbital variations are obvious; it can also be seen that this is a single-sided pulsator. The ellipsoidal light variations (ELV) caused by the tidal deformation of the primary star are obvious with a typical double-wave; the alternating maxima are caused by Doppler boosting, as discussed in Section~\ref{sec:lcmodels} below. Of the two ELV minima, the dominant (pulsating) star is faintest when the line of sight is along the line of apsides and the ${\rm L}_1$ side of the primary is visible to the observer. The radial velocities discussed in Section~\ref{sec:cocam} above confirm this. The brighter ELV minimum is enhanced by irradiation effects on the fainter companion star (see Section~\ref{sec:lcmodels}) at the same time we are viewing the L$_3$ side of the primary. 

The light curve in the top panel also shows clearly that the pulsation amplitudes are greatest at the orbital phase of the deeper ELV minimum (which we define below as orbital phase zero -- photometric minimum; this is the same as the superior conjunction phase derived from the radial velocities, as seen in Fig.\,\ref{fig:RV}), and that the pulsation amplitudes are much smaller at orbital phase 0.5 when we are looking at the opposite side of the primary star.

As we will see, the secondary does not contribute significant light to the system (see Section~\ref{sec:lcmodels} below), so our light curve discussion is only about the primary star. The orbital photometric variations along with some g-mode frequencies and a few low-frequency instrumental artefacts were removed from the data with a high-pass filter (in practice, automated sequential pre-whitening of sinusoids in the frequency range $0-6$\,d$^{-1}$ until the background noise level was reached at low frequency). The resulting light curve of the residuals to that process is shown in the bottom panel of Fig.\,\ref{fig:lc1} where the concentration of pulsation amplitude in one hemisphere -- the ${\rm L}_1$ side of the star -- is evident. It is also evident from the light curve of the high-pass filtered data that the star is multi-periodic, as is shown by the clear beating of the pulsations. 

\subsection{The orbital frequency}
\label{sec:orbfreq}

We claimed in Section\,\ref{sec:intro} that CO~Cam is an oblique pulsator with the pulsation axis aligned with the tidal axis -- the line of apsides. To prove this assertion, we need a precise determination of the orbital frequency. We obtained that using radial velocity measurements over a span of more than a century in Section~\ref{sec:cocam} above, but it is important that the analysis of the {\em TESS} light curve be independent, since it is possible for there to be small orbital period changes in that long time span. We want the orbital frequency purely from the photometry to prove the oblique pulsator model, and for independent comparison with the radial velocity analysis in Section~\ref{sec:cocam}. 

\begin{figure}
\centering
\includegraphics[width=1.0\linewidth,angle=0]{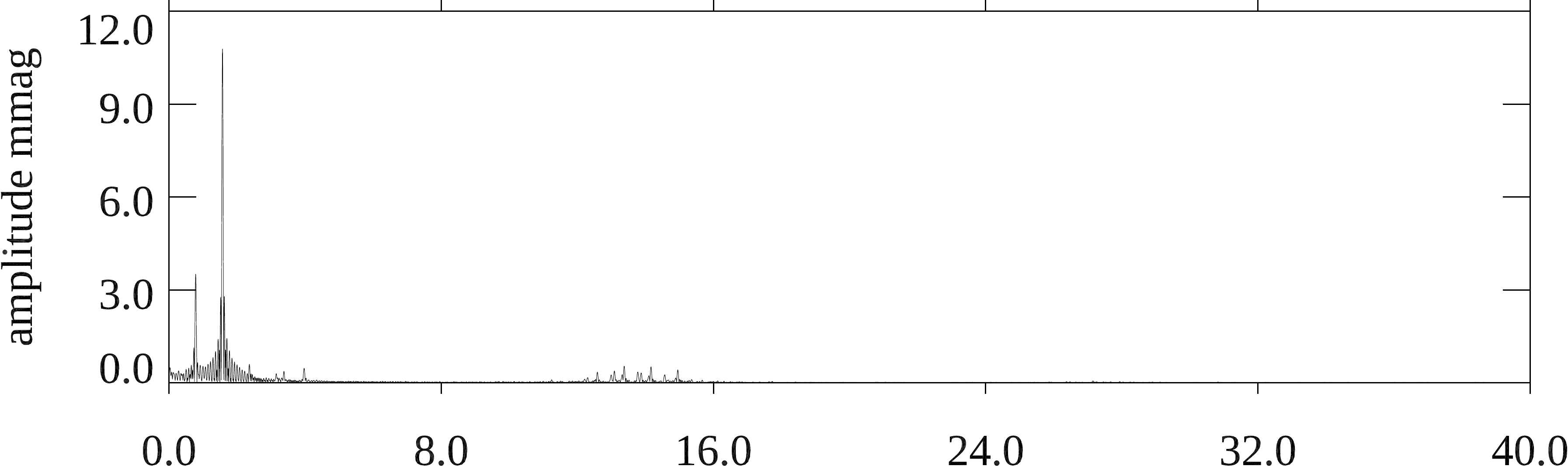} \vglue0.1cm
\includegraphics[width=1.0\linewidth,angle=0]{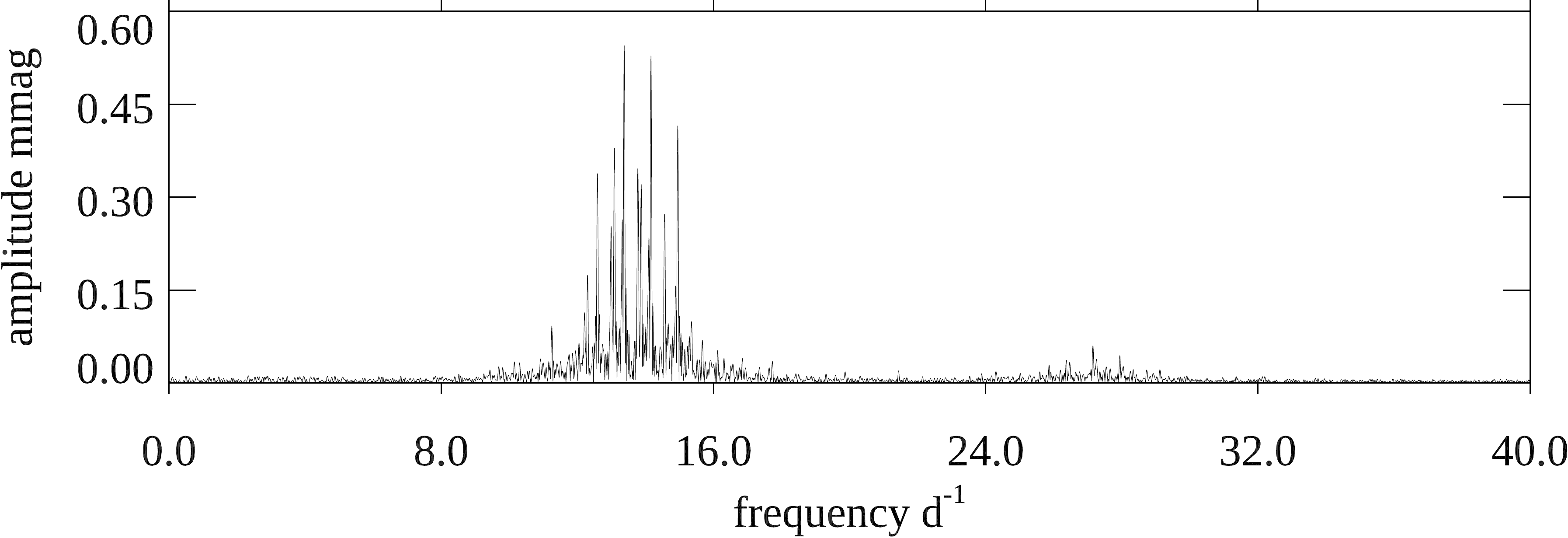}	
\caption{Top: The initial amplitude spectrum, where the highest peak is at $2\nu_{\rm orb}$; the orbital frequency, $\nu_{\rm orb} = 0.786832 \pm 0.000010$\,d$^{-1}$, was derived from a harmonic series of half of the frequency of the highest peak, $2\nu_{\rm orb}$. Second panel: The amplitude spectrum of the residuals after pre-whitening all low frequency variations -- orbital, g~modes, and artefacts. Pulsation multiplets and their harmonics and combinations are visible. Note the change in ordinate scale.}
\label{fig:ft1}
\end{figure}

The {\em TESS} data were analysed using a Discrete Fourier Transform \citep{1985MNRAS.213..773K} to produce amplitude spectra. The top panel in Fig.\,\ref{fig:ft1} shows the amplitude spectrum for the original data, where the orbital variations dominate. The bottom panel shows the amplitude spectrum after a high-pass filter has removed the low frequency variance, including the orbital variations, some g~modes, and instrumental artefacts. By inspection of the bottom panel it can be seen that there are multiplets split by the orbital frequency, and that the star is multi-periodic. It can also be seen that there are harmonics and/or combination frequencies of at least some of the p-mode frequencies. 

In this section we use data for which we have removed some low frequency g-mode peaks and some instrumental artefacts. It is obvious from the top panel of Fig.\,\ref{fig:ft1} that the highest peak is the second harmonic of the orbital frequency. We therefore fitted a harmonic series based on half that frequency by least-squares and nonlinear least-squares. The derived orbital frequency from this process yielded a value of $\nu_{\rm orb} = 0.786832 \pm 0.000010$\,d$^{-1}$, or an orbital period of $P_{\rm orb} = 1.270920 \pm 0.000015$\,d$^{-1}$. Our orbital frequency derived from 26.85\,d of {\em TESS} data agrees within 5$\sigma$ with the much higher precision value given in Table\,\ref{tbl:mags} derived from over a century of radial velocity measurements. We do not consider this to be a discrepancy, given the low-frequency {\em TESS} instrumental noise in the vicinity of the orbital frequency, and the possibility of some change in the orbital period over the century that this star has been studied. 

\subsection{The p-mode pulsation frequencies}
\label{sec:pmodefreqs}

To study the pulsations, we used the high-pass filtered data. The few g-mode frequencies that were found before this filter was applied are discussed in Section~\ref{sec:gmodefreqs}. The bottom panel  of Fig.\,\ref{fig:lc1} shows the light curve after the high pass filter has removed the orbital variations and other low frequencies; the bottom panel of Fig.\,\ref{fig:ft1} shows the amplitude spectrum for the residuals after the high-pass filtering. It is clear that the pulsation amplitudes are high when the L$_1$ side of the binary is towards the observer, and very low when the other side is most visible. This is the signature of a single-sided pulsator. 

Our first frequency analysis of these high-pass data used sequential pre-whitening. This involved finding the highest amplitude frequency in the data, then fitting it, and all previously determined frequencies, by least-squares and then nonlinear least-squares. During this process we checked for any significant changes in the amplitudes and phases of previously determined frequencies when a new frequency was added to the list. That is a useful check to see if the window functions of any of the frequencies are not fully resolved from each other. We then searched for patterns in the extracted frequencies and found that there are multiplets split, within the errors, by exactly the orbital frequency, and that the time of pulsation maximum is the time of passage of the ${\rm L}_1$ point across the line of sight. This is the signature of oblique pulsation. Knowing this from our first frequency analysis then allowed us to simplify the pre-whitening and the illustration of the pulsation frequencies in the description that follows. 

\begin{figure}
\centering
\includegraphics[width=1.0\linewidth,angle=0]{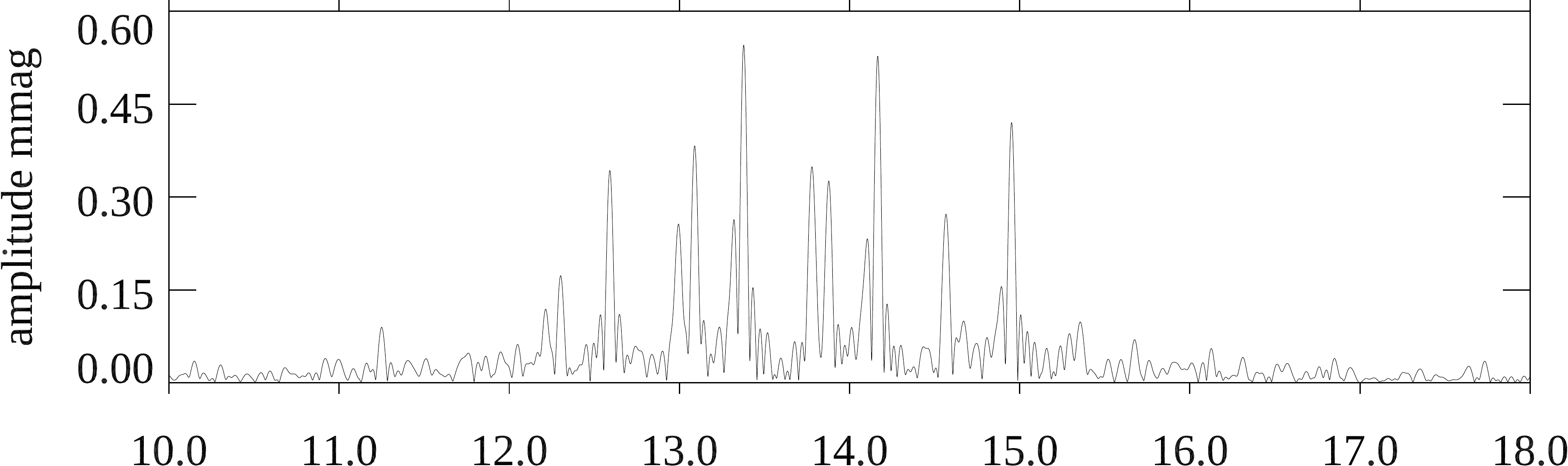}	
\includegraphics[width=1.0\linewidth,angle=0]{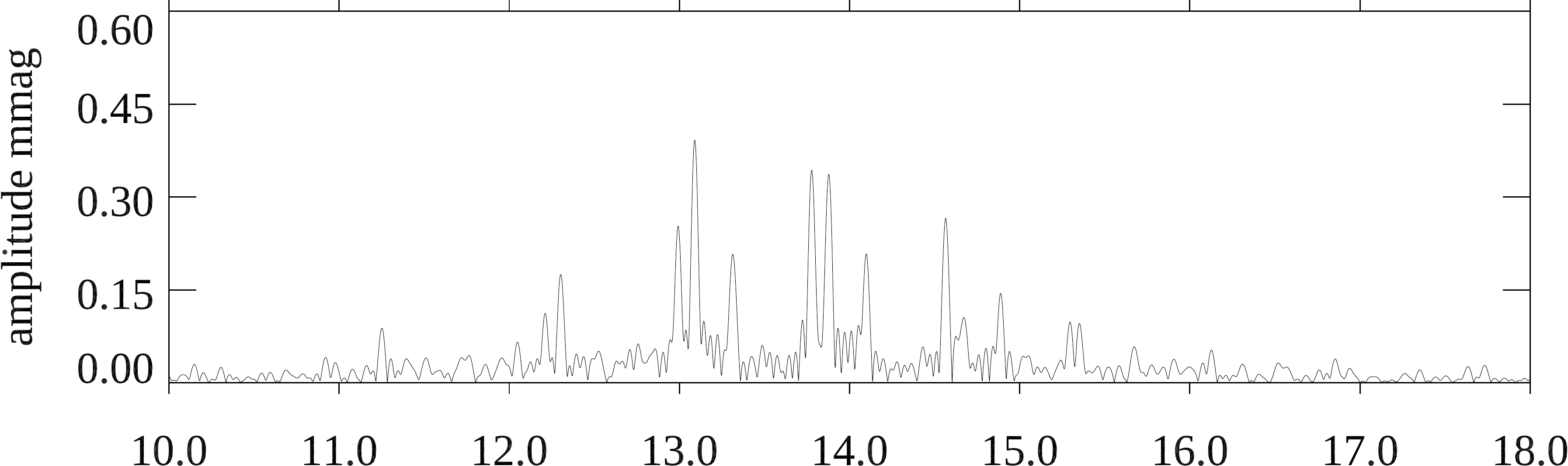}	
\includegraphics[width=1.0\linewidth,angle=0]{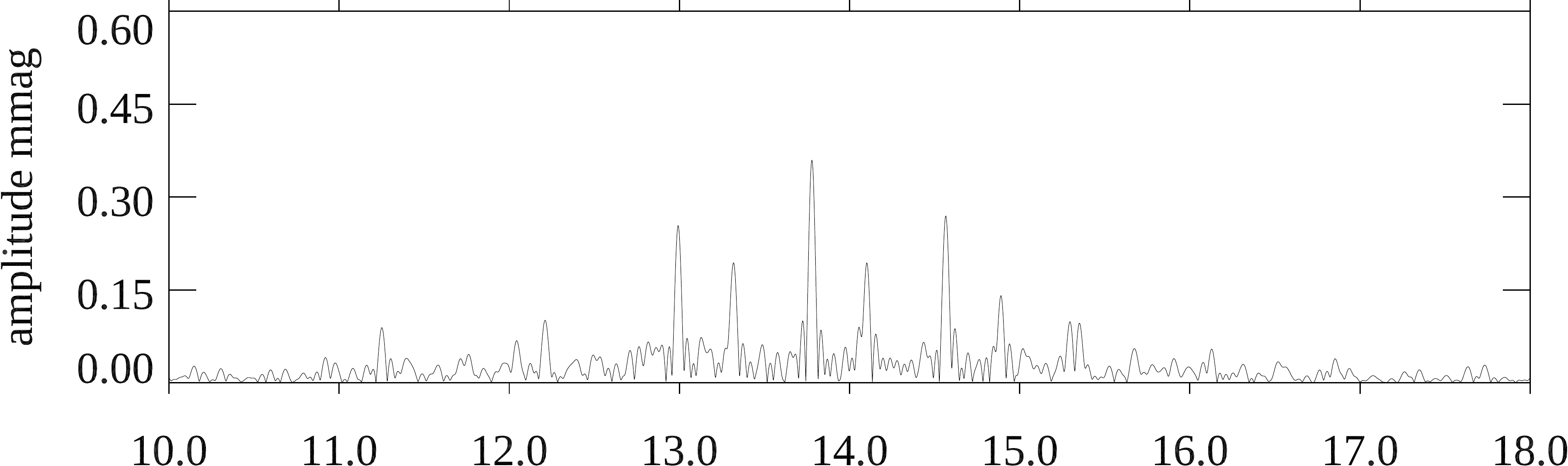}	
\includegraphics[width=1.0\linewidth,angle=0]{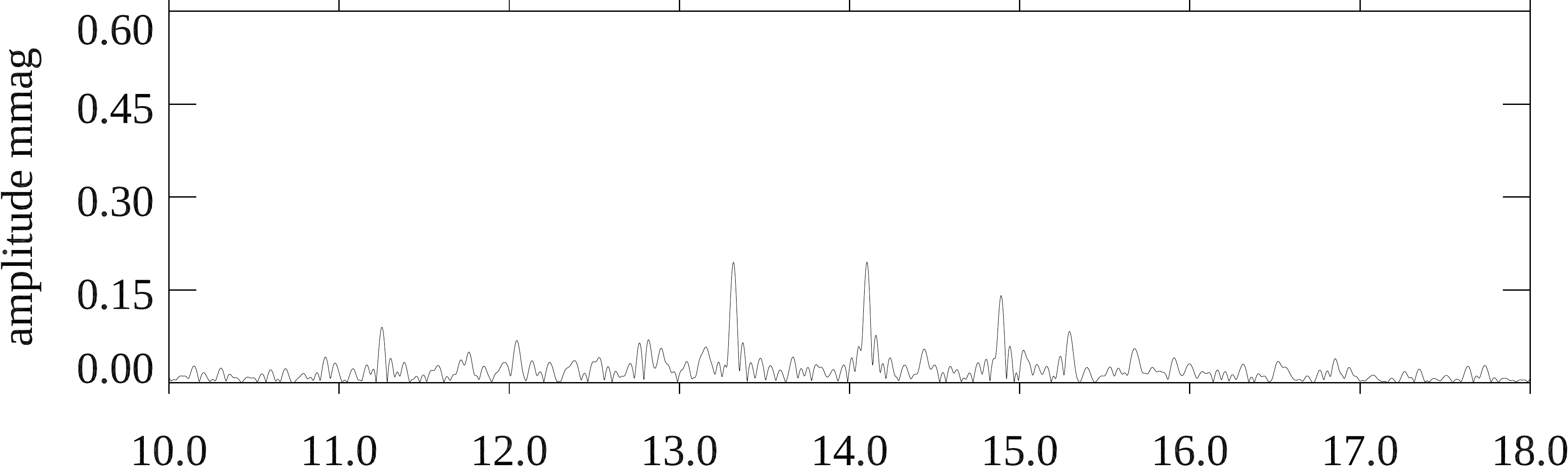}	
\includegraphics[width=1.0\linewidth,angle=0]{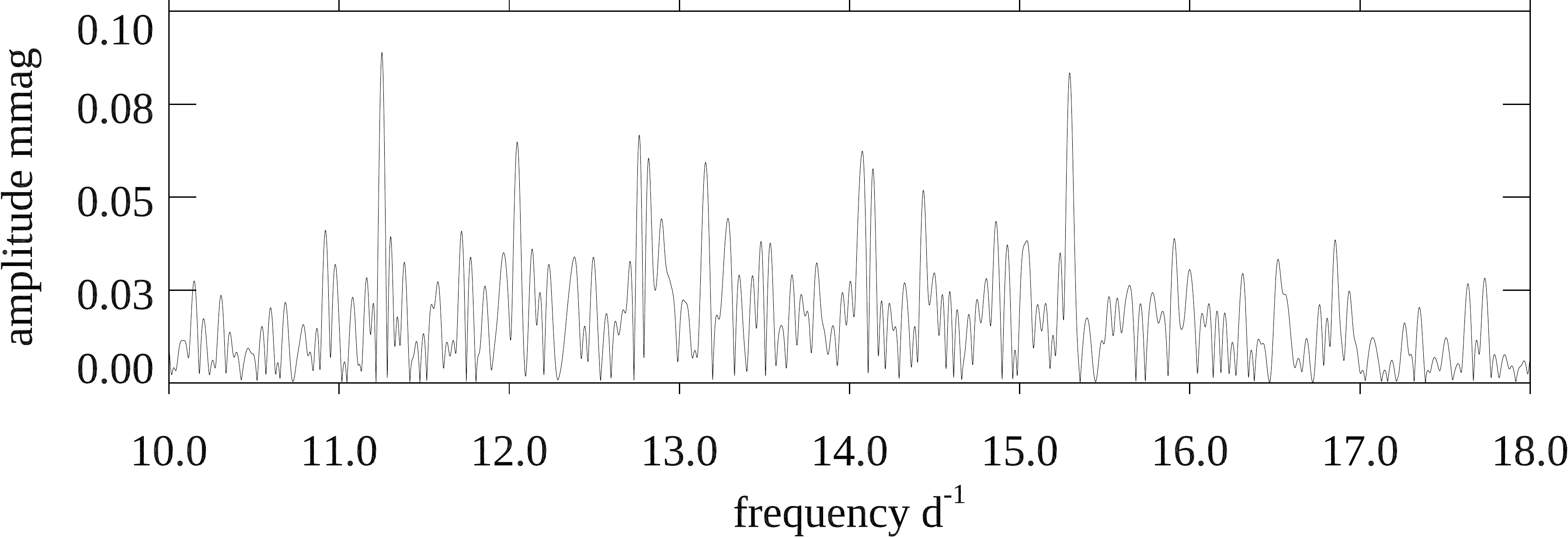}	
\caption{Top: The amplitude spectrum of the high-pass data, showing the pulsation frequencies and their orbital sidelobes. Sequential pre-whitening of multiplets based on the highest peaks is shown in the 5 panels. The last of those, the bottom panel, shows that there is significant amplitude left after pre-whitening the highest amplitude four multiplets.  The two highest peaks, $\nu_5 = 11.25135$\,d$^{-1}$ and $\nu_6 = 15.29294$\,d$^{-1}$, show some orbital sidelobes, but these are too close to the noise level to find the full multiplet, thus we stopped the analysis with four secure oblique pulsator multiplets. Note the change of ordinate scale in the bottom panel. }
\label{fig:ft2}
\end{figure}

The top panel of Fig.\,\ref{fig:ft2} shows a high-resolution view of the amplitude spectrum of the pulsation frequencies in the high-pass filtered data. There are four multiplets, each split by the orbital frequency; we know this from our first peak-by-peak frequency analysis. Each of these multiplets is generated by the amplitude and phase modulation of a single pulsation mode frequency.  The central peak of the multiplet is the pulsation frequency and the sidelobes describe the amplitude and phase modulation of the pulsation with the rotation of the mode with respect to our line of sight as the two stars orbit each other. This is a valuable opportunity that oblique pulsators provide: we are able to view the pulsation mode from varying aspects, thus giving important information about the mode geometry. No other type of pulsator provides this detailed view of the surface geometry of the pulsation mode. 

The pulsation axis is the tidal axis. That is a guide to understanding this frequency analysis. To prove this contention, we have extracted the pulsation mode frequencies and found the amplitudes and phases of all components of the multiplets. The method that we used was to identify the highest amplitude mode frequency, fit that frequency and its $\pm3$ orbital sidelobes, first by least-squares, then nonlinear least-squares. The frequencies of that multiplet were then pre-whitened from the data and the amplitude spectrum of the residuals was inspected; the highest peak was then selected as a mode frequency and its multiplet, plus all previous multiplets, were fitted simultaneously to the high-pass data. Importantly, the members of the multiplet were split by exactly the orbital frequency, as determined in Section\,\ref{sec:orbfreq} above. That equal splitting is necessary to exploit the information in the pulsation phases. 

We identified the highest peak in the top panel of Fig.\,\ref{fig:ft2} as a mode frequency, $\nu_1$, but the first orbital sidelobe of this frequency, $\nu_1 + \nu_{\rm orb}$, differs in amplitude by less than the highest noise peaks, hence there is some ambiguity. A least-squares fit of a frequency septuplet centred on $\nu_1$ shows four central larger amplitude peaks separated by $\nu_{\rm orb}$ with the outlying members of the multiplet of much lower amplitude, but formally significant. There is no obvious symmetry that helps to choose between $\nu_1$ and $\nu_1 + \nu_{\rm orb}$ as the mode frequency; we chose the highest peak as the most probable. 

Pre-whitening by the frequency septuplet for $\nu_1$ found in this manner gave the amplitude spectrum of the residuals in panel 2 of Fig.\,\ref{fig:ft2}. The highest peak here, $\nu_2$, has the same problem as we found with $\nu_1$: its $\nu_2 + \nu_{\rm orb}$ is almost as high in amplitude as $\nu_2$. Again, we selected the highest peak as the mode frequency and fitted a frequency septuplet split by $\nu_{\rm orb}$. 

\begin{figure}
\centering
\includegraphics[width=1.0\linewidth,angle=0]{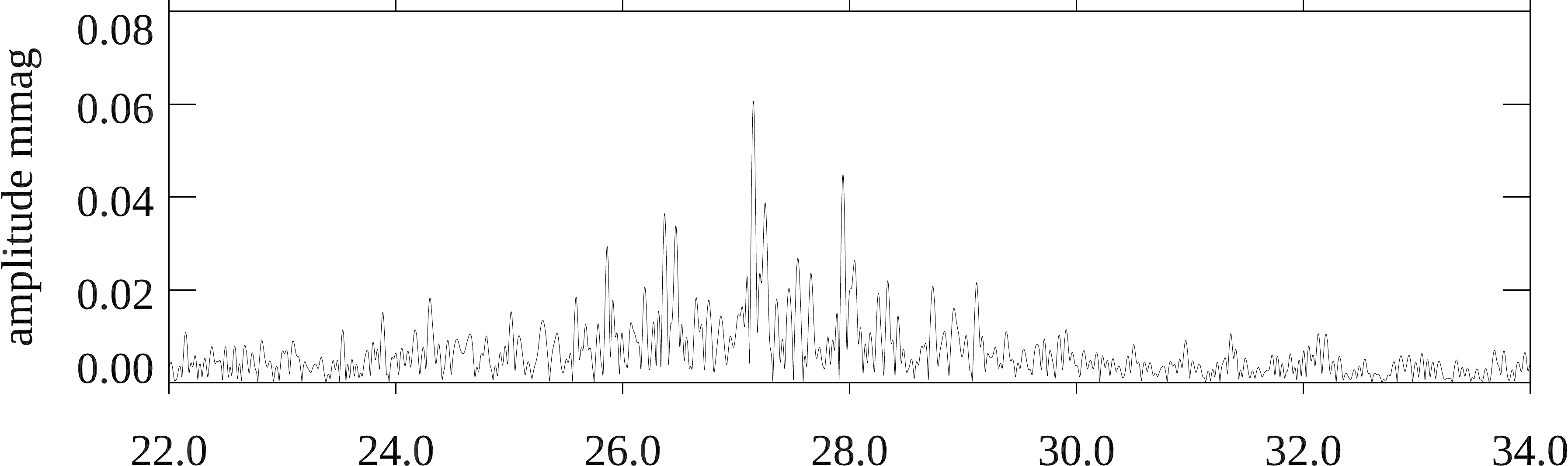}	\vglue0.1cm
\includegraphics[width=1.0\linewidth,angle=0]{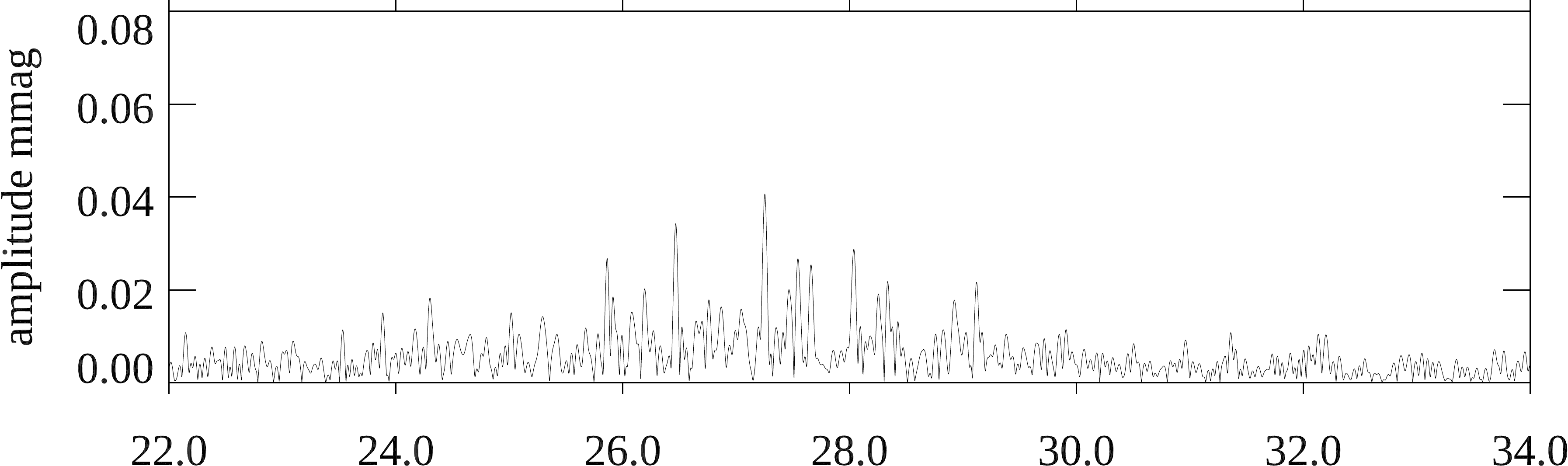}	\vglue0.1cm
\includegraphics[width=1.0\linewidth,angle=0]{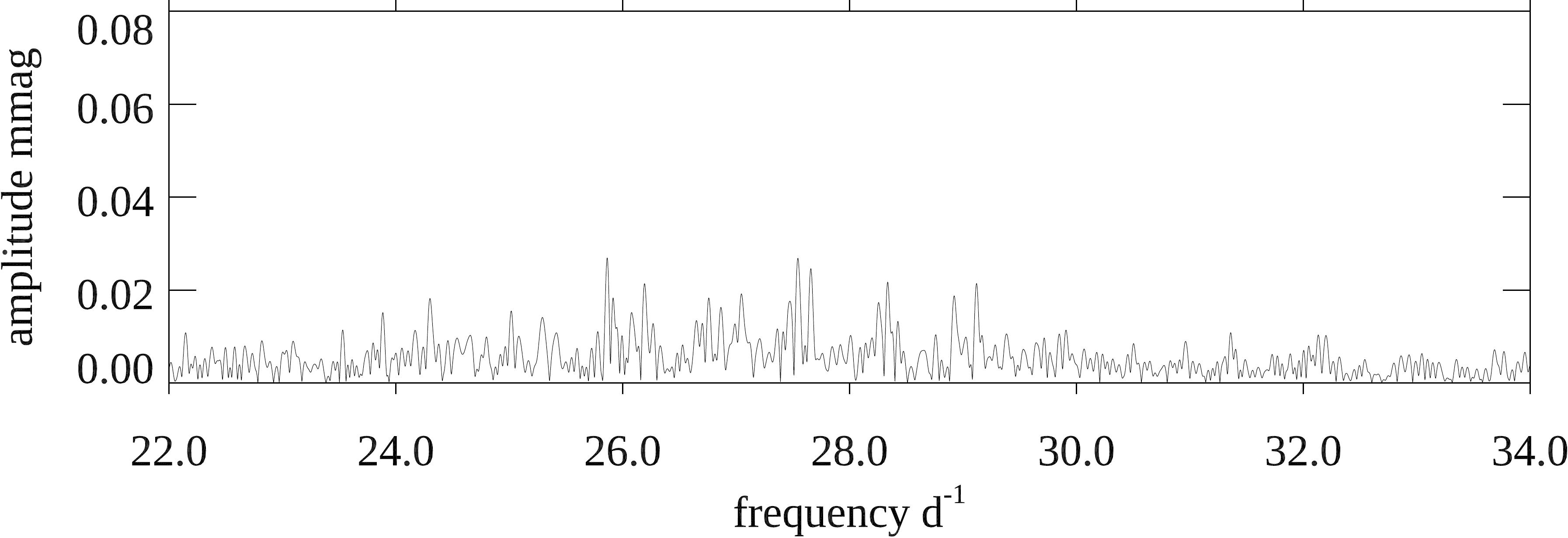}	
\caption{Top: The amplitude spectrum of the high-pass data, showing the frequencies and their orbital sidelobes in the frequency range of the second harmonics and combination frequencies. The highest peak is a combination term of $\nu_1 + \nu_2$, hence validates the choice of those frequencies as mode frequencies. In the middle panel the highest peak is not any combination or harmonic. Following pre-whitening of a multiplet for that frequency the bottom panel shows further significant amplitude, but not at any combination or harmonic frequencies. We did not pursue these harmonic and combination frequencies further. }
\label{fig:ft3}
\end{figure}

Panel 3 of Fig.\,\ref{fig:ft2} shows the amplitude spectrum of the new residuals, where $\nu_3$ appears unambiguous. Both least-squares and nonlinear least-squares fitting give a symmetrical multiplet for $\nu_3$. We can  now determine that we did select the correct frequency for $\nu_2$. The top panel of Fig.\,\ref{fig:ft3}, in the vicinity of the second harmonics of the four pulsation modes, shows the highest peak there to be a combination term at a frequency of 27.15204\,d$^{-1}$; that is, to within the frequency resolution of the data set of 0.037\,d$^{-1}$, equal to $\nu_1 + \nu_3 = 27.15521$\,d$^{-1}$. This supports the identification of $\nu_1$ as the mode frequency, since if we had chosen the higher frequency sidelobe for $\nu_1$ as the mode frequency, this combination term would not be $\nu_1 + \nu_3$.

Following pre-whitening of the frequencies $\nu_1$, $\nu_2$ and $\nu_3$ along with their orbital sidelobes, panel 4 of Fig.\,\ref{fig:ft2} shows another ambiguous peak. Following the precedent above, we selected the highest peak -- the central frequency of the three obvious members of the multiplet. The bottom, fifth panel, of Fig.\,\ref{fig:ft2} shows the amplitude spectrum of the residuals to the four mode frequencies plus orbital sidelobes fit. There are two more frequencies at $\nu_5 = 11.25135$\,d$^{-1}$ and $\nu_6 = 15.29294$\,d$^{-1}$ that show some orbital sidelobes, but we did not make further fits of septuplets for these frequencies because many of the orbital sidelobes are too close to the noise level.  

\subsection{Testing the oblique pulsator model for CO~Cam} 

When fitting the frequency septuplets, we fitted the mode frequencies simultaneously by nonlinear least-squares to optimise the frequency determinations, but we forced the orbital sidelobes to each of the four mode frequencies to be split by exactly the orbital frequency, which we also assume is the rotational frequency, since close binaries with orbital periods near 1.27\,d are normally synchronous rotators. This latter assumption is not critical to our arguments concerning oblique pulsation. That the orbital sidelobes were sequentially pre-whitened, as can be seen in the steps shown in Fig.\,\ref{fig:ft2}, demonstrates that the frequency splitting of the multiplets is equal to the orbital frequency within the frequency resolution of the data set. 

However, if it were instead conjectured that the multiplets are rotational septuplets for $\ell =3$ octupole modes, then the frequency splitting would be given by the standard rotational splitting formula
$\nu_{3,m} = \nu_{3,0} \pm m(1-C_{n,\ell}) \nu_{\rm rot}$. For low radial overtone modes, such as these seen in CO~Cam, the value of the Ledoux constant, $C_{n,\ell}$, is typically a few per cent. In that case the beat frequency among the members of the multiplet is slightly less than the orbital frequency, $\nu_{\rm orb}$, and the pulsation pattern precesses with respect to the line of apsides. We show in the next section that pulsation maximum for all four mode frequencies occurs when the ${\rm L}_1$ side of the star transits the line of sight. This would be pure coincidence for rotationally split mode frequencies, hence arguing strongly that the modes are obliquely pulsating along the tidal axis. 

We can extend this argument. For obliquely pulsating normal modes the pulsation phases of all of the members of the septuplet are equal at the time of pulsation maximum, and the amplitudes of the multiplet members have a symmetry about the mode frequency. However, CO~Cam is not a perfect sphere; it is tidally distorted. The modes are also distorted and are not simply obliquely pulsating normal modes. 

Let us look at the phases. We have chosen a zero point in time for the least-squares fitting of the four septuplets so that the pulsation phases of the first orbital sidelobes for the highest amplitude mode frequency, $\nu_1 = 13.37815$\,d$^{-1}$, are forced to be equal. The results are shown in Table~\ref{table:1} where it can be seen that the phases for $\nu_1$ and its first orbital sidelobes are essentially equal. The same is true for $\nu_4$, but less so for the other septuplets. However, it must be remembered that these pulsation modes are tidally distorted, hence are not simply described by a single spherical harmonic, as would be the case in a spherically symmetric star. This phase argument is to first order. Nevertheless, the zero point of the time scale chosen to force those first sidelobes to have equal phase for $\nu_1$ must be the time of pulsation maximum, and the time of orbital minimum light when the ${\rm L}_1$ point crosses the line of sight. Those are requirements of the oblique pulsator model. In the next section we further test to see if that is the case. 

We note that the Doppler shift of the pulsation frequency by the orbital motion also generates a frequency multiplet split by exactly the orbital frequency \citep{2012MNRAS.422..738S}. Applying Eqn. (21) of \citet{2012MNRAS.422..738S} yields a value of their parameter $\alpha = 0.004$, where they show to first order that $\alpha = \frac{A_{+1} +A_{-1}}{A_0}$, where $A_{+1}$ and $A_{-1}$ are the amplitudes of the first sidelobes, and $A_0$ is the amplitude of the mode frequency. Thus for the highest amplitude mode frequency of CO~Cam, $\nu_1 = 13.37815$\,d$^{-1}$, its amplitude of $A_0 = 0.548$\,mmag then gives the expected amplitudes for the first sidelobes of $A_{+1} = A_{-1} = 0.001$\,mmag. This is much less than the root-mean-square uncertainty in amplitude, as given in Table~3, of 0.003\,mmag, hence the effect of frequency modulation caused by the orbital motion is negligible in this analysis and has no discernible impact on the observed frequency multiplets, which are caused by the amplitude and phase modulation of oblique pulsation.

A larger effect comes from the changing background light from the ellipsoidal variability, which can be seen in Fig.\,8 to have a semi-amplitude of just over 1~per~cent. Thus, for a constant pulsation amplitude a multiplet is generated over the orbital period that describes amplitude modulation by 1~per~cent. This is tiny compared to the observed amplitude modulation, but is detectable and does contribute a small amount to the amplitudes of the sidelobes in the frequency multiplets. It does not affect the interpretation that the multiplets are generated by oblique pulsation.

\begin{table}
\centering
\caption{A least squares fit of the frequency septuplets for $\nu_1$, $\nu_2$, $\nu_3$ and $\nu_4$.  The frequencies are named in order of decreasing amplitude, as that is how they were selected in the sequential pre-whitening process. The zero point for the phases, $t_0 = 2458697.16295$,  has been chosen to be a time when the two first orbital sidelobes of $\nu_2$ have equal phase. For consistency, and to see the relative amplitudes of the multiplet components, all 7 members of each multiplet were fitted and are shown, even though two of them have significance in amplitude of $3\sigma$, or less. } 
\begin{tabular}{rrrr}
\hline
&\multicolumn{1}{c}{frequency} & \multicolumn{1}{c}{amplitude} &   
\multicolumn{1}{c}{phase}  \\
&\multicolumn{1}{c}{d$^{-1}$} & \multicolumn{1}{c}{mmag} &   
\multicolumn{1}{c}{radians}   \\
& & \multicolumn{1}{c}{$\pm 0.003$} &   
   \\
\hline
 $\nu_1 - 3\nu_{\rm orb}$ &  11.01766  &  0.021  & $  1.490  \pm  0.146$ \\ 
 $\nu_1 - 2\nu_{\rm orb}$ &  11.80449  &  0.028  & $  -2.065  \pm  0.114$ \\ 
$\nu_1- \phantom{1}\nu_{\rm orb}$  &  12.59132  &  0.359  & $  -1.622  \pm  0.009$ \\ 
$\nu_1$  &  13.37815  &  0.548  & $  -1.483  \pm  0.006$ \\ 
$\nu_1 + \phantom{1}\nu_{\rm orb}$  &  14.16499  &  0.522  & $  -1.620  \pm  0.006$ \\ 
$\nu_1 + 2\nu_{\rm orb}$  &  14.95182  &  0.427  & $  -1.853  \pm  0.008$ \\ 
 $\nu_1 + 3\nu_{\rm orb}$ &  15.73865  &  0.024  & $  -2.457  \pm  0.131$ \\ 
\hline
  $\nu_2 - 3\nu_{\rm orb}$ &  10.72904  &  0.009  & $  2.034  \pm  0.357$ \\ 
  $\nu_2 - 2\nu_{\rm orb}$ &  11.51587  &  0.039  & $  -1.973  \pm  0.080$ \\ 
  $\nu_2 - \phantom{1}\nu_{\rm orb}$&  12.30270  &  0.177  & $  -0.616  \pm  0.018$ \\ 
$\nu_2$  &  13.08953  &  0.391  & $  -0.437  \pm  0.008$ \\ 
 $\nu_2 + \phantom{1}\nu_{\rm orb}$ &  13.87636  &  0.349  & $  -0.390  \pm  0.009$ \\ 
  $\nu_2 + 2\nu_{\rm orb}$ &  14.66320  &  0.113  & $  -0.510  \pm  0.028$ \\ 
  $\nu_2 + 3\nu_{\rm orb}$ &  15.45003  &  0.031  & $  -0.996  \pm  0.099$ \\ 
  \hline
 $\nu_3 - 3\nu_{\rm orb}$  &  11.41656  &  0.031  & $  0.145  \pm  0.099$ \\ 
 $\nu_3 - 2\nu_{\rm orb}$  &  12.20339  &  0.094  & $  0.519  \pm  0.033$ \\ 
 $\nu_3 - \phantom{1}\nu_{\rm orb}$  &  12.99022  &  0.256  & $  1.120  \pm  0.012$ \\ 
$\nu_3$  &  13.77706  &  0.363  & $  1.711  \pm  0.009$ \\ 
  $\nu_3 + \phantom{1}\nu_{\rm orb}$ &  14.56389  &  0.275  & $  1.975  \pm  0.011$ \\ 
  $\nu_3 + 2\nu_{\rm orb}$ &  15.35072  &  0.099  & $  1.746  \pm  0.031$ \\ 
 $\nu_3 + 3\nu_{\rm orb}$  &  16.13755  &  0.049  & $  2.572  \pm  0.064$ \\ 
\hline
  $\nu_4 - 3\nu_{\rm orb}$ &  11.74757  &  0.035  & $  1.492  \pm  0.090$ \\ 
  $\nu_4 - 2\nu_{\rm orb}$ &  12.53440  &  0.041  & $  2.987  \pm  0.079$ \\ 
  $\nu_4 - \phantom{1}\nu_{\rm orb}$ &  13.32123  &  0.192  & $  2.956  \pm  0.017$ \\ 
$\nu_4$  &  14.10806  &  0.186  & $  2.989  \pm  0.017$ \\ 
  $\nu_4 + \phantom{1}\nu_{\rm orb}$ &  14.89490  &  0.138  & $  3.048  \pm  0.023$ \\ 
 $\nu_4 + 2\nu_{\rm orb}$  &  15.68173  &  0.054  & $  2.706  \pm  0.059$ \\ 
  $\nu_4 + 3\nu_{\rm orb}$ &  16.46856  &  0.002  & $  0.127  \pm  1.926$ \\ 
 \hline
\hline
\end{tabular}
\label{table:1}
\end{table}

\section{The phase diagrams}

\begin{figure}
\centering
\includegraphics[width=1.0\linewidth,angle=0]{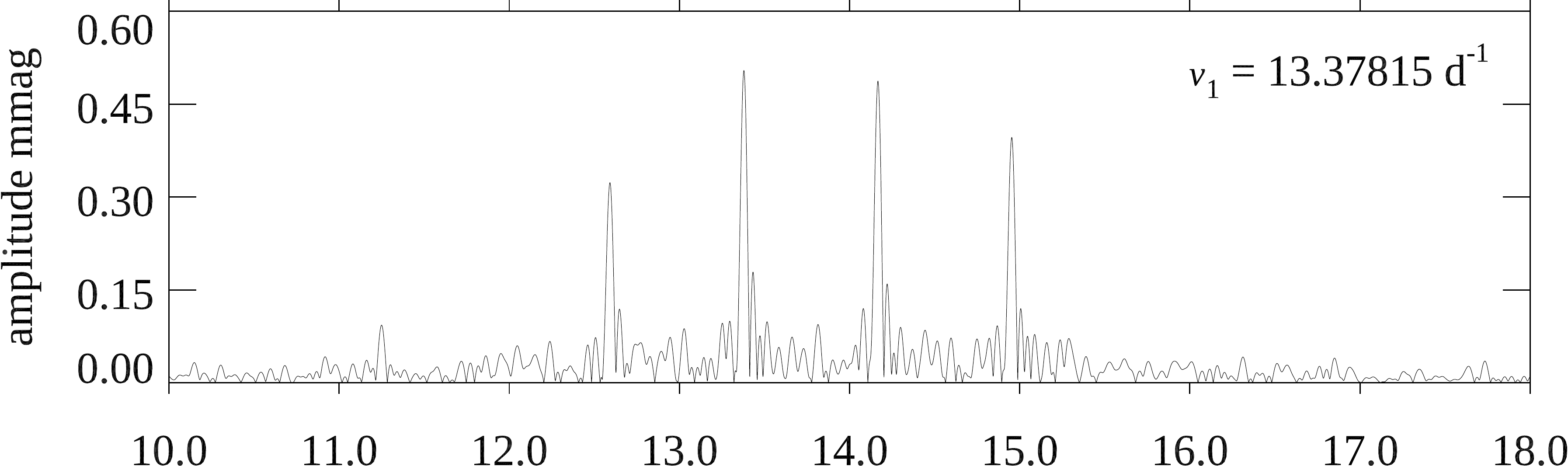}	
\includegraphics[width=1.0\linewidth,angle=0]{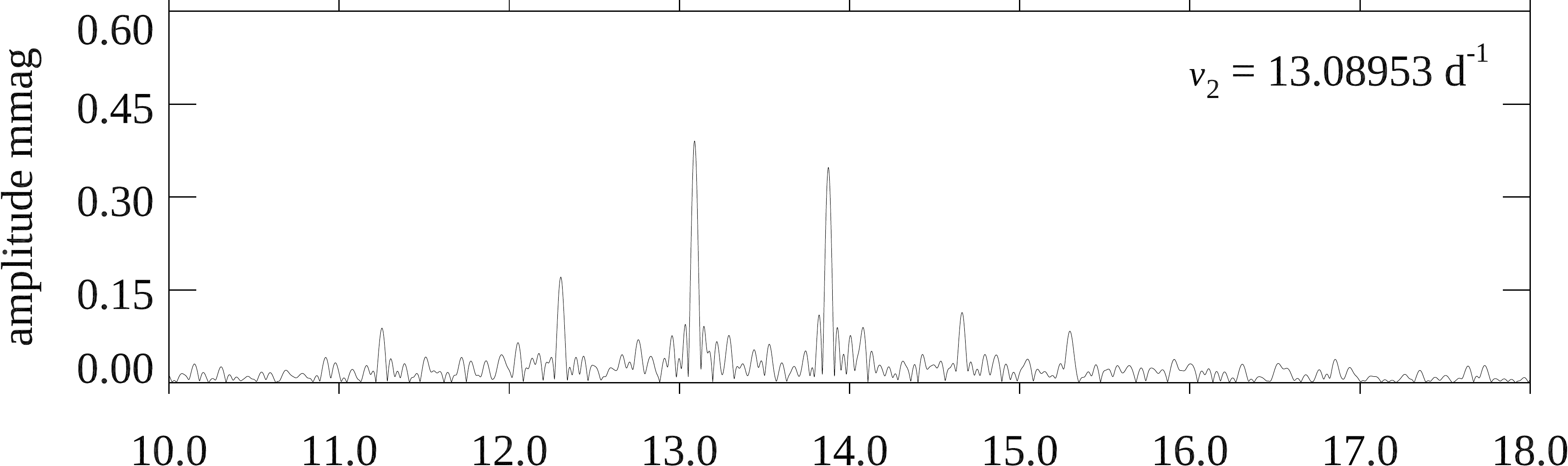}	
\includegraphics[width=1.0\linewidth,angle=0]{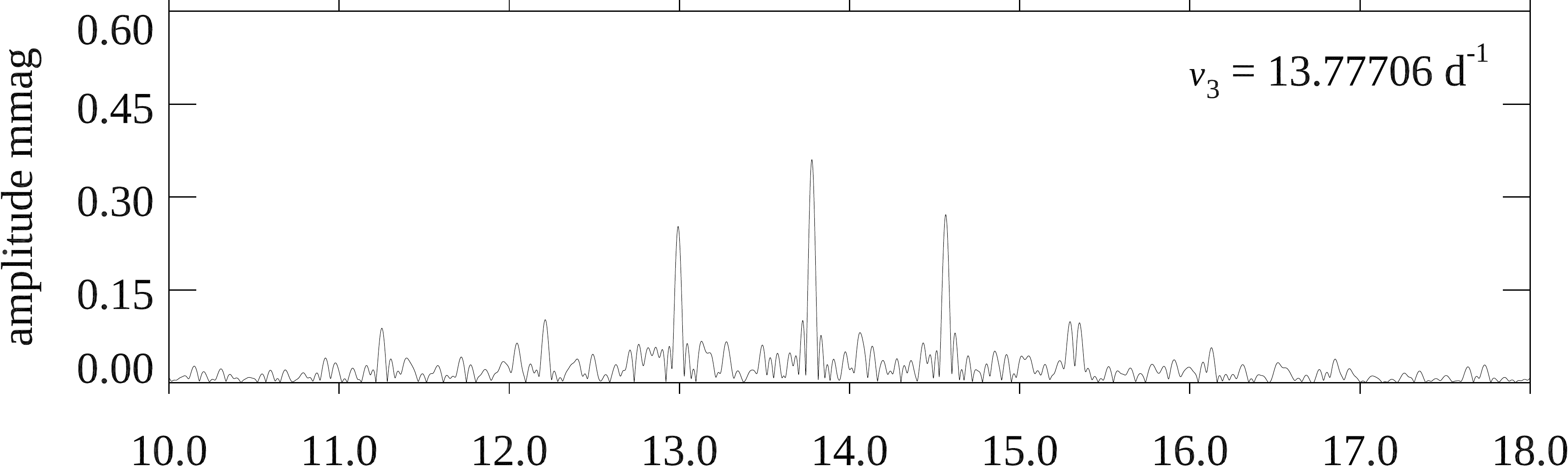}	
\includegraphics[width=1.0\linewidth,angle=0]{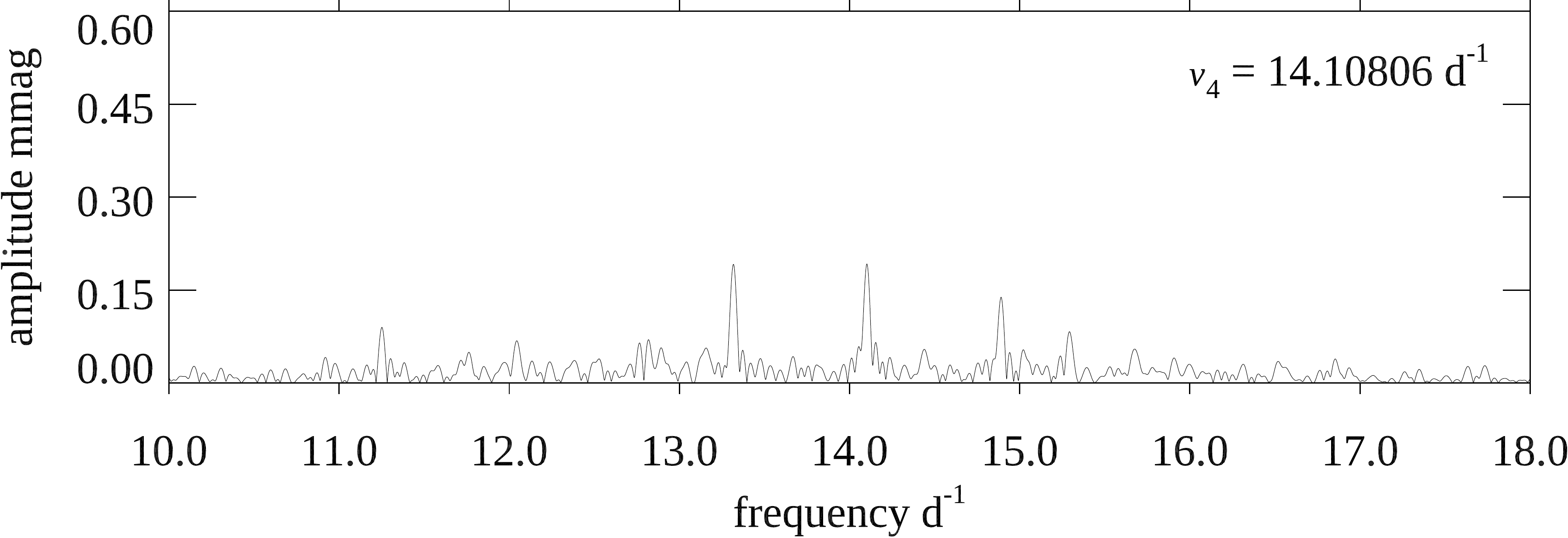}	
\caption{The four mode frequency septuplets isolated. For each of the septuplets some of the lower amplitude peaks are not visible, although the least-squares fits show that they are statistically significant in most cases; see Table~3. The low amplitudes of the $\pm 3\nu_{\rm rot}$ components indicates that quintuplets would be a good fit to most of these multiplets. We fitted septuplets for completeness. Pure spherical harmonic dipoles would give triplets, quadrupoles would give quintuplets, etc. The tidal distortion of the pulsation modes from simple spherical harmonics generates more multiplet components to describe the amplitude and phase variations over the orbit, as we observe the pulsation from varying aspect.}
\label{fig:ft4}
\end{figure}

The oblique pulsator model requires that pulsation maximum coincides with orbital light minimum for zonal modes ($m=0$) pulsating along the tidal axis. We first isolated each frequency multiplet by pre-whitening the other three from the high-pass data. Fig.\,\ref{fig:ft4} gives a graphical view of the amplitude spectra for each mode. These isolated data sets were then used to determine the pulsation amplitudes and phases for each of the four mode pulsation frequencies. 

Each of the mode frequencies was fitted to sections of the isolated data 0.3\,d long. This is about 1/4 of an orbital cycle, a compromise between orbital phase resolution (shorter sections) and better signal-to-noise ratio for the amplitudes and phases (longer sections). Within the oblique pulsator model the frequency septuplet is a consequence of amplitude and phase modulation as a function of orbital phase; there is only one mode frequency for each multiplet. We chose a zero point in time such that the phases of the first orbital sidelobes of the highest amplitude pulsation mode frequency were equal. That is, by choice of $t_0$ we set $\phi(\nu_1 - \nu_{\rm orb})$ = $\phi (\nu_1 + \nu_{\rm orb})$. 

\begin{figure}
\centering
\includegraphics[width=0.95\linewidth,angle=0]{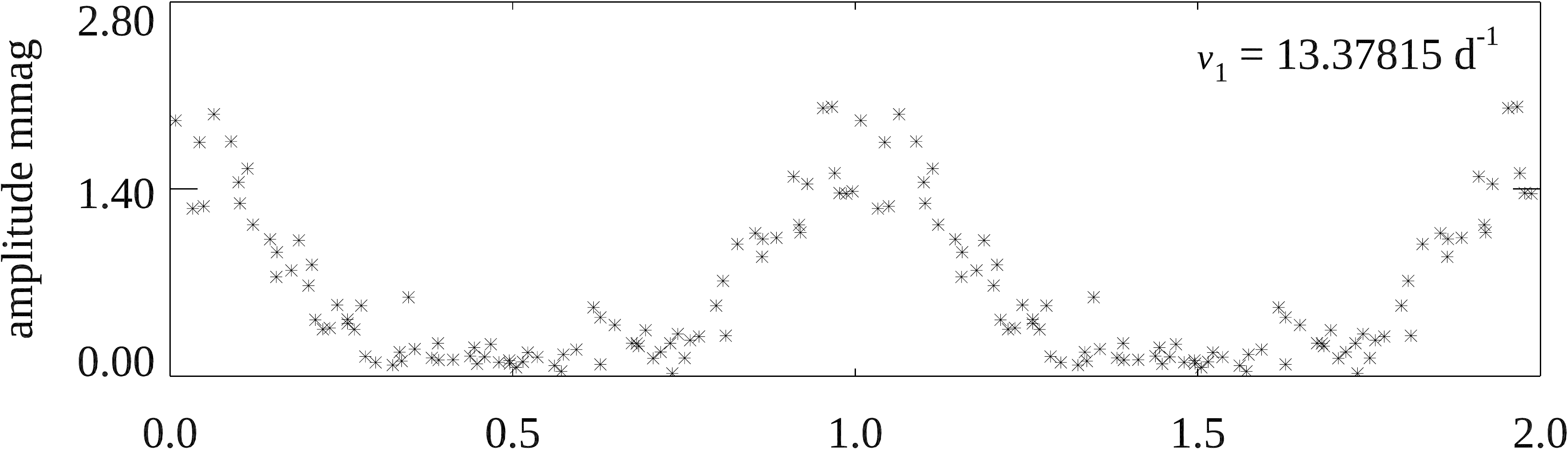}	 \vglue0.1cm
\includegraphics[width=0.95\linewidth,angle=0]{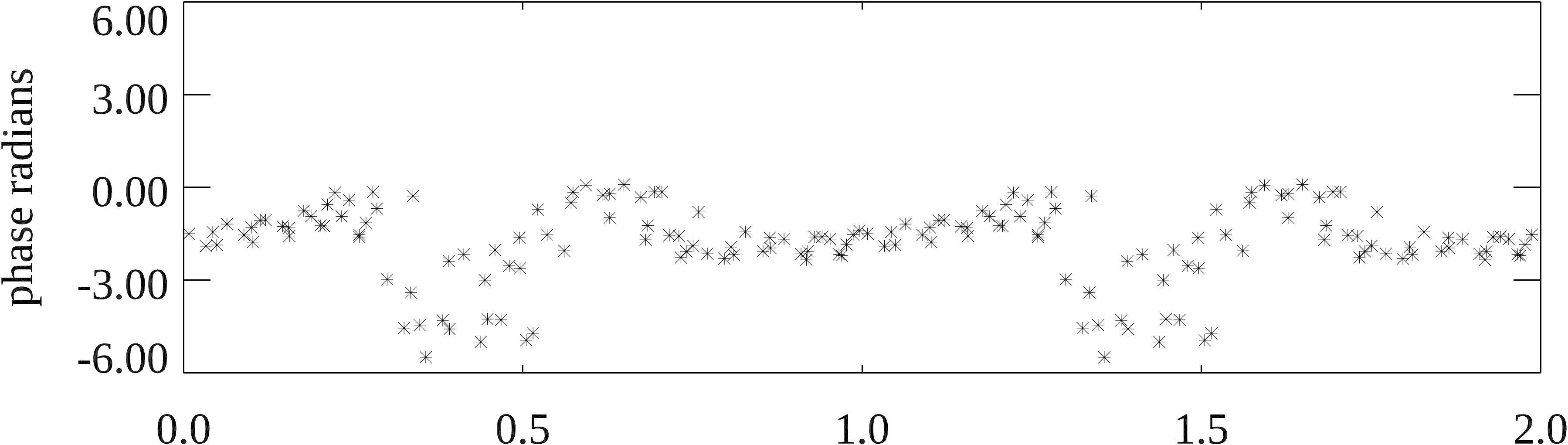}	 \vglue0.1cm	
\includegraphics[width=0.95\linewidth,angle=0]{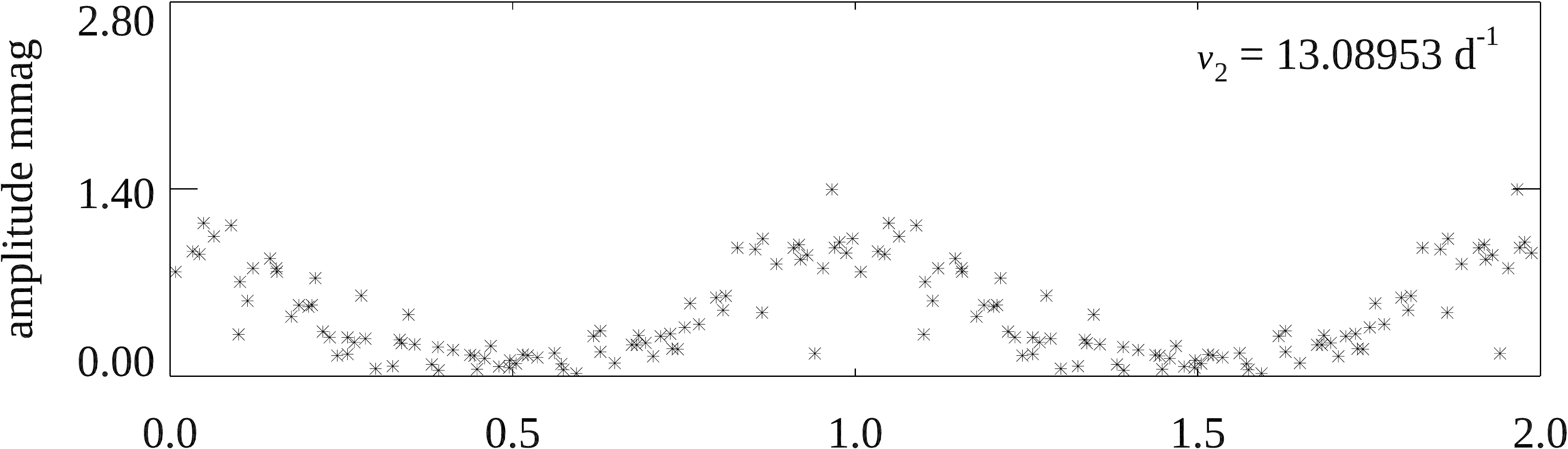}	 \vglue0.1cm
\includegraphics[width=0.95\linewidth,angle=0]{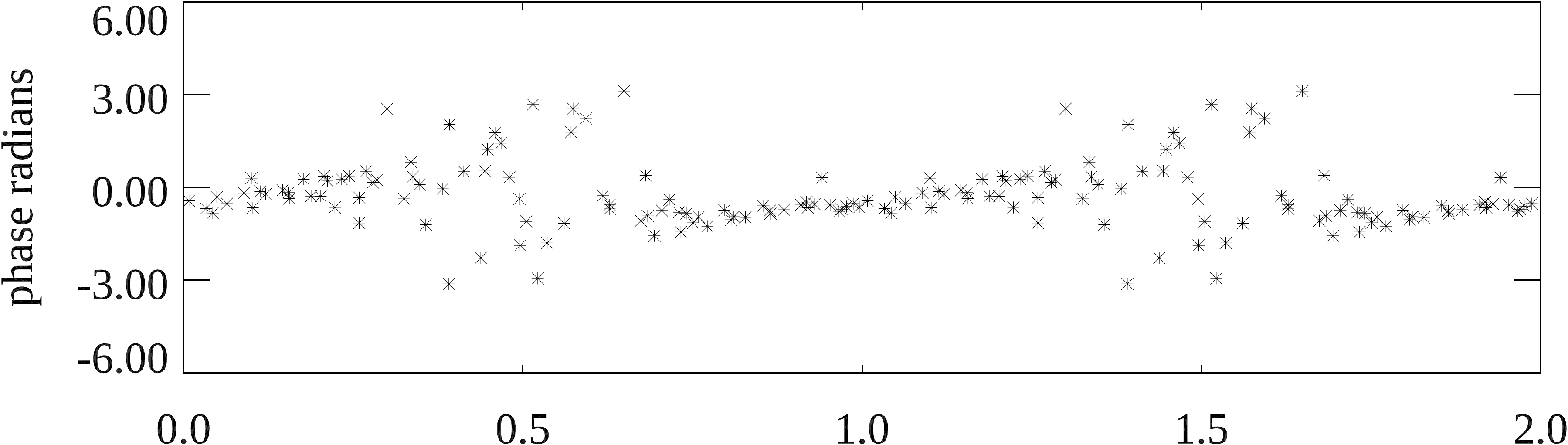}     \vglue0.1cm
\includegraphics[width=0.95\linewidth,angle=0]{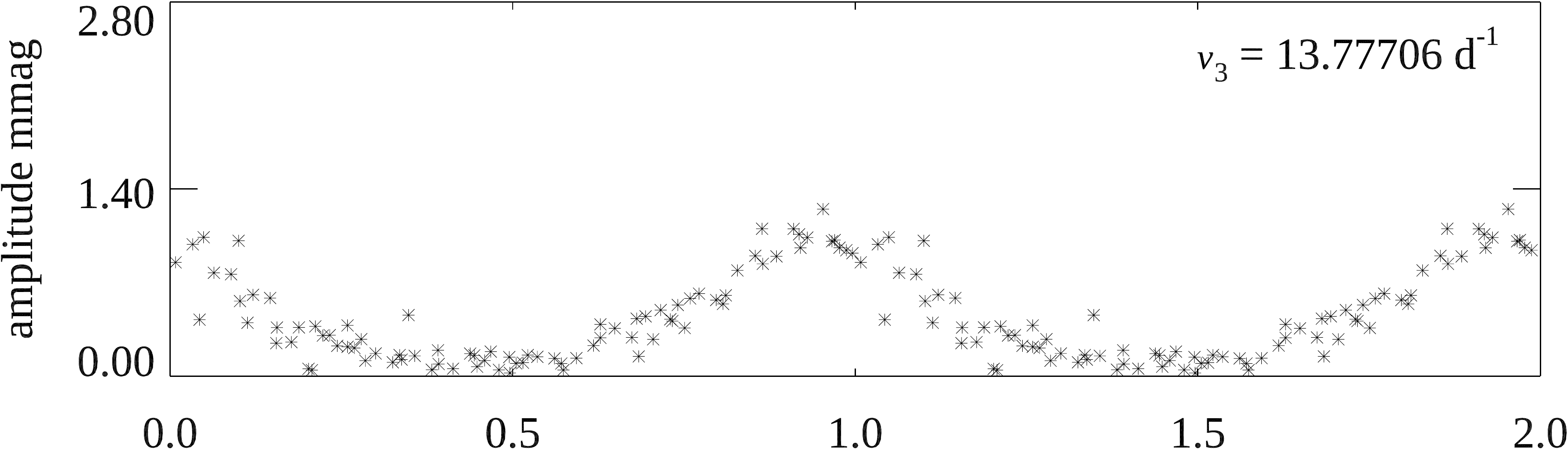}	 \vglue0.1cm
\includegraphics[width=0.95\linewidth,angle=0]{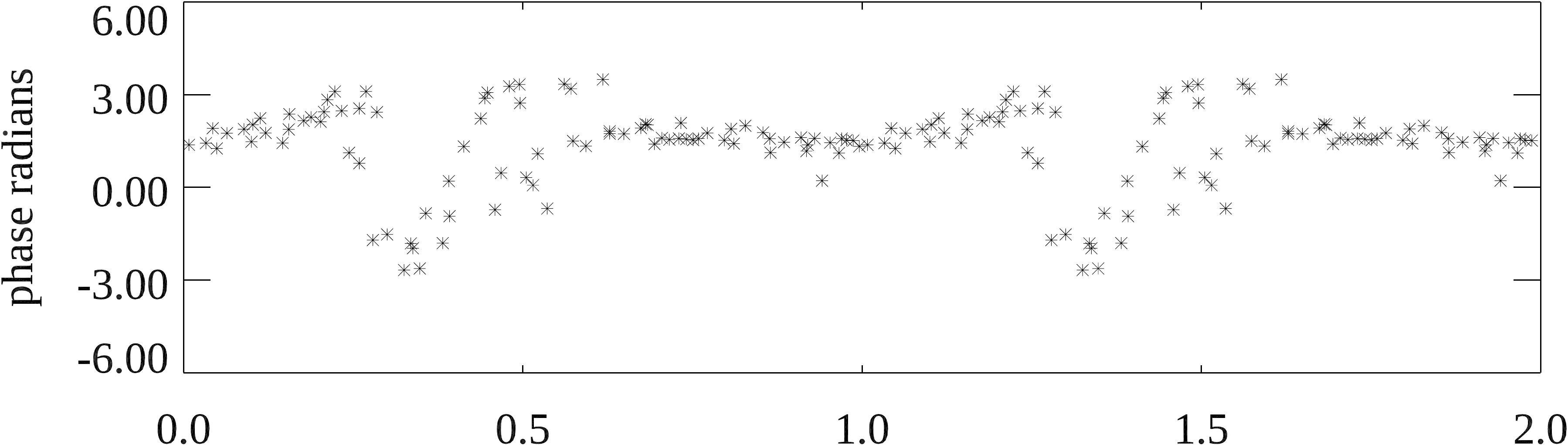}	 \vglue0.1cm
\includegraphics[width=0.95\linewidth,angle=0]{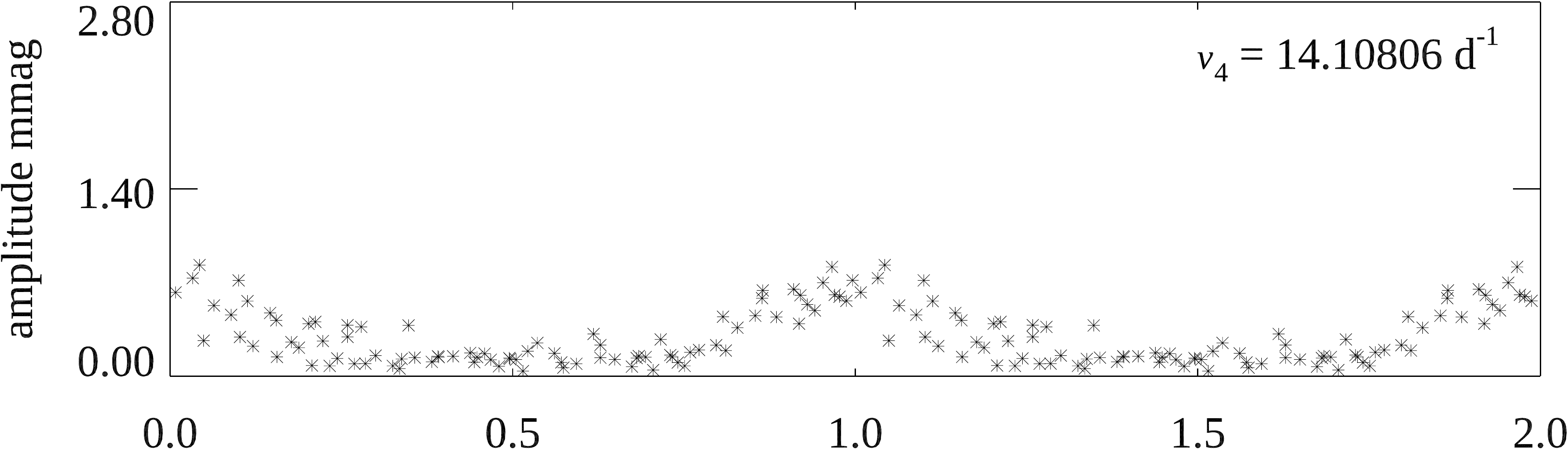}	 \vglue0.1cm
\includegraphics[width=0.95\linewidth,angle=0]{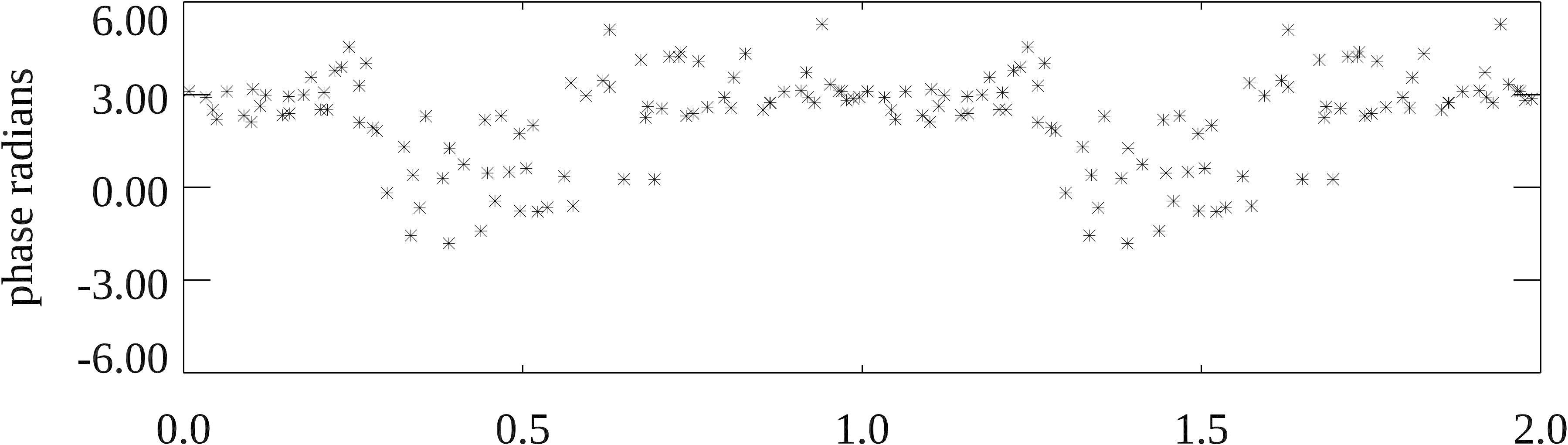}	 \vglue0.1cm
\includegraphics[width=0.95\linewidth,angle=0]{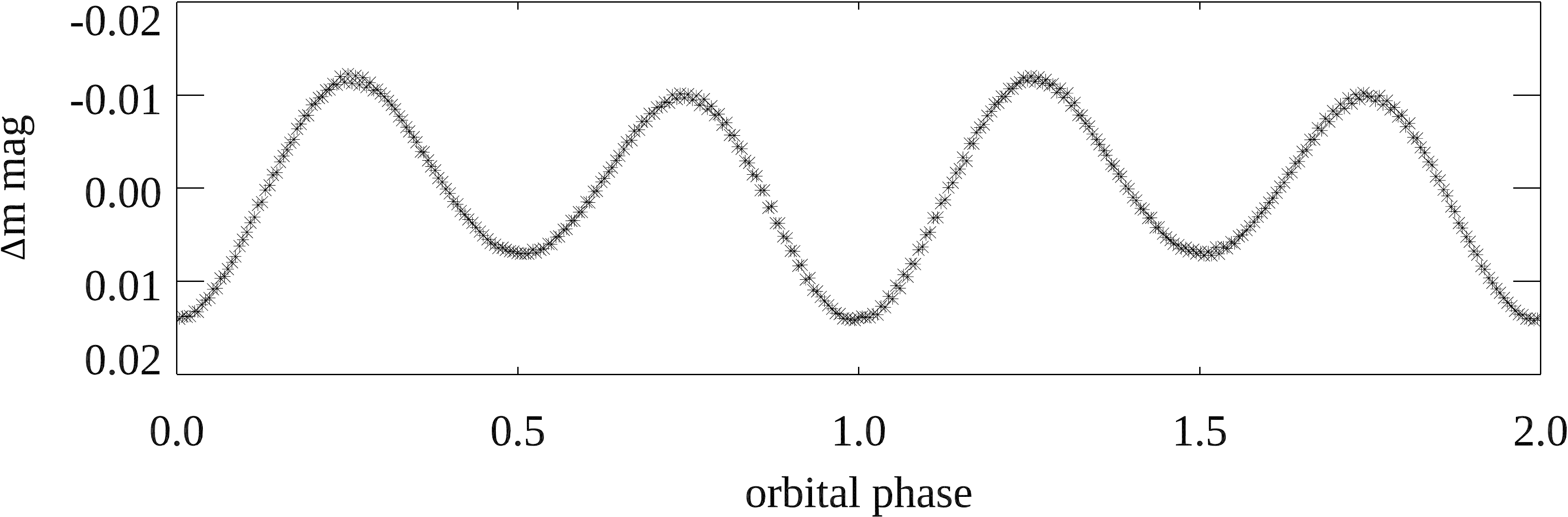}
\caption{Each pair of panels shows the pulsation amplitude and phase variation as a function of orbital phase for the isolated data (cf. Fig.\,\ref{fig:ft4}). The bottom panel shows the orbital light variations as a function of orbital phase. It can be seen that pulsation maxima and orbital light minimum coincide, as required by the oblique pulsator model.}
\label{fig:phamp}
\end{figure}

We plotted the pulsation amplitude and pulsation phase as a function of orbital phase for each of the four mode frequencies in Fig.\,\ref{fig:phamp}. We also plotted the orbital light variations with the larger amplitude pulsations removed and the data binned. It is clear from these figures that the pulsation amplitude maximum for each mode coincides with orbital minimum light when the ${\rm L}_1$ side of the primary star crosses the line of sight. This is the proof that these modes are pulsating along the tidal axis. Fig.\,\ref{fig:phamp} also shows clearly that the amplitude on the ${\rm L}_1$ side of the primary star is much larger than on the other side. This is the reason for calling the star a ``single-sided pulsator''. 

When the amplitudes are small, the errors on the phase become large. Nevertheless, it can be seen that there is a range of orbital phase between about $0.3 - 0.6$ where the pulsation phase reverses by $\sim$ $\upi$\,rad for $\nu_1$, $\nu_3$ and $\nu_4$, whereas the uncertainties are too great to say whether this is also the case for $\nu_2$. For $\nu_1$, $\nu_3$ and $\nu_4$ this shows that the line of sight has crossed a surface pulsation node. Whether these nonradial distorted modes are primarily dipole modes, or some higher degree, is not certain. Their surface flux perturbation requires a superposition of several spherical harmonics to reproduce the data. 

For CO~Cam the latitudinal change in flux perturbation may be confined to superficial layers where tidal effects are significant, so that ``single-sided'' pulsation may occur for all mode degrees, even for a radial mode whose amplitude in deeper layers is essentially spherical symmetric. Currently, this ``tidal trapping" phenomenon is not totally understood, but the physical causes and theoretical implications are discussed in Section \ref{sec:theory} below, and in Fuller et al. (in preparation). 

\subsection{The g-mode pulsation frequencies}
\label{sec:gmodefreqs}

\begin{figure}
\centering
\includegraphics[width=1.0\linewidth,angle=0]{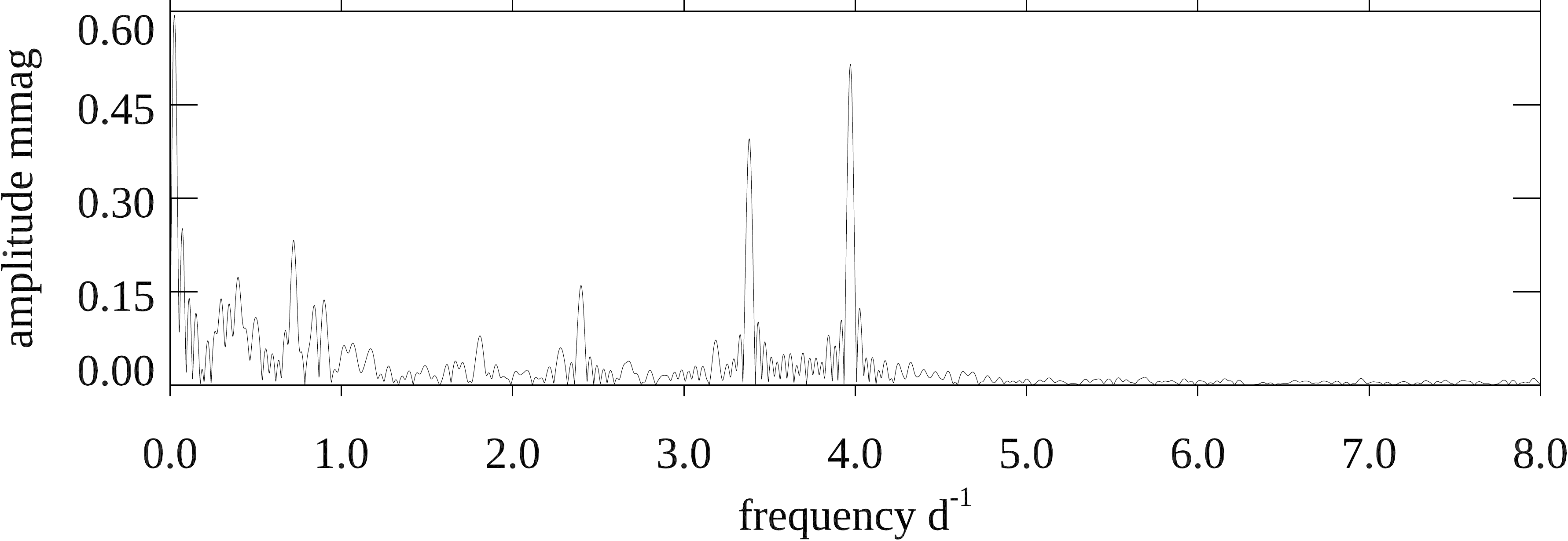}	
\caption{Amplitude spectrum after pre-whitening the orbital harmonic series, showing peaks for at least 3 g~modes at $2.398509, 
3.380613 ~{\rm and}~ 3.971885$\,d$^{-1}$. The lowest frequency peaks below 1\,d$^{-1}$ are probably instrumental artefacts.}
\label{fig:gmodes}
\end{figure}

With the original SAP data, once the orbital harmonic series is pre-whitened, at low frequency there are some g~mode peaks visible among the instrumental noise peaks. There are three low frequencies that are probably due to g modes. Some modelling is needed to try to find mode identifications for these. The mode frequencies for the three clear g-mode peaks are $2.398509, 
3.380613 ~{\rm and}~ 3.971885$\,d$^{-1}$; these can be seen in Fig.\,\ref{fig:gmodes}. The first and third of these are separated by the twice orbital frequency, but are not harmonics of that frequency, hence they may be non-linearly coupled. Without a richer set of g-mode frequencies, little more can be concluded from this.

\section{Zeroth order p-mode constraints}

The standard simple relation for a toy model pulsator relating the pulsation period and mean density is:
\begin{equation}
P \sqrt{\frac{\overline\rho}{\overline\rho_\odot}} = Q,
\end{equation}
where $Q$ is a constant for a given pulsation mode, defined by this equation.
This can be rewritten in terms of observables as
\begin{equation}
\log Q = \log P + \frac{1}{2} \log g + \frac{1}{10} M_{\rm bol} +\log T_{\rm eff} -6.454,
\end{equation}

\noindent where $P$ is in days and $\log g$ uses cgs units. Taking $T_{\rm eff} = 7070$\,K (from Table\,\ref{tbl:mags}), $\log g = 4.0$ from the TIC, and estimating $M_{\rm bol} = 2.5$\,mag from the Gaia parallax and $V$ magnitude then gives for $\nu_1 = 13.37815$\,d$^{-1}$ a value of $Q = 0.033$. For standard $\delta$~Sct models this is what is expected for the radial fundamental mode. That there are multiple modes close in frequency suggests that at least some of them are nonradial modes, and that those have low radial overtones, a useful constraint.    

\section{System Parameters}
\label{sec:systemparms}

\subsection{Methods}
\label{sec:methods}

In order to evaluate the binary system parameters of CO~Cam, we utilised two essentially independent approaches to the analysis.  In the first, we find the stellar masses, inclination and system age that best yield a match to the existing measurements of the spectral energy distribution (SED) and the measured radial velocity of the primary star.  In the second approach, we model the {\em TESS} light curve and simultaneously the radial velocity curve with the {\tt phoebe2} binary light curve emulator \citep{2016ApJS..227...29P}.  Both methods utilise a Markov chain Monte Carlo (MCMC) approach to evaluate the uncertainties in the parameters.  

\subsection{Modelling the SED plus Radial Velocity}
\label{sec:sedplusRV}

The principal input ingredients for this approach are: (1) the known $K$ velocity for the primary star (see Fig.~\ref{fig:RV} and \citealt{2017AN....338..671B}); (2) the measured SED points\footnote{http://viz-beta.u-strasbg.fr/vizier/sed/doc/} between 0.15 and 25 $\mu$m; and (3) the Gaia distance \citep{2018A&A...616A...2L}.  Another more minor ingredient is the constraint imposed by various spectral estimates of the primary's surface gravity which we combine to yield $\log g = 3.95 \pm 0.1$ (cgs units; \citealt{2007ApJ...658.1289T}; \citealt{2009A&A...493.1099S}). In support of the literature data, two consecutive high-resolution spectra ($R=55000$) of CO Cam were taken with the CAOS spectrograph \citep{2016AJ....151..116L} attached to the 91-cm telescope at Catania Observatory on the night of 21/22 September 2019. Integration times of 600 and 900 s were employed to yield a S/N of the combined spectrum of 224. After standard reductions, this spectrum was analysed following the methods outlined by \cite{2016MNRAS.458.2307K}, resulting in $T_{\rm eff}=7000 \pm 100$\,K, log\,$g= 4.0 \pm 0.1$, microturbulence velocity $\xi= 2.9 \pm 0.2$ km s$^{-1}$ and a projected rotational velocity $v \sin i = 63 \pm 4$ km s$^{-1}$.

We also make use of the {\tt MIST} ({\tt MESA} Isochrones \& Stellar Tracks; \citealt{2016ApJS..222....8D};  \citealt{2016ApJ...823..102C};  \citealt{2011ApJS..192....3P}; \citealt{2015ApJS..220...15P}) evolution tracks for stellar masses between 0.7 and 3.0\,M$_\odot$ with solar composition\footnote{We have chosen solar metallicity for lack of better information about the interior composition of the primary star in CO~Cam.  Am spectral peculiarities are confined to a thin surface layer.}, in steps of 0.1\,M$_\odot$.  And, finally, we utilise the \citet{2003IAUS..210P.A20C} model stellar atmospheres for $4000 < T_{\rm eff} < 10,000$ K in steps of 250 K. 

Our approach follows that of \citet{2006ApJ...653..647D}, \citet{2013ApJ...778...95M,2015ApJ...810...61M}, \citet{2018A&A...616A..38M}, \citet{2019MNRAS.489.1644W} and Borkovits et al.~(2019), but we outline our procedure here for completeness because it differs in some aspects.  We use an MCMC code (see, e.g.,\,\citealt{2005AJ....129.1706F}) that evaluates four parameters: the primary mass, $M_1$, secondary mass, $M_2$, system inclination angle, $i$, and the {\tt MIST} equivalent evolutionary phase (EEP) of the primary star.  

The steps of the {\tt MIST} evolutionary tracks are numbered such that the ten `principal EEP'  values (Roman numerals) and corresponding `subdivided EEP' numbers (Arabic numerals) are: \mbox{I-1}  giving the first point on the pre-main-sequence (PMS) phase; \mbox{II-202} the start of the zero-age main sequence (ZAMS); \mbox{III-353} the intermediate-age main-sequence (IAMS) phase; and \mbox{IV-454} the terminal-age main-sequence (TAMS) and the prelude to the subgiant branch (or Hertzsprung gap).  Note that the subdivided EEP numbers continuously progress to higher values (\mbox{X-1710} at the relic white dwarf cooling sequence), but these latter values include evolutionary phases that do not concern us in the CO~Cam system. While the subdivided EEP numbers are continuous within a principal EEP set, they do not generally represent equal steps in the evolutionary age, $\tau$, stellar radius, $R$, or effective temperature, $T_{\rm eff}$, and that this part of the `non-uniformity' in sampling must be handled separately.

For each link in the MCMC chain we use the value of $M_1$ and the EEP value for the primary to find $R_1$ and $T_{\rm eff,1}$ from the corresponding {\tt MIST} tracks, using interpolation for masses between those that are tabulated. That also provides an age, $\tau$, for the star.  Since we tentatively assume that the two stars in the binary are coeval, we use the value of $\tau$ to find the EEP for the secondary star.  From that, we determine the values of $R_2$ and $T_{\rm eff,2}$. 

At this point we check to see that neither star overfills its Roche lobe and that no eclipses should be seen (since there are no eclipses observed in CO~Cam).  If either of these is not satisfied, then that step in the MCMC chain is rejected.

We then utilise the two masses and the inclination to determine what the $K$ velocity of the primary should be.  This is compared to the measured value, and the uncertainty is used to determine the contribution to $\chi^2$ due to the RV evaluation.

Finally, we use $R_1$ and $T_{\rm eff,1}$, as well as $R_2$ and $T_{\rm eff,2}$, along with interpolated Kurucz model spectra, to fit the 29 available SED points.  The value of $\chi^2$ for this part of the analysis is added to the contribution from the RV match, and a decision is made in the usual way via the Metropolis-Hastings jump condition (\citealt{1953JChPh..21.1087M}; \citealt{hastings}) as to whether to accept the new step or not.  

This is done $10^7$ times and the posterior system parameters are collected. The parameter posterior distributions are further weighted according the derivative of the age with respect to the primary EEP number: $d\tau/d({\rm EEP})$. This corrects for the unevenly spaced EEP points within a larger evolutionary category, and across their boundaries. 

\begin{table}
\centering
\caption{Derived Parameters for the CO~Cam System}
\begin{tabular}{lcc}
\hline
\hline
Input Constraints & SED + RV$^a$ & Light curve + RV$^b$   \\
\hline
$K_1$ (km~s$^{-1}$)$^c$ & $72.3 \pm 0.3$ & $72.3 \pm 0.3$ \\  
$\log$ g$_1$ (cgs)$^d$ & $3.95 \pm 0.1$ & ... \\  
$v\,\sin i$ (km s$^{-1}$)$^e$ & $63 \pm 4$ & ... \\
Spectral & 29 SED points$^f$ & ...\\
Light curve modeling & ... & {\em TESS}$^g$ \\
Luminosity & $L_2 < 0.2 \, L_1$$^g$ & ... \\
Distance (pc) & $33.56 \pm 0.17$$^h$ & ... \\
\hline
Derived  Parameter & SED + RV$^a$  & Light curve + RV$^b$\\
\hline
$M_1$ (M$_\odot$) &  $1.52 \pm 0.02$ &  $1.48^{+0.02}_{-0.01}$  \\
$M_2$ (M$_\odot$) & $0.83 \pm 0.08$ & $0.86 \pm 0.02$ \\
$R_1$ (R$_\odot$) & $1.79 \pm 0.07$ & $1.83 \pm 0.01$ \\
$R_2$ (R$_\odot$) & $0.75 \pm 0.07$ & $0.84 \pm 0.02$ \\
$T_{\rm eff,1}$ (K) & $7075 \pm 120$ & $7080 \pm 80$ \\
$T_{\rm eff,2}$ (K) & $5085 \pm 325$ & $5050 \pm 150$ \\
$i$ (deg) & $53.6 \pm 5$ & $48.9 \pm 1.0$   \\
$a$ (R$_\odot$) & $6.54 \pm 0.06$ & $6.55 \pm 0.03$ \\
$R_1/R_L$ & $0.63 \pm 0.03$ & $0.65 \pm 0.02$ \\
$K_2$ (km~s$^{-1}$)$^i$ & $135 \pm 11$ & $124 \pm 2$ \\
age (Gyr) & $1.36 \pm 0.16$ & ... \\
$\beta_1$$^j$ & ... & $0.88 \pm 0.04$ \\
$A_2$$^k$ & ... & $0.81 \pm 0.05$ \\
\hline

\label{tbl:parms}  
\end{tabular}

{\bf Notes.}  (a) See Section~\ref{sec:sedplusRV} for details. (b) See Section~\ref{sec:lcmodels} for details. (c) \citet{2017AN....338..671B} (d) See Section~\ref{sec:sedplusRV} for references.  (e) See Section~\ref{sec:sedplusRV} for the source. (f) The SED points with $\lambda < 0.3 \, \mu$m are taken from VizieR (http://viz-beta.u-strasbg.fr/vizier/sed/doc/), while the four UV points are from \citet{1978csuf.book.....T}. (g) This work. (h) Gaia DR2 \citep{2018A&A...616A...2L}. (i) Predicted value. (j) Gravity brightening parameter. (k) Bolometric albedo.
\end{table}

The results of this analysis are summarised in the middle column of Table \ref{tbl:parms}. We find that the primary star has a mass of 1.5\,M$_\odot$, and has evolved somewhat off the ZAMS, i.e., is near the IAMS with a radius of 1.8\,R$_\odot$.  The unseen secondary star is low enough in mass ($0.8-0.9$\,M$_\odot$) that it is still near the ZAMS\footnote{This assumes no mass loss or transfer in CO~Cam during its prior history.  We address this issue in Section~\ref{sec:mdot}.}.  If the mass turns out to be as large as 0.9\,M$_\odot$, then the secondary could ultimately be directly detectable in future more sensitive spectroscopic studies.  

The distribution of allowed $T_{\rm eff,1}$ values for the primary star is well constrained to $7070 \pm 100$ K. By contrast, the allowed values for $T_{\rm eff,2}$ are $\simeq 5050 \pm 150$\,K.  Additional ground-based spectroscopic constraints on $T_{\rm eff,2}$ would greatly help in determining the mass of the secondary. 

\begin{figure}
\centering
\includegraphics[width=1.0\linewidth,angle=0]{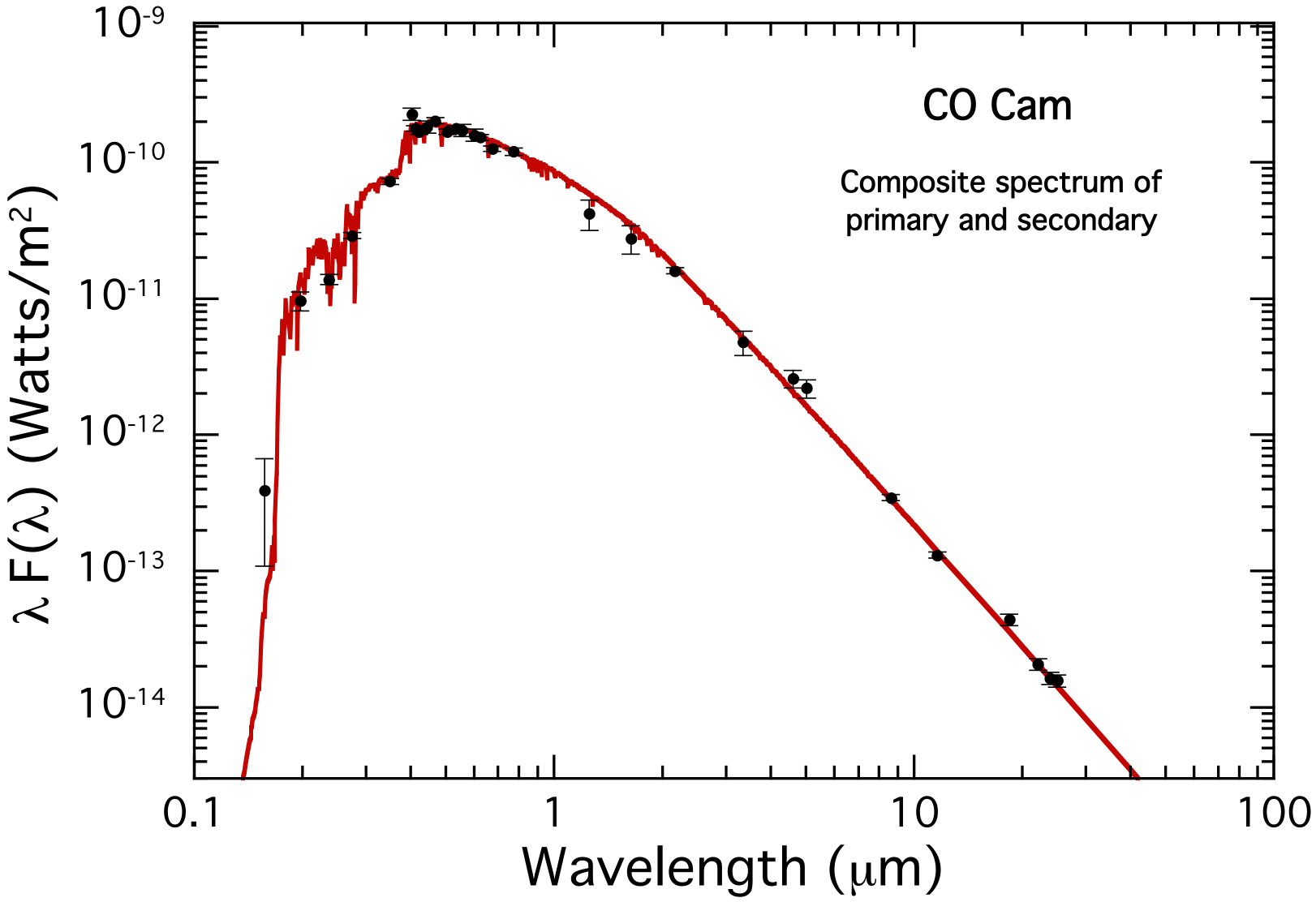}	
\caption{Spectral fit to the 29 available SED data using the combined flux from the two stars. This emerges directly from the MCMC analysis where the SED is fitted during each link of the MCMC.}
\label{fig:sedfit}   
\end{figure}

As an important part of the process of evaluating the system parameters in this first approach, we fitted the SED for CO~Cam during each step of the MCMC analysis.  An illustrative fit to a composite spectrum (due to both stars) is shown in Fig.~\ref{fig:sedfit}. The fitted model is a composite of the corresponding two \citet{2003IAUS..210P.A20C} model spectra.  

\begin{figure}
\centering
\includegraphics[width=1.0\linewidth,angle=0]{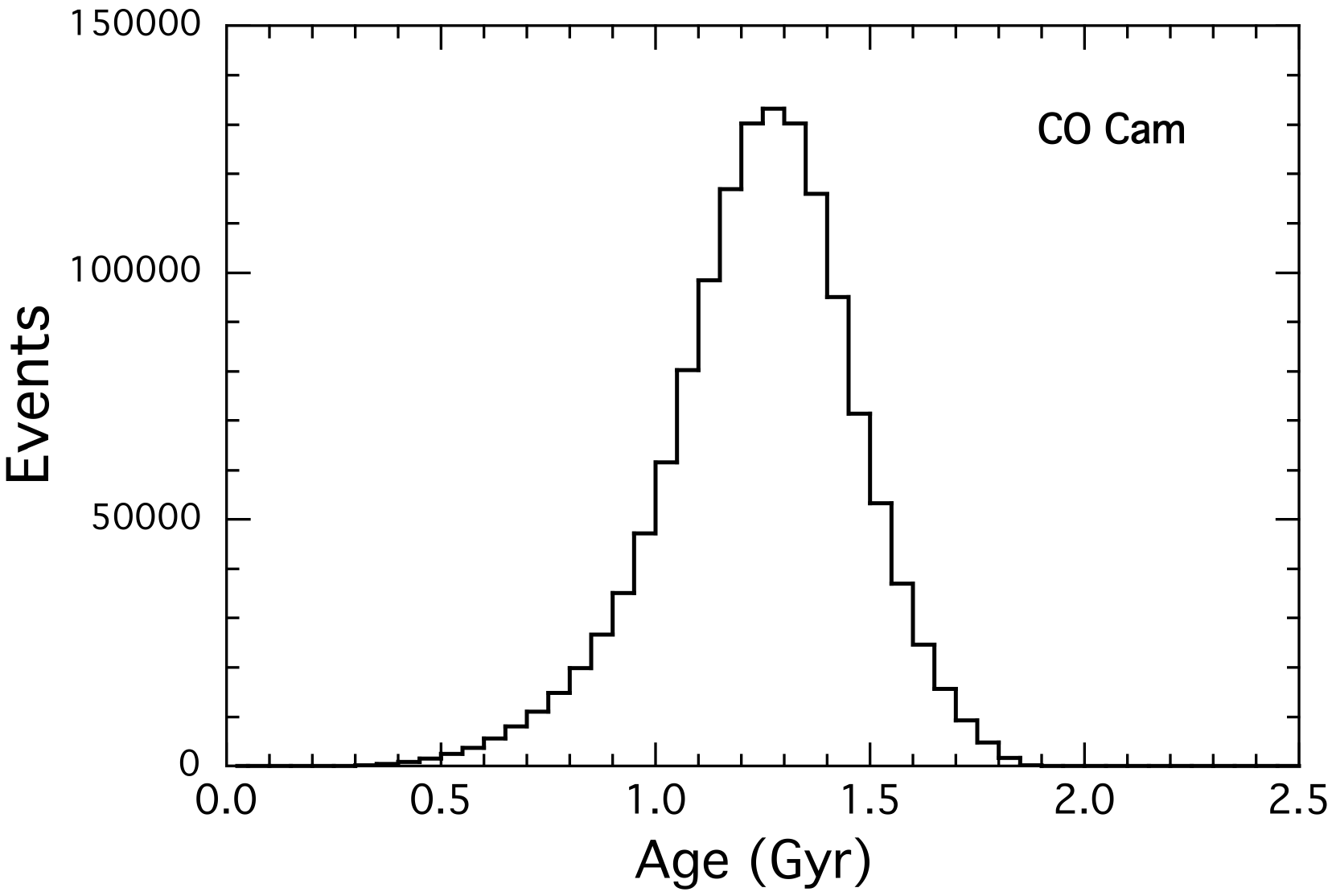}	
\caption{The age probability distribution for the system deduced from the MCMC analysis of the system parameters and the {\tt MIST}  evolution tracks (see Section~\ref{sec:sedplusRV} for references).}
\label{fig:age}   
\end{figure}

The age distribution inferred from the EEP values in the MCMC analysis is shown in Fig.~\ref{fig:age}.  The likely age of CO~Cam is $1.36 \pm 0.16$ Gyr.  This is indeed the amount of time required for a 1.5\,M$_\odot$ star to evolve to the IAMS.

The inclination angle is well constrained to be $50 \pm 2^\circ$.  However, this is refined even further in Section~\ref{sec:lcmodels}.

\begin{figure}
\centering
\includegraphics[width=1.0\linewidth,angle=0]{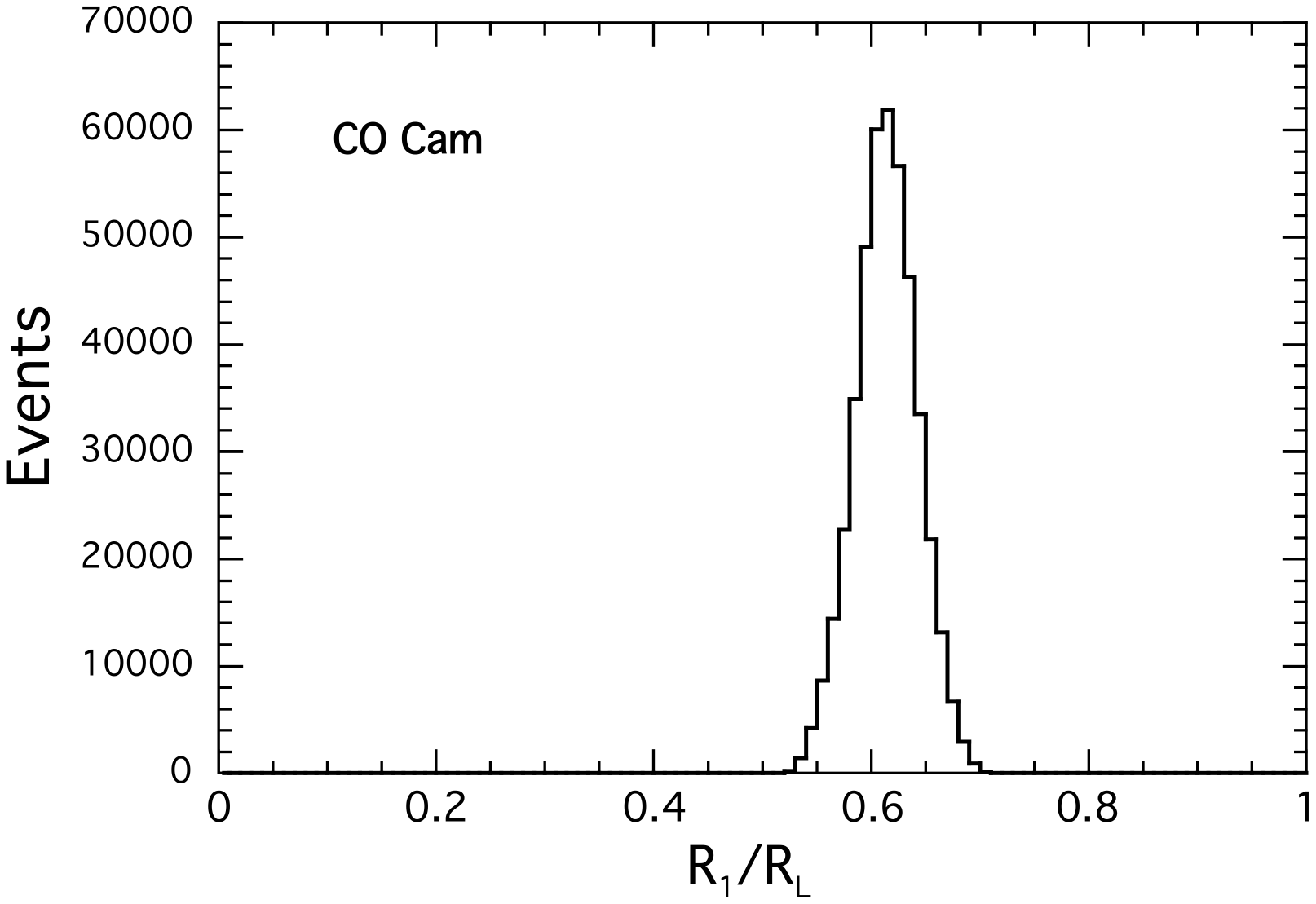}	
\caption{Distribution of Roche-lobe filling factors for the primary star (the pulsator).  It appears that the primary fills no more than 2/3 of its Roche lobe.}
\label{fig:lobe_filling}   
\end{figure}

Finally, in Fig.\,\ref{fig:lobe_filling} we show the distribution of Roche-lobe `filling factors' for the primary star, formally $R_1/R_L$.  This quantity seems to be robustly near 2/3, well below unity. In contrast, the stars in HD~74423 are very near-Roche lobe filling \citep{2020NatAs.tmp...45H}. It is a challenge to understand why the pulsations are so strongly modulated with orbital phase in both systems, as we discuss further in Section~\ref{sec:theory}. 

\subsection{Modelling the {\em TESS} Light Curve plus RV Curve}
\label{sec:lcmodels}

\begin{figure}
\centering
\includegraphics[width=1.0\linewidth,angle=0]{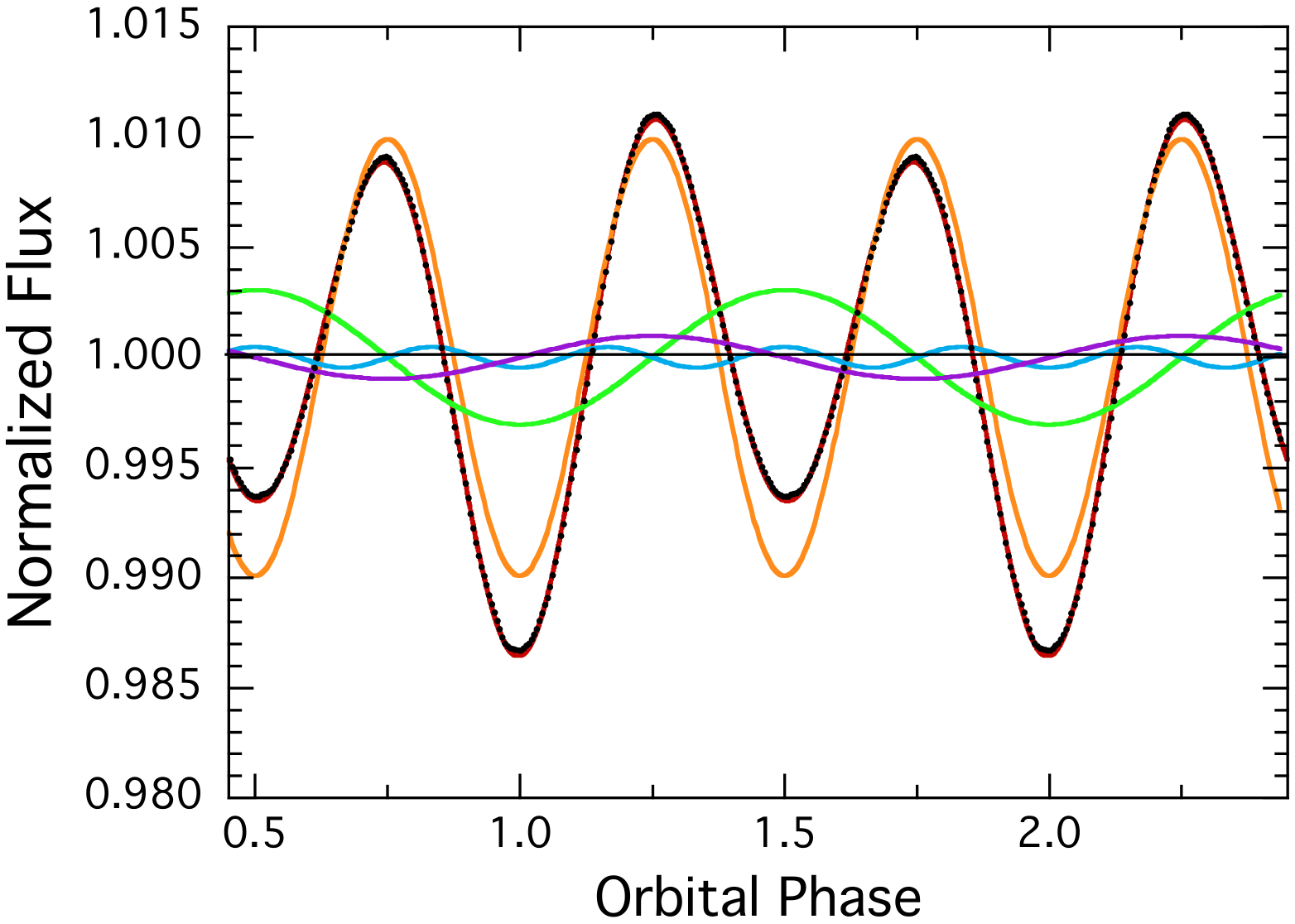}	
\caption{Simple decomposition of the orbital light curve of CO~Cam into four orthogonal sinusoids representing ellipsoidal light variation, the illumination effect, and Doppler boosting (see text and Eqn.~[\ref{eqn:lc_function}] for details). The black dots are the data points and the red curve is the fitted function, i.e., the sum of the four terms in Eqn.~(\ref{eqn:lc_function}).  To first order, the orange and green curves represent the ELV and illumination effects, respectively, while the purple curve is the Doppler boosting term.}
\label{fig:orbital_lc}   
\end{figure}

In the previous subsection we found the basic system parameters from an MCMC evaluation of the two masses, the inclination angle, and the evolutionary phases of the two stars.  The fitted parameters were $K_1$, $\log g_1$, $v\sin i$, and 29 SED points, coupled with the Gaia distance.

We now proceed to test and refine these parameters via simultaneous fitting of the {\em TESS} orbital light curve as well as the radial velocity curve \citep{2017AN....338..671B} using the next-generation Wilson-Devinney code {\tt phoebe2}  \citep{2016ApJS..227...29P}.  First, we removed the pulsations from the light curve as shown in the bottom panel of Fig.~\ref{fig:phamp}.  Then we decomposed the flux, $F(t)$, into its lowest three orthogonal frequencies (i.e., multiples of 1, 2, and 3 times the orbital frequency):
\begin{equation}
F(t) = 1 + B_1 \cos\omega t + B_2 \cos 2\omega t + B_3 \cos3 \omega t +  A_1 \sin\omega t \ \ \ \ \
\label{eqn:lc_function}
\end{equation}
(\citealt{1959cbs..book.....K}; \citealt{2011ApJ...728..139C}), where the $B_2$ term, at twice the orbital frequency, represents the bulk of the ellipsoidal variability (ELV) amplitude, the $B_1$ term represents most of the irradiation effect (at the orbital frequency) and the $A_1$ term provides a good representation of the Doppler boosting effect (\citealt{2003ApJ...588L.117L}; \citealt{2010ApJ...715...51V}).  The ELV also contributes to $B_1$ and $B_3$, while the irradiation effect contributes to $B_2$ as well.  However, only the Doppler boosting effect is expected to  contribute significantly to $A_1$, especially if the stars are co-rotating with the orbit, which is almost certainly the case here.  

In Fig.~\ref{fig:orbital_lc} we decompose the orbital light curve into these four terms shown respectively as $B_2$ (orange; $-9.91$ mmag), $B_1$ (green; $-3.05$ mmag), $B_3$ (blue; $-0.47$ mmag) and $A_1$ (purple; +0.97 mmag).  The fit to the orbital light curve is nearly perfect, indicating that all three physical effects account for all the discernible orbital modulations.

As of the version 2.2 release \citep{2019arXiv191209474J}, {\tt phoebe2} does not include a treatment of the Doppler boosting effect.  Thus, we elected to subtract off the $A_1$ term from the light curve before fitting via the MCMC methodology outlined in \citet{2018A&A...619A..84B} and \citet{2019MNRAS.482L..75J}.  The component masses, radii and temperatures, and the orbital inclination were allowed to vary freely over the ranges determined by the analysis presented in Section~\ref{sec:sedplusRV}, with the 1-$\sigma$ uncertainty ranges taken as uniform priors for the {\tt phoebe2} MCMC analysis. The only additional free parameters were the gravity brightening exponent, $\beta_1$ (where $T^4_\mathrm{eff,local}=T^4_\mathrm{eff,pole} (g_\mathrm{local}/g_\mathrm{pole})^\beta$), of the primary and the (Bond) bolometric albedo \citep{2019ApJS..240...36H}, $A_2$, of the secondary, which are critical for constraining the ELV and irradiation effect amplitudes, respectively.  

\begin{figure}
\centering
\includegraphics[width=1.0\linewidth,angle=0]{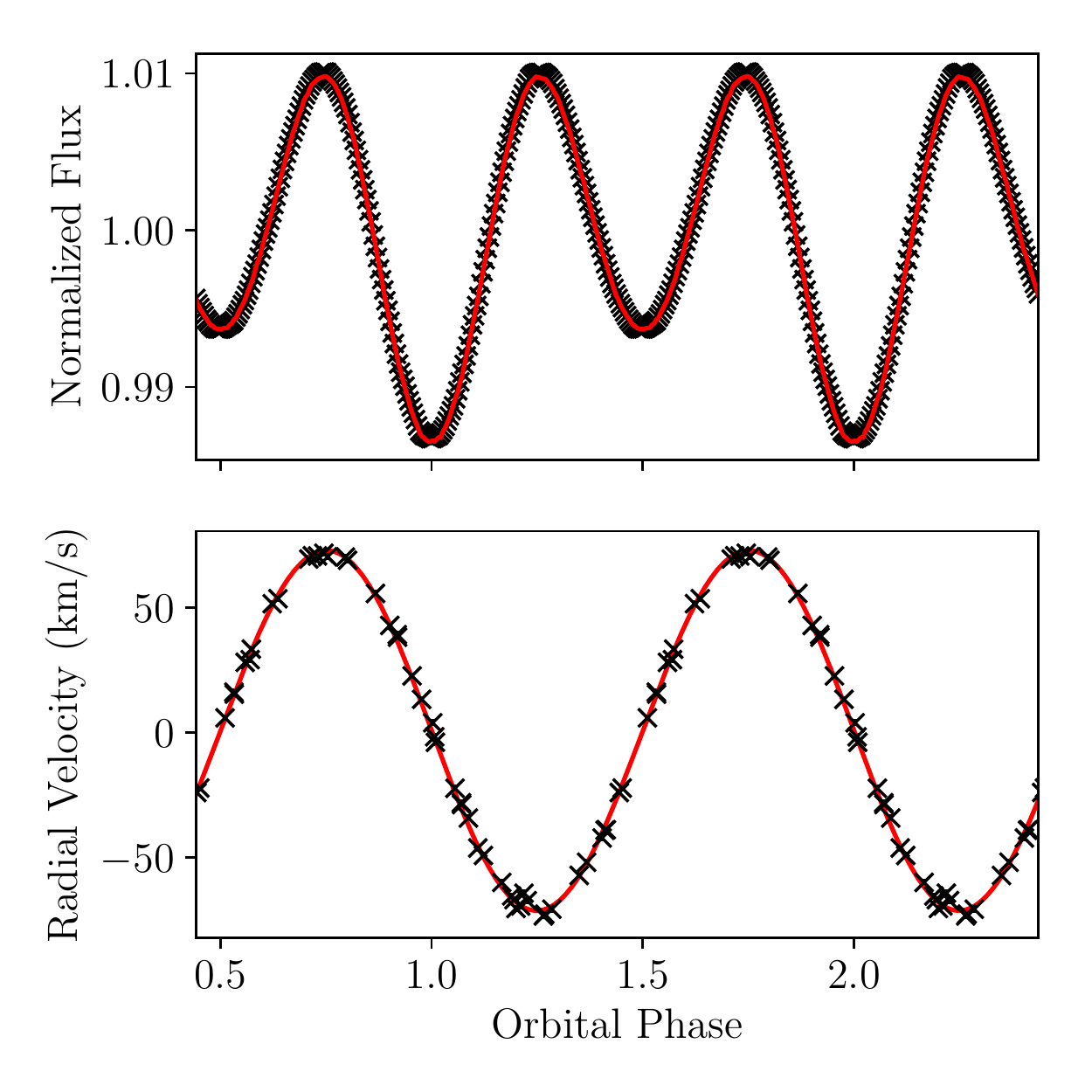}	
\caption{The model {\tt phoebe2} light curve (top) and radial velocity curve (bottom) overlaid on the observations.  In each panel the {\tt phoebe2} model is in red and the data points are in black. The Doppler boosting term has been removed from the {\em TESS} lightcurve (top panel) before doing the fit.}
\label{fig:phoebe_fit}   
\end{figure}

The best-fitting light and radial velocity curves are presented in Fig.\,\ref{fig:phoebe_fit} while the model parameters are listed in the last column of Table \ref{tbl:parms}.

It is clear that the {\tt phoebe2} model provides a remarkably good fit to both the observed light and radial velocity curves, with all model variables extremely well constrained.  While the properties of the primary were already relatively well constrained by the analysis presented in Section \ref{sec:sedplusRV}, the {\tt phoebe2} fitting also provides relatively strong constraints on the temperature and radius of the secondary even though it contributes minimally at most orbital phases.  This is principally due to the contribution of the irradiation effect around the shallower light curve minimum, which is a function of the albedo, radius and temperature of the secondary, as well as the emergent spectrum of the primary which is a function of primary mass, radius and temperature (which are in turn well-constrained by the shape of the light curve).  Ultimately, the strongly inter-related nature of all the model parameters and their impact on the observations, means that a unique solution could be derived which further refines the parameters from the analysis in Section \ref{sec:sedplusRV}.

\subsection{Prior Possible History of Mass Transfer}
\label{sec:mdot}

Thus far in our analysis of the system parameters in CO~Cam we have made at least the implicit assumption that the secondary star is coeval with the primary and that there has been no prior episode of mass transfer.  In the SED plus RV fitting in  Section.~\ref{sec:sedplusRV},  we in fact made explicit use of the coeval and no-mass-transfer assumptions in employing {\tt MIST} evolution tracks for the two stars.  It is hardly surprising, then, that the secondary of mass $\la 1$\,M$_\odot$ ends up having the properties of a low-mass main-sequence star, given that the more massive primary star is only slightly evolved. 

By contrast, the light curve plus RV fitting in Section~\ref{sec:lcmodels} made no such assumptions about the evolutionary status of the two stars. That analysis places both the mass and radius of the secondary star in the MS region of the diagram.  This result does not prove that the star is on the MS, nor that it has never lost any mass previously, but it does suggest that mass transfer is unnecessary to bring the CO~Cam system into its current configuration.  

In that regard, we carried out an extensive exploration of parameter space to try to identify a range of initial conditions that would allow the primordial binary to evolve into a system resembling CO~Cam at the current epoch. Specifically, we considered primary masses between 0.5 and 1.5\,M$_\odot$, secondary masses between 1 and 3\,M$_\odot$, and orbital periods between the minimum required for the immediate onset of Roche lobe overflow ($P_{\rm RLOF}$) and ten times that value.  Varying the orbital period is equivalent to varying the degree to which the donor has evolved from the ZAMS at the onset of mass transfer (e.g., Case A, AB, or B mass transfer).  

In addition, we tried different combinations of systemic mass loss (fraction of mass expelled from the binary), and angular momentum loss prescriptions (e.g., whether magnetic stellar wind braking was operative\footnote{We adopted the prescription for mass and angular momentum loss described by \citet{2006csxs.book..623T} where the parameters $\alpha$ and $\beta$ were allowed to vary between 0 and 1.}).  A grid of about 800 evolutionary tracks was computed using the {\tt MESA} stellar evolution code (\citealt{2011ApJS..192....3P}; \citealt{2015ApJS..220...15P}), where the evolution of both components (donor and accretor) was computed simultaneously. We found that once mass transfer started, it was very difficult for the binary to enter a detached phase with properties similar to those inferred for CO~Cam.  The detached phases were of very short duration (on the order of 10\,Myr) and during this period the donor's surface was in close proximity to its Roche lobe.

While we have not carried out an exhaustive exploration of parameter space and thus cannot rule out the possibility of a prior history of mass transfer in CO~Cam, we do not consider it very likely.

\section{Pulsation Models}
\label{sec:pulsmodels}

In this section we discuss stellar models that have pulsation frequencies that agree with the highest amplitude p-mode frequencies detected in CO~Cam. Main-sequence models of  masses ranging from 1.4\,M$_\odot$ to 1.7\,M$_\odot$ with a solar abundance $(X,Z)=(0.72,0.014)$ were computed by the {\tt MESA} code \citep[ver.7184;][]{paxton2013,2015ApJS..220...15P}.
In some models overshooting from the convective core was included, adopting the exponentially decaying mixing scheme invented by \citet{herwig00}, where the scale length of the mixing  is given as $h_{\rm os}\cdot H_p$, with $H_p$ being the pressure scale length.  The extent of mixing is controlled by the free parameter $h_{\rm os}$. We consider the cases of $h_{\rm os}=0$ (no overshooting), $0.005$, $0.01$, and $0.02$.

For each assumed value of the parameter $h_{\rm os}$, we obtained evolutionary models for various masses, and found evolutionary stages where the frequency of the radial fundamental mode lies within the range of the observed four main frequencies, 13.0 to 14.1~d$^{-1}$. For those selected models, we have obtained adiabatic pulsation frequencies of nonradial modes with low latitudinal degree ($\ell \le 3$). Among them we looked for a model that reproduces the observed main frequencies $\nu_1, \nu_2, \nu_3$ and $\nu_4$ with radial ($\ell=0$) and low-degree nonradial ($\ell \le 3$) modes. While these modes are well-described by single spherical harmonics in the interior of the star, the tidal distortion of the surface requires a summation of low-degree spherical harmonics to describe the observations, hence even the radial modes will be amplitude modulated and generate a frequency multiplet in the oblique pulsator model (see \citealt{1992MNRAS.259..701K} for a discussion).
We found a best model for each of the parameters $h_{\rm os}= 0.01$ and $0.02$ (these correspond approximately to step-wise overshooting mixings of $0.1H_p$ and $0.2H_p$, respectively), while no satisfactory models were found for $h_{\rm os} \le 0.005$.

\begin{figure}
\includegraphics[width=0.49\textwidth]{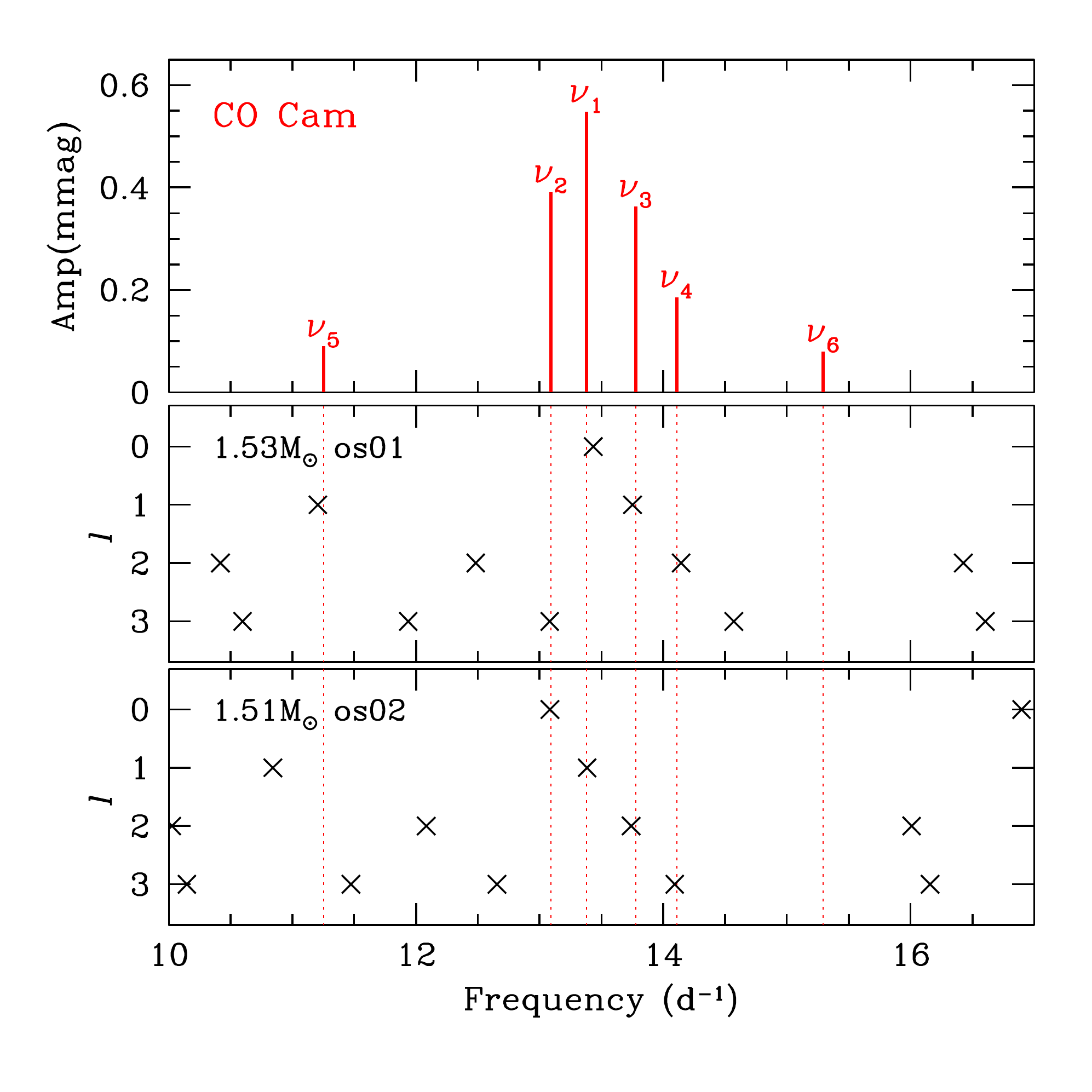}
\caption{Central pulsation frequencies of the multiplets in the p-mode range of CO~Cam (top panel) are compared with pulsation frequencies (crosses) of two models of 1.53\,M$_\odot$ and 1.51\,M$_\odot$. The vertical axes of the middle and bottom panels represent latitudinal degree, $\ell$. Each model is the best model for a given overshooting parameter, where os01 and os02 mean that $h_{\rm os}= 0.01$ and $0.02$, respectively.   
}
\label{fig:freqfit}   
\end{figure}

Pulsation frequencies of the two best models are compared with the observed p-mode frequencies of CO~Cam in Fig.~\ref{fig:freqfit}. In the 1.53-M$_\odot$ model with $h_{\rm os}=0.01$ (middle panel), the radial fundamental mode is fitted with the largest amplitude frequency $\nu_1$, while in the 1.51-M$_\odot$ model with $h_{\rm os}=0.02$ (bottom panel) the radial fundamental mode is fitted with $\nu_2$.  We refer to $\nu_1 \ldots \nu_4$ p modes to distinguish them from the lower frequency g modes, but the radial mode is actually the fundamental mode. In fact, the non-radial modes have the mixed character; i.e., a g-mode characteristic in the core region (where the Brunt-V\"ais\"al\"a frequency is higher than the pulsation frequencies) and the f-mode characteristic in the envelope, where the radial displacement has no node despite the fact that frequencies are above the Brunt-V\"ais\"al\"a  frequency and the Lamb frequency (see Fig. \ref{prop} below).

A low amplitude frequency $\nu_5 = 11.252$\,d$^{-1}$ (see Fig.\,\ref{fig:ft2}) is consistent with an $\ell=1$ mode of the 1.53-M$_\odot$ model, while no corresponding mode is present in the 1.51-M$_\odot$ model. Another low amplitude frequency $\nu_6 = 15.293$\,d$^{-1}$  cannot be fitted by either of the two models, although $\nu_6 - \nu_{\rm orb}= 14.506$\,d$^{-1}$ is close to a frequency of $\ell=3$ in the 1.53-M$_\odot$ model.

\begin{table}
\centering
\caption{Models that best reproduce the main p modes of CO~Cam}
\begin{tabular}{lcccccc}
\hline
\hline
Mass & $h_{\rm os}$ & $L$ & $T_{\rm eff}$ & R 
& age & XH$_{\rm c}$   \\
(M$_\odot)$ & & (L$_\odot$) & (K)  & (R$_\odot$) & ($10^9$yr) \\
\hline
1.53 & 0.01 & 7.44 & 6770 & 1.99 & 1.6 & 0.30 \\
1.51 & 0.02 & 7.38 & 6730 & 2.00 & 1.8 & 0.34 \\
\hline
\label{tbl:models} 
\end{tabular}
\end{table}   

Parameters of the two best models are listed in Table~\ref{tbl:models}. They are very similar to each other despite the difference in the overshooting parameters.  This is because they are mainly determined by the requirement that  the fundamental mode frequency, which is proportional to $M^{1/2}R^{-3/2}$, should be equal to $\nu_2$ or $\nu_1$.  However, we note here the best models given in Table \ref{tbl:models} have larger radii (by 3\,$\sigma$), are cooler $T_{\rm eff}$ (by 3\,$\sigma$), and have older ages (by 2\,$\sigma$) than we found from the analysis of the CO~Cam system parameters (see Table \ref{tbl:parms}).

In addition to the p modes, at least three g-mode frequencies (2.3985, 3.3806, 3.9719~d$^{-1}$) were detected. These frequencies correspond to intermediate order g modes ($n=-7$ to $-12$ for $\ell=1$ and $n=-18$ to $-30$ for $\ell=3$ ). Further comparison with models is difficult because  the effects of rotation (0.787~d$^{-1}$, if synchronised) are considerable.

\section{Theory}
\label{sec:theory}

{

\begin{figure}
\includegraphics[width=0.48\textwidth]{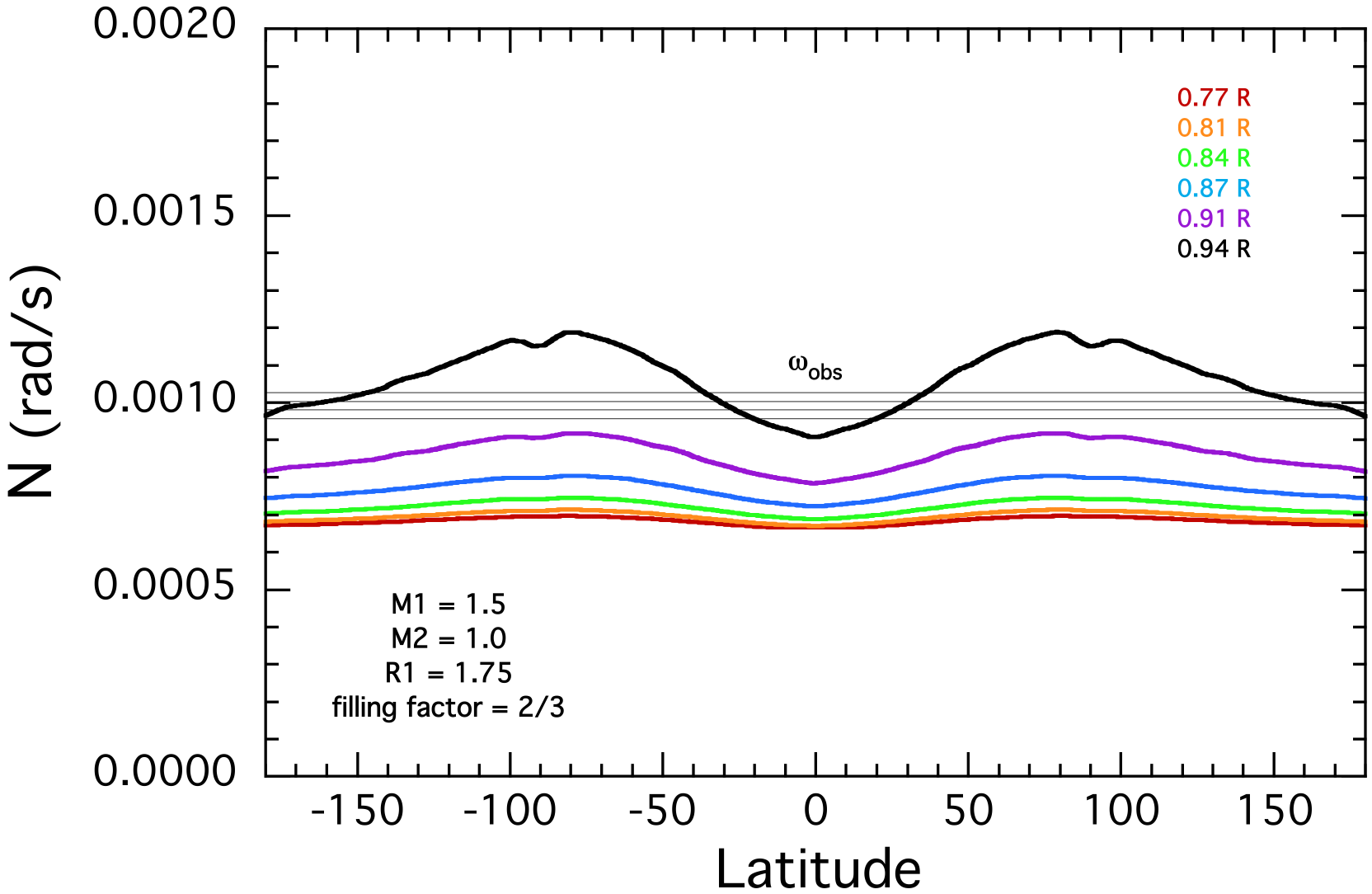}
\caption{\label{omegac} 
The Brunt-V\"ais\"al\"a frequency, $N$, as a function of tidal latitude (measured in the orbital plane) of a tidally distorted polytropic model with the same Roche geometry as CO~Cam. Different colours correspond to different radial coordinates within the star, and the horizontal black lines are the observed pulsation frequencies ($\omega_1 - \omega_4$) of CO~Cam.}
\end{figure}  

The pulsational amplitude modulation of CO~Cam is puzzling. Unlike the single-sided pulsator HD~74423 \citep{2020NatAs.tmp...45H}, CO~Cam appears to substantially underfill its Roche lobe, so the star is not particularly asymmetric, and it does not extend to near its L$_1$ point where the effective gravity vanishes. Fig.\,\ref{omegac} shows a model of the Brunt-V\"ais\"al\"a frequency $N$ for a tidally distorted polytropic model as a function of tidal latitude $\theta$ (i.e., angle away from the line of apsides on the L$_1$ side of the star). This simple model has a density profile corresponding to a polytrope of index 3 ($\gamma=4/3$ ), an adiabatic index $\Gamma_1=5/3$ appropriate for an ideal gas, and it extends to 0.67 of its Roche radius (i.e., the radius of a sphere with equal volume to its Roche lobe). For this model, the Brunt-V\"ais\"al\"a frequency is given by $N=2 g_{\rm eff}/(5 c_s)$, where $g_{\rm eff}$ is the effective gravity, and $c_s$ is the sound speed. Comparison with a more realistic (but undistorted) stellar model in Fig. \ref{prop} shows fairly good agreement in the value of $N$. While there is substantial latitudinal variation of $N$ near the star's surface, the asymmetry between $\theta=0^\circ$ (near the L$_1$ point) and $\theta=180^\circ$ (near the L$_3$ point) is very small. Hence, even if the tidal distortion is strong enough to align the pulsation axis with the line of apsides, it is not clear why the pulsations should have much larger amplitude on the L$_1$ side of the star compared to the L$_3$ side. 

\begin{figure}
\includegraphics[scale=0.36]{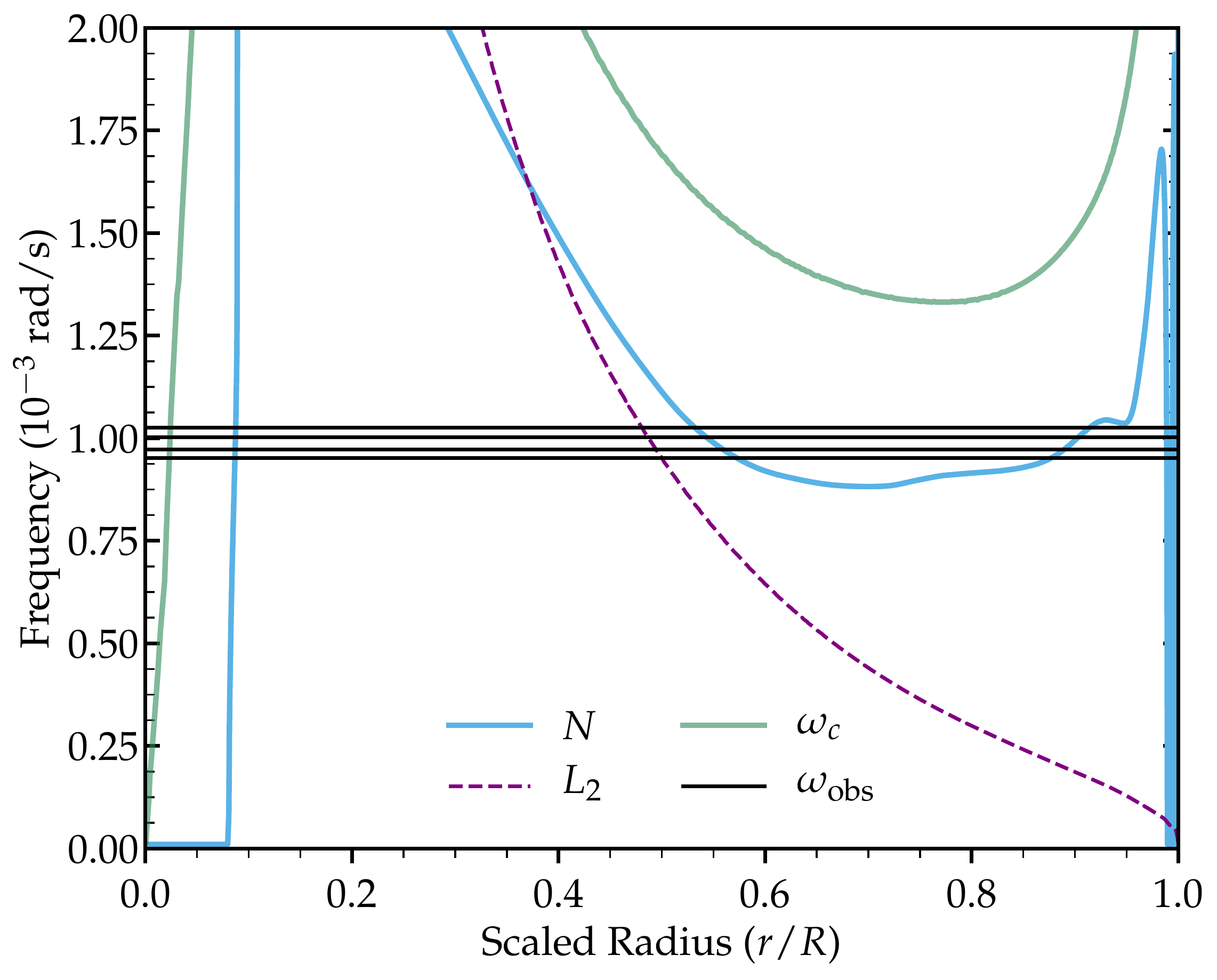}
\caption{\label{prop} 
Propagation diagram for the primary star in CO~Cam, showing the Brunt-V\"ais\"al\"a frequency $N$ (blue line), acoustic cutoff frequency $\omega_c$ (green line), $\ell=2$ Lamb frequency $L_2$ (purple dashed line), and observed pulsation frequencies $\omega_{\rm obs}$ (solid lines). The observed pulsations behave most like fundamental modes in the envelope, and gravity modes in the core.}
\end{figure}  

Another difference between CO~Cam and HD~74423 is that the modes in CO~Cam can best be described as fundamental modes (f modes) rather than pressure modes (p modes), as found by the modelling in Section~\ref{sec:pulsmodels}. In contrast, HD~74423 pulsates in a single first overtone p mode. This conclusion is confirmed from a stellar model of CO~Cam, shown in Fig.\,\ref{prop}. The model is constructed using the {\tt MESA} stellar evolution code \citep{2011ApJS..192....3P,paxton2013,2015ApJS..220...15P,paxton:18,paxton:19} with properties very similar to those in Table 4. The observed pulsations are comparable to the Brunt-V\"ais\"al\"a frequency in the envelope, above the envelope's low-$\ell$ Lamb frequencies, but below the acoustic cutoff frequency $\omega_c = c_s/(2H_p)$, where $H_p$ is a pressure scale height. Hence, the pulsations are evanescent f modes in the envelope, though note they have g mode character near the core.

It is not clear why these f modes should be trapped on one side of the star, but one possibility is that the latitudinal variation of $N$ may help confine the modes. Tidal confinement seems possible because the value of $N$ in the star's envelope is very similar to the observed pulsation frequencies, so pulsations will be sensitive to lateral variations in $N$, which can behave like a wave guide. Robust calculations are difficult because a WKB approximation is not good for f modes, but the value of $N$ can behave like an effective potential. Waves whose frequency is greater than $N$ at the potential's minimum, but lower than $N$ at the potential's maximum (as is the case at $r=0.94 \, R$ in Fig. \ref{omegac}) may be trapped at the deepest potential minimum at $\theta=0^\circ$. 

From the observed pulsation amplitude modulation, some sort of tidal trapping must be occurring. Fig.\,\ref{amp} shows the amplitude and phase modulation of $\ell=0$ and $\ell=1$ modes (with $m=0$) aligned with the tidal axis. We plot both ``normal" modes whose perturbed flux is given by the corresponding spherical harmonic, and ``trapped" modes for which we have multiplied by the flux perturbation by a factor of $\cos^2(\theta/2)$. This arbitrary trapping reduces the flux perturbation on the L$_3$ side of the star. While none of the models matches the data particularly well, a trapped $\ell=1$ mode comes closest. It is clear that some sort of tidal trapping is required in CO Cam in order to match the strong amplitude modulation over the orbital phase. Additionally, radial modes produce no phase variation throughout the orbit, so there must also be some latitudinal phase variation of the modes, as expected for non-radial modes. The smooth observed phase variation requires an imaginary component in the surface flux perturbation, because a purely real flux perturbation can only produce phase jumps of $\pi$. In other words, the perturbed flux pattern must laterally propagate across the stellar surface, rather than the usual case of a standing mode in the lateral direction.

In short, we currently cannot explain the observed amplitude or phase modulation of the pulsation modes of CO~Cam, but it is clear that a substantial amount of latitudinal tidal trapping and phase modulation is required. More detailed theoretical investigations should seek to explain these phenomena and to make predictions for future single-sided pulsators.

\begin{figure}
\includegraphics[scale=0.36]{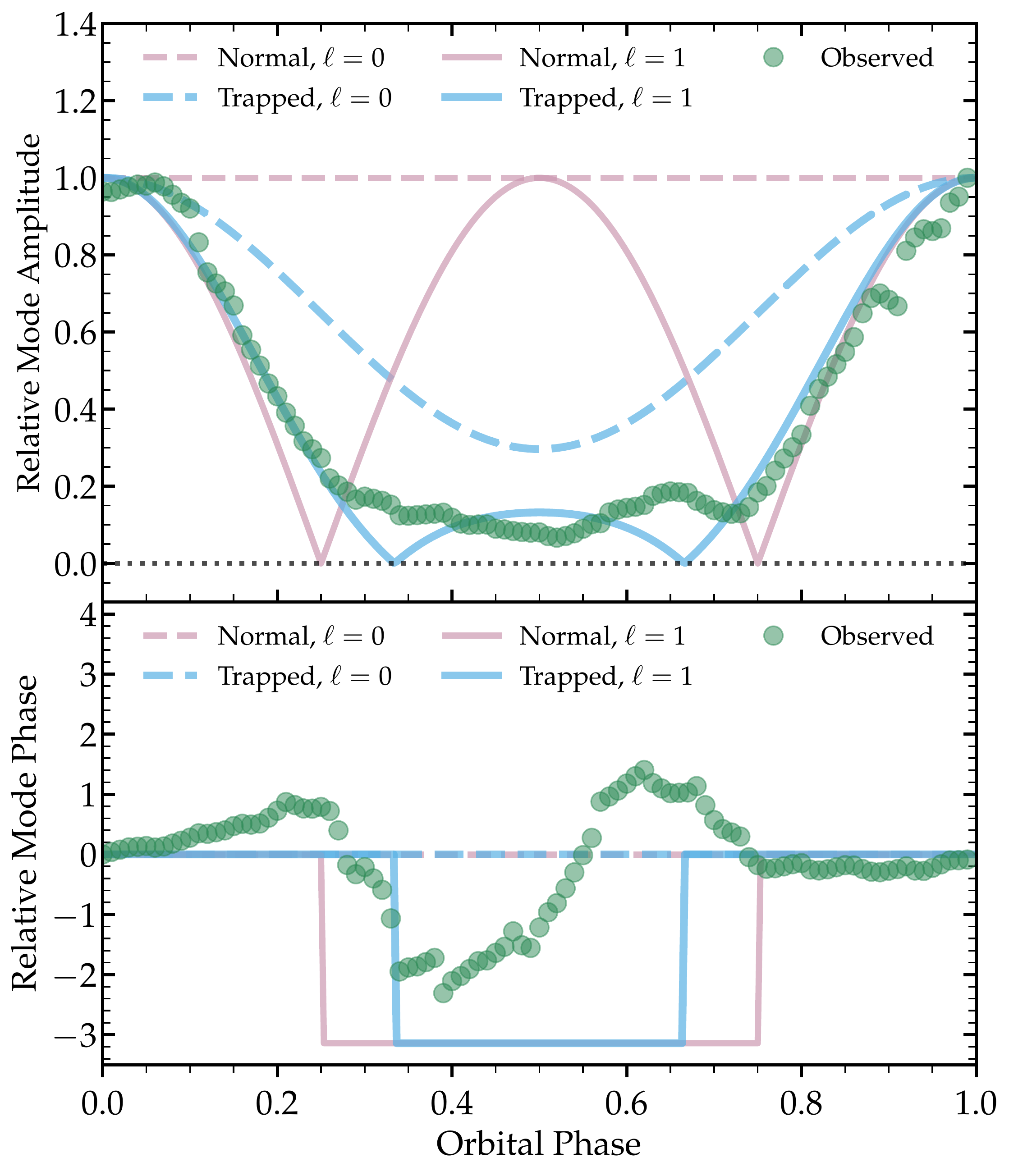}
\caption{\label{amp} 
Top: As a function of orbital phase, amplitude modulation of different types of modes, viewed at an inclination angle of $i=50^\circ$ to the orbital axis. Each mode is assumed to be aligned with the tidal axis, but the flux perturbation is enhanced near the L$_1$ point for the ``trapped" modes, while the ``normal" modes have the usual spherical harmonic dependence. Green dots are the (smoothed) observed phase variation of mode $\nu_2$. Bottom: Corresponding mode phase variations.}
\end{figure} 
}
\section{Discussion and conclusions}
\label{sec:conclusions}

CO~Cam is the second ``single-sided pulsator'' to be discovered. The first was HD\,74423 \citep{2020NatAs.tmp...45H}, which is composed of two stars of nearly identical mass in a 1.58-d orbit; both of those stars are $\lambda$\,Boo stars. CO~Cam also has its pulsation axis along the tidal axis, but in its case there are at least four oblique pulsation modes, and it is a marginal Am star. The orbital period is 1.2709927\,d and the secondary is a cooler main-sequence star (probably a G star). The system parameters for both components of CO~Cam are very well characterised by our modelling of the SED plus the radial velocities, and, independently, the light curve and radial velocities. 

Both CO~Cam and HD\,74423 show much higher pulsation amplitude in the hemisphere closest to the ${\rm L}_1$ point. CO~Cam only extends to $\approx 2/3$ of its Roche lobe radius and its tidal distortion is not highly asymmetric, so it is difficult to understand how its modes can be confined to one side of the star (Section \ref{sec:theory}). In contrast, HD~74423 is nearly Roche lobe filling and is very asymmetric, allowing acoustic waves to propagate very close to the photosphere near the L$_1$ point such that they cause much larger flux modulation on that side of the star (Fuller et al., in prep). More single-sided pulsators should be identified and examined in order to  understand better the physical causes of the tidal trapping phenomenon.

With these being the first known cases of ``single-sided pulsation'' we cannot yet answer a plethora of questions that they give rise to. We are confident that more of these stars will be found, and we are carrying out the search for them. Until those are found and we have an understanding of the range of behaviour possible for these stars, this discussion now will concentrate on the questions that they pose, and the physics they may address. 

HD\,74423 and CO~Cam are $\delta$~Sct stars; they both lie within the traditionally defined instability strip where it crosses the main-sequence. \citet{2018MNRAS.476.3169B} and \citet{2016MNRAS.460.1970B} give an extensive discussion of nearly 1000 $\delta$~Sct stars observed for four years by the {\it Kepler} mission, showing that the instability strip is not well-defined. In a more extensive study of 15000 {\it Kepler} A and F stars, \citet{2019MNRAS.485.2380M} showed that the fraction of $\delta$~Sct stars peaks in the middle of the instability strip at 70~per~cent, and drops off at hotter and cooler $T_{\rm eff}$. They find the main-sequence blue and red edges of the $\delta$~Sct instability strip to be 9100\,K and 7100\,K, corresponding to spectral types A3 to F0.  The reasons why stars of similar age and structure pulsate in some cases and not others are not known, whether they are in binary systems, or are single stars.

HD~74423 addresses this question directly.  One of the stars in that system pulsates with a single p~mode with the pulsation axis along the tidal axis. The star has no other pulsation modes, neither p~modes nor g~modes, and its companion shows no pulsations at all. In a coeval binary system with two, nearly identical $\lambda$\,Boo stars, why does one star pulsate and the other does not? What is the difference in driving and damping in these stars? Are some pulsating close binary stars oblique pulsators, while others are not? As more single-sided pulsators are found, we will begin to understand better the astrophysics needed to answer these questions.

This problem of why some stars pulsate and others do not is not understood for close binary stars \citep{2019A&A...630A.106G}. New techniques for studying binary stars using pulsations as the frequency standard have been very successful in characterising $\delta$~Sct stars in binaries.  \citet{2018MNRAS.474.4322M} characterise the orbits of 341 new binaries with $\delta$~Sct pulsators using {\it Kepler} data alone. And the ``heartbeat stars'', eccentric binaries with pulsating components that have strong tidal interaction and excitation at periastron passage, now number around 200 examples; see \citet{2019arXiv191108687G} for recent studies and further references. These stars mostly have relatively long orbital periods where the stars are only tidally distorted at periastron, although there are now some known with orbital periods of only a few days. The Am stars were once thought not to show $\delta$~Sct pulsation \citep{1976ApJS...32..651K}, but with higher precision photometry it is now known that many of them are $\delta$~Sct stars \citep{2017MNRAS.465.2662S}. Most Am stars are in low eccentricity binaries with orbital periods in the 1 to 10 d range, hence have equatorial rotation velocities usually less than 100\,km~s$^{-1}$. It is thus perhaps not surprising that we have found single-sided pulsation in CO~Cam, which is an Am star in a short orbital period binary. Given that the Am phenomenon arises from atomic diffusion, which needs stability against turbulent mixing that is facilitated by slow rotation, and given that short orbital period binaries tend to be synchronous rotators, leading to relatively slow rotation, we expect to find more single-sided pulsators among the Am $\delta$~Sct stars. 

While \citet{2019A&A...630A.106G} have presented the first systematic study of pulsation in close binary stars in the {\it Kepler} and {\em TESS} data sets, there is as yet no detailed study of all of those stars. Such a study is needed to address questions that arise from HD\,74423 and CO~Cam. Why does HD\,74423 show only one pulsation mode? Most $\delta$~Sct stars are highly multi-periodic; often $\delta$~Sct stars with only a single pulsation mode are found among the High Amplitude $\delta$~Sct (HADS) stars. Does the tidal distortion of HD\,74423 select a single pulsation mode? Or is it a selection effect that the first single-sided pulsator has only one mode, since that is easier to see, hence to discover? CO~Cam shows four principal modes, plus probably others, hence single-sided pulsators can be multi-periodic. Do short orbital period binaries with components in the instability strip often, or even usually, pulsate obliquely with the tidal axis as the pulsation axis, but are difficult to recognise with confusion among all the oblique pulsation multiplets? A systematic survey for these stars will shed light on these questions.

As mentioned above, CO~Cam is an Am star. Is that relevant to its single-sided pulsation, or is it simply that close binary stars in the instability strip tend to be Am stars because of rotational synchronisation? Is is relevant to the single-sided pulsation that HD74423 has two $\lambda$\,Boo stars? We do not know. 

Pulsating stars in close binaries offer the combination of astrophysical inference from both asteroseismology and traditional binary star physics. Add to that the advantage conferred by oblique pulsation, which gives a view of the pulsation modes from varying aspect, and it can be seen that these single-sided pulsators offer new opportunities in stellar astrophysics. 

\section*{acknowledgements}

DWK acknowledges financial support from the UK Science and Technology Facilities Council (STFC) via grant ST/M000877/1. GH and FKA acknowledge financial support by the Polish National Science Centre (NCN) under grant 2015/18/A/ST9/00578, and thanks Elizaveta Kozlova for making the paper by Wojtkiewicz-Okulicz available. JF acknowledges support from an Innovator Grant from The Rose Hills Foundation, and the Sloan Foundation through grant FG-2018-10515. DJ acknowledges support from the State Research Agency (AEI) of the Spanish Ministry of Science, Innovation and Universities (MCIU) and the European Regional Development Fund (FEDER) under grant AYA2017-83383-P.  DJ also acknowledges support under grant P/308614 financed by funds transferred from the Spanish Ministry of Science, Innovation and Universities, charged to the General State Budgets and with funds transferred from the General Budgets of the Autonomous Community of the Canary Islands by the Ministry of Economy, Industry, Trade and Knowledge. 
L.N. thanks the Natural Sciences and Engineering Research Council (NSERC) of Canada for funding and thanks J. Aiken for his technical assistance. We thank Barry Smalley, Simon Murphy and Daniel Holdsworth for useful suggestions and discussions, and we thank the anonymous referee for a thoughtful and useful report. M.G. acknowledges financial support from the 5th National Committee of the Italian National Institute for Nuclear Physics (INFN) under the grant MAPS\_3D. We thank Richard Bischoff for supplying his team's CO~Cam spectra. This paper includes data collected by the {\em TESS} mission. Funding for the {\em TESS} mission is provided by the NASA Explorer Program. Funding for the {\em TESS} Asteroseismic Science Operations Centre is provided by the Danish National Research Foundation (grant agreement no. DNRF106), ESA PRODEX (PEA 4000119301), and Stellar Astrophysics Centre (SAC) at Aarhus University.

\bibliography{cocam.bib}

\begin{thebibliography}{85}
\expandafter\ifx\csname natexlab\endcsname\relax\def\natexlab#1{#1}\fi

\bibitem[{{Abt}(1961)}]{1961ApJS....6...37A}
{Abt} H.~A., 1961, \apjs, 6, 37

\bibitem[{{Aerts}, {Christensen-Dalsgaard} \& {Kurtz}(2010){Aerts},
  {Christensen-Dalsgaard}, \& {Kurtz}}]{2010aste.book.....A}
{Aerts} C., {Christensen-Dalsgaard} J., {Kurtz} D.~W., 2010, {Asteroseismology}

\bibitem[{{Bigot} \& {Dziembowski}(2002)}]{2002A&A...391..235B}
{Bigot} L., {Dziembowski} W.~A., 2002, \aap, 391, 235

\bibitem[{{Bigot} \& {Kurtz}(2011)}]{2011A&A...536A..73B}
{Bigot} L., {Kurtz} D.~W., 2011, \aap, 536, A73

\bibitem[{{Bischoff} {et~al}\mbox{.}(2017){Bischoff}, {Mugrauer}, {Zehe},
  {W{\"o}ckel}, {Pannicke}, {Lux}, {Wagner}, {Heyne}, {Adam}, \&
  {Neuh{\"a}user}}]{2017AN....338..671B}
{Bischoff} R. {et~al.}, 2017, Astronomische Nachrichten, 338, 671

\bibitem[{{Boffin} {et~al}\mbox{.}(2018){Boffin}, {Jones}, {Wesson},
  {Beletsky}, {Miszalski}, {Saviane}, {Monaco}, {Corradi}, {Santand er
  Garc{\'\i}a}, \& {Rodr{\'\i}guez-Gil}}]{2018A&A...619A..84B}
{Boffin} H. M.~J. {et~al.}, 2018, \aap, 619, A84

\bibitem[{{Bowman} \& {Kurtz}(2018)}]{2018MNRAS.476.3169B}
{Bowman} D.~M., {Kurtz} D.~W., 2018, \mnras, 476, 3169

\bibitem[{{Bowman} {et~al}\mbox{.}(2016){Bowman}, {Kurtz}, {Breger}, {Murphy},
  \& {Holdsworth}}]{2016MNRAS.460.1970B}
{Bowman} D.~M., {Kurtz} D.~W., {Breger} M., {Murphy} S.~J., {Holdsworth} D.~L.,
  2016, \mnras, 460, 1970

\bibitem[{{Cantiello}, {Fuller} \& {Bildsten}(2016){Cantiello}, {Fuller}, \&
  {Bildsten}}]{2016ApJ...824...14C}
{Cantiello} M., {Fuller} J., {Bildsten} L., 2016, \apj, 824, 14

\bibitem[{{Carter}, {Rappaport} \& {Fabrycky}(2011){Carter}, {Rappaport}, \&
  {Fabrycky}}]{2011ApJ...728..139C}
{Carter} J.~A., {Rappaport} S., {Fabrycky} D., 2011, \apj, 728, 139

\bibitem[{{Castelli} \& {Kurucz}(2003)}]{2003IAUS..210P.A20C}
{Castelli} F., {Kurucz} R.~L., 2003, in IAU Symposium, Vol. 210, Modelling of
  Stellar Atmospheres, {Piskunov} N., {Weiss} W.~W., {Gray} D.~F., eds., p. A20

\bibitem[{{Choi} {et~al}\mbox{.}(2016){Choi}, {Dotter}, {Conroy}, {Cantiello},
  {Paxton}, \& {Johnson}}]{2016ApJ...823..102C}
{Choi} J., {Dotter} A., {Conroy} C., {Cantiello} M., {Paxton} B., {Johnson}
  B.~D., 2016, \apj, 823, 102

\bibitem[{{Clemens} \& {Rosen}(2004)}]{2004ApJ...609..340C}
{Clemens} J.~C., {Rosen} R., 2004, \apj, 609, 340

\bibitem[{{Crawford}(1979)}]{1979AJ.....84.1858C}
{Crawford} D.~L., 1979, \aj, 84, 1858

\bibitem[{{Cutri} \& {et al.}(2013)}]{2013yCat.2328....0C}
{Cutri} R.~M., {et al.}, 2013, VizieR Online Data Catalog, II/328

\bibitem[{{Devor} \& {Charbonneau}(2006)}]{2006ApJ...653..647D}
{Devor} J., {Charbonneau} D., 2006, \apj, 653, 647

\bibitem[{{Dotter}(2016)}]{2016ApJS..222....8D}
{Dotter} A., 2016, \apjs, 222, 8

\bibitem[{ESA(1997)}]{1997ESASP1200.....E}
ESA, 1997, in ESA Special Publication, Vol. 1200, ESA Special Publication

\bibitem[{{Ford}(2005)}]{2005AJ....129.1706F}
{Ford} E.~B., 2005, \aj, 129, 1706

\bibitem[{{Gaia Collaboration} {et~al}\mbox{.}(2018){Gaia Collaboration},
  {Brown}, {Vallenari}, {Prusti}, {de Bruijne}, {Babusiaux}, {Bailer-Jones},
  {Biermann}, {Evans}, {Eyer}, {Jansen}, {Jordi}, {Klioner}, {Lammers},
  {Lindegren}, {Luri}, {Mignard}, {Panem}, {Pourbaix}, {Randich}, {Sartoretti},
  {Siddiqui}, {Soubiran}, {van Leeuwen}, {Walton}, {Arenou}, {Bastian},
  {Cropper}, {Drimmel}, {Katz}, {Lattanzi}, {Bakker}, {Cacciari},
  {Casta{\~n}eda}, {Chaoul}, {Cheek}, {De Angeli}, {Fabricius}, {Guerra},
  {Holl}, {Masana}, {Messineo}, {Mowlavi}, {Nienartowicz}, {Panuzzo},
  {Portell}, {Riello}, {Seabroke}, {Tanga}, {Th{\'e}venin}, {Gracia-Abril},
  {Comoretto}, {Garcia-Reinaldos}, {Teyssier}, {Altmann}, {Andrae}, {Audard},
  {Bellas-Velidis}, {Benson}, {Berthier}, {Blomme}, {Burgess}, {Busso},
  {Carry}, {Cellino}, {Clementini}, {Clotet}, {Creevey}, {Davidson}, {De
  Ridder}, {Delchambre}, {Dell'Oro}, {Ducourant},
  {Fern{\'a}ndez-Hern{\'a}ndez}, {Fouesneau}, {Fr{\'e}mat}, {Galluccio},
  {Garc{\'\i}a-Torres}, {Gonz{\'a}lez-N{\'u}{\~n}ez}, {Gonz{\'a}lez-Vidal},
  {Gosset}, {Guy}, {Halbwachs}, {Hambly}, {Harrison}, {Hern{\'a}ndez},
  {Hestroffer}, {Hodgkin}, {Hutton}, {Jasniewicz}, {Jean-Antoine-Piccolo},
  {Jordan}, {Korn}, {Krone-Martins}, {Lanzafame}, {Lebzelter}, {L{\"o}ffler},
  {Manteiga}, {Marrese}, {Mart{\'\i}n-Fleitas}, {Moitinho}, {Mora}, {Muinonen},
  {Osinde}, {Pancino}, {Pauwels}, {Petit}, {Recio-Blanco}, {Richards},
  {Rimoldini}, {Robin}, {Sarro}, {Siopis}, {Smith}, {Sozzetti}, {S{\"u}veges},
  {Torra}, {van Reeven}, {Abbas}, {Abreu Aramburu}, {Accart}, {Aerts},
  {Altavilla}, {{\'A}lvarez}, {Alvarez}, {Alves}, {Anderson}, {Andrei},
  {Anglada Varela}, {Antiche}, {Antoja}, {Arcay}, {Astraatmadja}, {Bach},
  {Baker}, {Balaguer-N{\'u}{\~n}ez}, {Balm}, {Barache}, {Barata}, {Barbato},
  {Barblan}, {Barklem}, {Barrado}, {Barros}, {Barstow}, {Bartholom{\'e}
  Mu{\~n}oz}, {Bassilana}, {Becciani}, {Bellazzini}, {Berihuete}, {Bertone},
  {Bianchi}, {Bienaym{\'e}}, {Blanco-Cuaresma}, {Boch}, {Boeche}, {Bombrun},
  {Borrachero}, {Bossini}, {Bouquillon}, {Bourda}, {Bragaglia}, {Bramante},
  {Breddels}, {Bressan}, {Brouillet}, {Br{\"u}semeister}, {Brugaletta},
  {Bucciarelli}, {Burlacu}, {Busonero}, {Butkevich}, {Buzzi}, {Caffau},
  {Cancelliere}, {Cannizzaro}, {Cantat-Gaudin}, {Carballo}, {Carlucci},
  {Carrasco}, {Casamiquela}, {Castellani}, {Castro-Ginard}, {Charlot},
  {Chemin}, {Chiavassa}, {Cocozza}, {Costigan}, {Cowell}, {Crifo}, {Crosta},
  {Crowley}, {Cuypers}, {Dafonte}, {Damerdji}, {Dapergolas}, {David}, {David},
  {de Laverny}, {De Luise}, {De March}, {de Martino}, {de Souza}, {de Torres},
  {Debosscher}, {del Pozo}, {Delbo}, {Delgado}, {Delgado}, {Di Matteo},
  {Diakite}, {Diener}, {Distefano}, {Dolding}, {Drazinos}, {Dur{\'a}n},
  {Edvardsson}, {Enke}, {Eriksson}, {Esquej}, {Eynard Bontemps}, {Fabre},
  {Fabrizio}, {Faigler}, {Falc{\~a}o}, {Farr{\`a}s Casas}, {Federici},
  {Fedorets}, {Fernique}, {Figueras}, {Filippi}, {Findeisen}, {Fonti},
  {Fraile}, {Fraser}, {Fr{\'e}zouls}, {Gai}, {Galleti}, {Garabato},
  {Garc{\'\i}a-Sedano}, {Garofalo}, {Garralda}, {Gavel}, {Gavras}, {Gerssen},
  {Geyer}, {Giacobbe}, {Gilmore}, {Girona}, {Giuffrida}, {Glass}, {Gomes},
  {Granvik}, {Gueguen}, {Guerrier}, {Guiraud}, {Guti{\'e}rrez-S{\'a}nchez},
  {Haigron}, {Hatzidimitriou}, {Hauser}, {Haywood}, {Heiter}, {Helmi}, {Heu},
  {Hilger}, {Hobbs}, {Hofmann}, {Holland}, {Huckle}, {Hypki}, {Icardi},
  {Jan{\ss}en}, {Jevardat de Fombelle}, {Jonker}, {Juh{\'a}sz}, {Julbe},
  {Karampelas}, {Kewley}, {Klar}, {Kochoska}, {Kohley}, {Kolenberg},
  {Kontizas}, {Kontizas}, {Koposov}, {Kordopatis}, {Kostrzewa-Rutkowska},
  {Koubsky}, {Lambert}, {Lanza}, {Lasne}, {Lavigne}, {Le Fustec}, {Le
  Poncin-Lafitte}, {Lebreton}, {Leccia}, {Leclerc}, {Lecoeur-Taibi},
  {Lenhardt}, {Leroux}, {Liao}, {Licata}, {Lindstr{\o}m}, {Lister}, {Livanou},
  {Lobel}, {L{\'o}pez}, {Managau}, {Mann}, {Mantelet}, {Marchal}, {Marchant},
  {Marconi}, {Marinoni}, {Marschalk{\'o}}, {Marshall}, {Martino}, {Marton},
  {Mary}, {Massari}, {Matijevi{\v{c}}}, {Mazeh}, {McMillan}, {Messina},
  {Michalik}, {Millar}, {Molina}, {Molinaro}, {Moln{\'a}r}, {Montegriffo},
  {Mor}, {Morbidelli}, {Morel}, {Morris}, {Mulone}, {Muraveva}, {Musella},
  {Nelemans}, {Nicastro}, {Noval}, {O'Mullane}, {Ord{\'e}novic},
  {Ord{\'o}{\~n}ez-Blanco}, {Osborne}, {Pagani}, {Pagano}, {Pailler},
  {Palacin}, {Palaversa}, {Panahi}, {Pawlak}, {Piersimoni}, {Pineau}, {Plachy},
  {Plum}, {Poggio}, {Poujoulet}, {Pr{\v{s}}a}, {Pulone}, {Racero}, {Ragaini},
  {Rambaux}, {Ramos-Lerate}, {Regibo}, {Reyl{\'e}}, {Riclet}, {Ripepi}, {Riva},
  {Rivard}, {Rixon}, {Roegiers}, {Roelens}, {Romero-G{\'o}mez}, {Rowell},
  {Royer}, {Ruiz-Dern}, {Sadowski}, {Sagrist{\`a} Sell{\'e}s}, {Sahlmann},
  {Salgado}, {Salguero}, {Sanna}, {Santana-Ros}, {Sarasso}, {Savietto},
  {Schultheis}, {Sciacca}, {Segol}, {Segovia}, {S{\'e}gransan}, {Shih},
  {Siltala}, {Silva}, {Smart}, {Smith}, {Solano}, {Solitro}, {Sordo}, {Soria
  Nieto}, {Souchay}, {Spagna}, {Spoto}, {Stampa}, {Steele},
  {Steidelm{\"u}ller}, {Stephenson}, {Stoev}, {Suess}, {Surdej}, {Szabados},
  {Szegedi-Elek}, {Tapiador}, {Taris}, {Tauran}, {Taylor}, {Teixeira},
  {Terrett}, {Teyssand ier}, {Thuillot}, {Titarenko}, {Torra Clotet}, {Turon},
  {Ulla}, {Utrilla}, {Uzzi}, {Vaillant}, {Valentini}, {Valette}, {van Elteren},
  {Van Hemelryck}, {van Leeuwen}, {Vaschetto}, {Vecchiato}, {Veljanoski},
  {Viala}, {Vicente}, {Vogt}, {von Essen}, {Voss}, {Votruba}, {Voutsinas},
  {Walmsley}, {Weiler}, {Wertz}, {Wevers}, {Wyrzykowski}, {Yoldas},
  {{\v{Z}}erjal}, {Ziaeepour}, {Zorec}, {Zschocke}, {Zucker}, {Zurbach}, \&
  {Zwitter}}]{2018A&A...616A...1G}
{Gaia Collaboration} {et~al.}, 2018, \aap, 616, A1

\bibitem[{{Gaia Collaboration} {et~al}\mbox{.}(2016){Gaia Collaboration},
  {Brown}, {Vallenari}, {Prusti}, {de Bruijne}, {Mignard}, {Drimmel},
  {Babusiaux}, {Bailer-Jones}, {Bastian}, {Biermann}, {Evans}, {Eyer},
  {Jansen}, {Jordi}, {Katz}, {Klioner}, {Lammers}, {Lindegren}, {Luri},
  {O'Mullane}, {Panem}, {Pourbaix}, {Randich}, {Sartoretti}, {Siddiqui},
  {Soubiran}, {Valette}, {van Leeuwen}, {Walton}, {Aerts}, {Arenou}, {Cropper},
  {H{\o}g}, {Lattanzi}, {Grebel}, {Holland}, {Huc}, {Passot}, {Perryman},
  {Bramante}, {Cacciari}, {Casta{\~n}eda}, {Chaoul}, {Cheek}, {De Angeli},
  {Fabricius}, {Guerra}, {Hern{\'a}ndez}, {Jean-Antoine-Piccolo}, {Masana},
  {Messineo}, {Mowlavi}, {Nienartowicz}, {Ord{\'o}{\~n}ez-Blanco}, {Panuzzo},
  {Portell}, {Richards}, {Riello}, {Seabroke}, {Tanga}, {Th{\'e}venin},
  {Torra}, {Els}, {Gracia-Abril}, {Comoretto}, {Garcia-Reinaldos}, {Lock},
  {Mercier}, {Altmann}, {Andrae}, {Astraatmadja}, {Bellas-Velidis}, {Benson},
  {Berthier}, {Blomme}, {Busso}, {Carry}, {Cellino}, {Clementini}, {Cowell},
  {Creevey}, {Cuypers}, {Davidson}, {De Ridder}, {de Torres}, {Delchambre},
  {Dell'Oro}, {Ducourant}, {Fr{\'e}mat}, {Garc{\'\i}a-Torres}, {Gosset},
  {Halbwachs}, {Hambly}, {Harrison}, {Hauser}, {Hestroffer}, {Hodgkin},
  {Huckle}, {Hutton}, {Jasniewicz}, {Jordan}, {Kontizas}, {Korn}, {Lanzafame},
  {Manteiga}, {Moitinho}, {Muinonen}, {Osinde}, {Pancino}, {Pauwels}, {Petit},
  {Recio-Blanco}, {Robin}, {Sarro}, {Siopis}, {Smith}, {Smith}, {Sozzetti},
  {Thuillot}, {van Reeven}, {Viala}, {Abbas}, {Abreu Aramburu}, {Accart},
  {Aguado}, {Allan}, {Allasia}, {Altavilla}, {{\'A}lvarez}, {Alves},
  {Anderson}, {Andrei}, {Anglada Varela}, {Antiche}, {Antoja}, {Ant{\'o}n},
  {Arcay}, {Bach}, {Baker}, {Balaguer-N{\'u}{\~n}ez}, {Barache}, {Barata},
  {Barbier}, {Barblan}, {Barrado y Navascu{\'e}s}, {Barros}, {Barstow},
  {Becciani}, {Bellazzini}, {Bello Garc{\'\i}a}, {Belokurov}, {Bendjoya},
  {Berihuete}, {Bianchi}, {Bienaym{\'e}}, {Billebaud}, {Blagorodnova},
  {Blanco-Cuaresma}, {Boch}, {Bombrun}, {Borrachero}, {Bouquillon}, {Bourda},
  {Bouy}, {Bragaglia}, {Breddels}, {Brouillet}, {Br{\"u}semeister},
  {Bucciarelli}, {Burgess}, {Burgon}, {Burlacu}, {Busonero}, {Buzzi}, {Caffau},
  {Cambras}, {Campbell}, {Cancelliere}, {Cantat-Gaudin}, {Carlucci},
  {Carrasco}, {Castellani}, {Charlot}, {Charnas}, {Chiavassa}, {Clotet},
  {Cocozza}, {Collins}, {Costigan}, {Crifo}, {Cross}, {Crosta}, {Crowley},
  {Dafonte}, {Damerdji}, {Dapergolas}, {David}, {David}, {De Cat}, {de Felice},
  {de Laverny}, {De Luise}, {De March}, {de Martino}, {de Souza}, {Debosscher},
  {del Pozo}, {Delbo}, {Delgado}, {Delgado}, {Di Matteo}, {Diakite},
  {Distefano}, {Dolding}, {Dos Anjos}, {Drazinos}, {Duran}, {Dzigan},
  {Edvardsson}, {Enke}, {Evans}, {Eynard Bontemps}, {Fabre}, {Fabrizio},
  {Faigler}, {Falc{\~a}o}, {Farr{\`a}s Casas}, {Federici}, {Fedorets},
  {Fern{\'a}ndez-Hern{\'a}ndez}, {Fernique}, {Fienga}, {Figueras}, {Filippi},
  {Findeisen}, {Fonti}, {Fouesneau}, {Fraile}, {Fraser}, {Fuchs}, {Gai},
  {Galleti}, {Galluccio}, {Garabato}, {Garc{\'\i}a-Sedano}, {Garofalo},
  {Garralda}, {Gavras}, {Gerssen}, {Geyer}, {Gilmore}, {Girona}, {Giuffrida},
  {Gomes}, {Gonz{\'a}lez-Marcos}, {Gonz{\'a}lez-N{\'u}{\~n}ez},
  {Gonz{\'a}lez-Vidal}, {Granvik}, {Guerrier}, {Guillout}, {Guiraud},
  {G{\'u}rpide}, {Guti{\'e}rrez-S{\'a}nchez}, {Guy}, {Haigron},
  {Hatzidimitriou}, {Haywood}, {Heiter}, {Helmi}, {Hobbs}, {Hofmann}, {Holl},
  {Holland }, {Hunt}, {Hypki}, {Icardi}, {Irwin}, {Jevardat de Fombelle},
  {Jofr{\'e}}, {Jonker}, {Jorissen}, {Julbe}, {Karampelas}, {Kochoska},
  {Kohley}, {Kolenberg}, {Kontizas}, {Koposov}, {Kordopatis}, {Koubsky},
  {Krone-Martins}, {Kudryashova}, {Kull}, {Bachchan}, {Lacoste-Seris}, {Lanza},
  {Lavigne}, {Le Poncin-Lafitte}, {Lebreton}, {Lebzelter}, {Leccia}, {Leclerc},
  {Lecoeur-Taibi}, {Lemaitre}, {Lenhardt}, {Leroux}, {Liao}, {Licata},
  {Lindstr{\o}m}, {Lister}, {Livanou}, {Lobel}, {L{\"o}ffler}, {L{\'o}pez},
  {Lorenz}, {MacDonald}, {Magalh{\~a}es Fernandes}, {Managau}, {Mann},
  {Mantelet}, {Marchal}, {Marchant}, {Marconi}, {Marinoni}, {Marrese},
  {Marschalk{\'o}}, {Marshall}, {Mart{\'\i}n-Fleitas}, {Martino}, {Mary},
  {Matijevi{\v{c}}}, {Mazeh}, {McMillan}, {Messina}, {Michalik}, {Millar},
  {Mirand a}, {Molina}, {Molinaro}, {Molinaro}, {Moln{\'a}r}, {Moniez},
  {Montegriffo}, {Mor}, {Mora}, {Morbidelli}, {Morel}, {Morgenthaler},
  {Morris}, {Mulone}, {Muraveva}, {Musella}, {Narbonne}, {Nelemans},
  {Nicastro}, {Noval}, {Ord{\'e}novic}, {Ordieres-Mer{\'e}}, {Osborne},
  {Pagani}, {Pagano}, {Pailler}, {Palacin}, {Palaversa}, {Parsons}, {Pecoraro},
  {Pedrosa}, {Pentik{\"a}inen}, {Pichon}, {Piersimoni}, {Pineau}, {Plachy},
  {Plum}, {Poujoulet}, {Pr{\v{s}}a}, {Pulone}, {Ragaini}, {Rago}, {Rambaux},
  {Ramos-Lerate}, {Ranalli}, {Rauw}, {Read}, {Regibo}, {Reyl{\'e}}, {Ribeiro},
  {Rimoldini}, {Ripepi}, {Riva}, {Rixon}, {Roelens}, {Romero-G{\'o}mez},
  {Rowell}, {Royer}, {Ruiz-Dern}, {Sadowski}, {Sagrist{\`a} Sell{\'e}s},
  {Sahlmann}, {Salgado}, {Salguero}, {Sarasso}, {Savietto}, {Schultheis},
  {Sciacca}, {Segol}, {Segovia}, {Segransan}, {Shih}, {Smareglia}, {Smart},
  {Solano}, {Solitro}, {Sordo}, {Soria Nieto}, {Souchay}, {Spagna}, {Spoto},
  {Stampa}, {Steele}, {Steidelm{\"u}ller}, {Stephenson}, {Stoev}, {Suess},
  {S{\"u}veges}, {Surdej}, {Szabados}, {Szegedi-Elek}, {Tapiador}, {Taris},
  {Tauran}, {Taylor}, {Teixeira}, {Terrett}, {Tingley}, {Trager}, {Turon},
  {Ulla}, {Utrilla}, {Valentini}, {van Elteren}, {Van Hemelryck}, {van
  Leeuwen}, {Varadi}, {Vecchiato}, {Veljanoski}, {Via}, {Vicente}, {Vogt},
  {Voss}, {Votruba}, {Voutsinas}, {Walmsley}, {Weiler}, {Weingrill}, {Wevers},
  {Wyrzykowski}, {Yoldas}, {{\v{Z}}erjal}, {Zucker}, {Zurbach}, {Zwitter},
  {Alecu}, {Allen}, {Allende Prieto}, {Amorim}, {Anglada-Escud{\'e}},
  {Arsenijevic}, {Azaz}, {Balm}, {Beck}, {Bernstein}, {Bigot}, {Bijaoui},
  {Blasco}, {Bonfigli}, {Bono}, {Boudreault}, {Bressan}, {Brown}, {Brunet},
  {Bunclark}, {Buonanno}, {Butkevich}, {Carret}, {Carrion}, {Chemin},
  {Ch{\'e}reau}, {Corcione}, {Darmigny}, {de Boer}, {de Teodoro}, {de Zeeuw},
  {Delle Luche}, {Domingues}, {Dubath}, {Fodor}, {Fr{\'e}zouls}, {Fries},
  {Fustes}, {Fyfe}, {Gallardo}, {Gallegos}, {Gardiol}, {Gebran}, {Gomboc},
  {G{\'o}mez}, {Grux}, {Gueguen}, {Heyrovsky}, {Hoar}, {Iannicola}, {Isasi
  Parache}, {Janotto}, {Joliet}, {Jonckheere}, {Keil}, {Kim}, {Klagyivik},
  {Klar}, {Knude}, {Kochukhov}, {Kolka}, {Kos}, {Kutka}, {Lainey}, {LeBouquin},
  {Liu}, {Loreggia}, {Makarov}, {Marseille}, {Martayan}, {Martinez-Rubi},
  {Massart}, {Meynadier}, {Mignot}, {Munari}, {Nguyen}, {Nordlander}, {Ocvirk},
  {O'Flaherty}, {Olias Sanz}, {Ortiz}, {Osorio}, {Oszkiewicz}, {Ouzounis},
  {Palmer}, {Park}, {Pasquato}, {Peltzer}, {Peralta}, {P{\'e}turaud},
  {Pieniluoma}, {Pigozzi}, {Poels}, {Prat}, {Prod'homme}, {Raison}, {Rebordao},
  {Risquez}, {Rocca-Volmerange}, {Rosen}, {Ruiz-Fuertes}, {Russo}, {Sembay},
  {Serraller Vizcaino}, {Short}, {Siebert}, {Silva}, {Sinachopoulos}, {Slezak},
  {Soffel}, {Sosnowska}, {Strai{\v{z}}ys}, {ter Linden}, {Terrell}, {Theil},
  {Tiede}, {Troisi}, {Tsalmantza}, {Tur}, {Vaccari}, {Vachier}, {Valles}, {Van
  Hamme}, {Veltz}, {Virtanen}, {Wallut}, {Wichmann}, {Wilkinson}, {Ziaeepour},
  \& {Zschocke}}]{2016A&A...595A...2G}
---, 2016, \aap, 595, A2

\bibitem[{{Gaulme} \& {Guzik}(2019)}]{2019A&A...630A.106G}
{Gaulme} P., {Guzik} J.~A., 2019, \aap, 630, A106

\bibitem[{{Gizon} {et~al}\mbox{.}(2016){Gizon}, {Sekii}, {Takata}, {Kurtz},
  {Shibahashi}, {Bazot}, {Benomar}, {Birch}, \&
  {Sreenivasan}}]{2016SciA....2E1777G}
{Gizon} L. {et~al.}, 2016, Science Advances, 2, e1601777

\bibitem[{{Gray} {et~al}\mbox{.}(2003){Gray}, {Corbally}, {Garrison},
  {McFadden}, \& {Robinson}}]{2003AJ....126.2048G}
{Gray} R.~O., {Corbally} C.~J., {Garrison} R.~F., {McFadden} M.~T., {Robinson}
  P.~E., 2003, \aj, 126, 2048

\bibitem[{{Guo} {et~al}\mbox{.}(2019){Guo}, {Shporer}, {Hambleton}, \&
  {Isaacson}}]{2019arXiv191108687G}
{Guo} Z., {Shporer} A., {Hambleton} K., {Isaacson} H., 2019, arXiv e-prints,
  arXiv:1911.08687

\bibitem[{{Handler} {et~al}\mbox{.}(2020){Handler}, {Kurtz}, {Rappaport},
  {Saio}, {Fuller}, {Jones}, {Guo}, {Chowdhury}, {Sowicka},
  {Ali{\c{c}}avu{\textcommabelow s}}, {Streamer}, {Murphy}, {Gagliano},
  {Jacobs}, \& {Vanderburg}}]{2020NatAs.tmp...45H}
{Handler} G. {et~al.}, 2020, Nature Astronomy

\bibitem[{{Hastings}(1970)}]{hastings}
{Hastings} W., 1970, Biometrica, 57, 97

\bibitem[{{Hatta} {et~al}\mbox{.}(2019){Hatta}, {Sekii}, {Takata}, \&
  {Kurtz}}]{2019ApJ...871..135H}
{Hatta} Y., {Sekii} T., {Takata} M., {Kurtz} D.~W., 2019, \apj, 871, 135

\bibitem[{{Herwig}(2000)}]{herwig00}
{Herwig} F., 2000, \aap, 360, 952

\bibitem[{{Horvat} {et~al}\mbox{.}(2019){Horvat}, {Conroy}, {Jones}, \&
  {Pr{\v{s}}a}}]{2019ApJS..240...36H}
{Horvat} M., {Conroy} K.~E., {Jones} D., {Pr{\v{s}}a} A., 2019, \apjs, 240, 36

\bibitem[{{Jones} {et~al}\mbox{.}(2019{\natexlab{a}}){Jones}, {Boffin},
  {Sowicka}, {Miszalski}, {Rodr{\'\i}guez-Gil}, {Santand er-Garc{\'\i}a}, \&
  {Corradi}}]{2019MNRAS.482L..75J}
{Jones} D., {Boffin} H. M.~J., {Sowicka} P., {Miszalski} B.,
  {Rodr{\'\i}guez-Gil} P., {Santand er-Garc{\'\i}a} M., {Corradi} R. L.~M.,
  2019{\natexlab{a}}, \mnras, 482, L75

\bibitem[{{Jones} {et~al}\mbox{.}(2019{\natexlab{b}}){Jones}, {Conroy},
  {Horvat}, {Giammarco}, {Kochoska}, {Pablo}, {Brown}, {Sowicka}, \&
  {Prsa}}]{2019arXiv191209474J}
{Jones} D. {et~al.}, 2019{\natexlab{b}}, arXiv e-prints, arXiv:1912.09474

\bibitem[{{Kahraman Ali\c{c}avu\c{s}} {et~al}\mbox{.}(2016){Kahraman
  Ali\c{c}avu\c{s}}, {Niemczura}, {De Cat}, {Soydugan}, {Ko{\l}aczkowski},
  {Ostrowski}, {Telting}, {Uytterhoeven}, {Poretti}, {Rainer}, {Su{\'a}rez},
  {Mantegazza}, {Kilmartin}, \& {Pollard}}]{2016MNRAS.458.2307K}
{Kahraman Ali\c{c}avu\c{s}} F. {et~al.}, 2016, \mnras, 458, 2307

\bibitem[{{Kamiaka}, {Benomar} \& {Suto}(2018){Kamiaka}, {Benomar}, \&
  {Suto}}]{2018MNRAS.479..391K}
{Kamiaka} S., {Benomar} O., {Suto} Y., 2018, \mnras, 479, 391

\bibitem[{{Kopal}(1959)}]{1959cbs..book.....K}
{Kopal} Z., 1959, {Close binary systems}

\bibitem[{{Kurtz}(1976)}]{1976ApJS...32..651K}
{Kurtz} D.~W., 1976, \apjs, 32, 651

\bibitem[{{Kurtz}(1982)}]{1982MNRAS.200..807K}
---, 1982, \mnras, 200, 807

\bibitem[{{Kurtz}(1985)}]{1985MNRAS.213..773K}
---, 1985, \mnras, 213, 773

\bibitem[{{Kurtz}(1992)}]{1992MNRAS.259..701K}
---, 1992, \mnras, 259, 701

\bibitem[{{Kurtz} {et~al}\mbox{.}(2014){Kurtz}, {Saio}, {Takata}, {Shibahashi},
  {Murphy}, \& {Sekii}}]{2014MNRAS.444..102K}
{Kurtz} D.~W., {Saio} H., {Takata} M., {Shibahashi} H., {Murphy} S.~J., {Sekii}
  T., 2014, \mnras, 444, 102

\bibitem[{{Ledoux}(1951)}]{1951ApJ...114..373L}
{Ledoux} P., 1951, \apj, 114, 373

\bibitem[{{Lee}(1916)}]{1916ApJ....43..320L}
{Lee} O.~J., 1916, \apj, 43, 320

\bibitem[{{Leone} {et~al}\mbox{.}(2016){Leone}, {Avila}, {Bellassai}, {Bruno},
  {Catalano}, {Di Benedetto}, {Di Stefano}, {Gangi}, {Giarrusso}, {Greco},
  {Martinetti}, {Miraglia}, {Munari}, {Pontoni}, {Scalia}, {Scuderi}, \&
  {Span{\'o}}}]{2016AJ....151..116L}
{Leone} F. {et~al.}, 2016, \aj, 151, 116

\bibitem[{{Liakos} \& {Niarchos}(2017)}]{2017MNRAS.465.1181L}
{Liakos} A., {Niarchos} P., 2017, \mnras, 465, 1181

\bibitem[{{Lindegren} {et~al}\mbox{.}(2018){Lindegren}, {Hern{\'a}ndez},
  {Bombrun}, {Klioner}, {Bastian}, {Ramos-Lerate}, {de Torres},
  {Steidelm{\"u}ller}, {Stephenson}, {Hobbs}, {Lammers}, {Biermann}, {Geyer},
  {Hilger}, {Michalik}, {Stampa}, {McMillan}, {Casta{\~n}eda}, {Clotet},
  {Comoretto}, {Davidson}, {Fabricius}, {Gracia}, {Hambly}, {Hutton}, {Mora},
  {Portell}, {van Leeuwen}, {Abbas}, {Abreu}, {Altmann}, {Andrei}, {Anglada},
  {Balaguer-N{\'u}{\~n}ez}, {Barache}, {Becciani}, {Bertone}, {Bianchi},
  {Bouquillon}, {Bourda}, {Br{\"u}semeister}, {Bucciarelli}, {Busonero},
  {Buzzi}, {Cancelliere}, {Carlucci}, {Charlot}, {Cheek}, {Crosta}, {Crowley},
  {de Bruijne}, {de Felice}, {Drimmel}, {Esquej}, {Fienga}, {Fraile}, {Gai},
  {Garralda}, {Gonz{\'a}lez-Vidal}, {Guerra}, {Hauser}, {Hofmann}, {Holl},
  {Jordan}, {Lattanzi}, {Lenhardt}, {Liao}, {Licata}, {Lister}, {L{\"o}ffler},
  {Marchant}, {Martin-Fleitas}, {Messineo}, {Mignard}, {Morbidelli}, {Poggio},
  {Riva}, {Rowell}, {Salguero}, {Sarasso}, {Sciacca}, {Siddiqui}, {Smart},
  {Spagna}, {Steele}, {Taris}, {Torra}, {van Elteren}, {van Reeven}, \&
  {Vecchiato}}]{2018A&A...616A...2L}
{Lindegren} L. {et~al.}, 2018, \aap, 616, A2

\bibitem[{Lindegren {et~al}\mbox{.}(2018)Lindegren {et~al.}}]{ruwe}
Lindegren L., {et~al.}, 2018, Gaia Technical Note: GAIA-C3-TN-LU-LL-124-01

\bibitem[{{Loeb} \& {Gaudi}(2003)}]{2003ApJ...588L.117L}
{Loeb} A., {Gaudi} B.~S., 2003, \apjl, 588, L117

\bibitem[{{Luri} {et~al}\mbox{.}(2018){Luri}, {Brown}, {Sarro}, {Arenou},
  {Bailer-Jones}, {Castro-Ginard}, {de Bruijne}, {Prusti}, {Babusiaux}, \&
  {Delgado}}]{2018A&A...616A...9L}
{Luri} X. {et~al.}, 2018, \aap, 616, A9

\bibitem[{{Maceroni} {et~al}\mbox{.}(2015){Maceroni}, {Lehmann}, {Da Silva}, \&
  {Montalb{\'a}n}}]{2015EPJWC.10104003M}
{Maceroni} C., {Lehmann} H., {Da Silva} R., {Montalb{\'a}n} J., 2015, in
  European Physical Journal Web of Conferences, Vol. 101, European Physical
  Journal Web of Conferences, p. 04003

\bibitem[{{Margoni}, {Munari} \& {Stagni}(1992){Margoni}, {Munari}, \&
  {Stagni}}]{1992A&AS...93..545M}
{Margoni} R., {Munari} U., {Stagni} R., 1992, \aaps, 93, 545

\bibitem[{{Maxted} \& {Hutcheon}(2018)}]{2018A&A...616A..38M}
{Maxted} P.~F.~L., {Hutcheon} R.~J., 2018, \aap, 616, A38

\bibitem[{{Metropolis} {et~al}\mbox{.}(1953){Metropolis}, {Rosenbluth},
  {Rosenbluth}, {Teller}, \& {Teller}}]{1953JChPh..21.1087M}
{Metropolis} N., {Rosenbluth} A.~W., {Rosenbluth} M.~N., {Teller} A.~H.,
  {Teller} E., 1953, \jcp, 21, 1087

\bibitem[{{Mirouh} {et~al}\mbox{.}(2019){Mirouh}, {Angelou}, {Reese}, \&
  {Costa}}]{2019MNRAS.483L..28M}
{Mirouh} G.~M., {Angelou} G.~C., {Reese} D.~R., {Costa} G., 2019, \mnras, 483,
  L28

\bibitem[{{Moe} \& {Di Stefano}(2013)}]{2013ApJ...778...95M}
{Moe} M., {Di Stefano} R., 2013, \apj, 778, 95

\bibitem[{{Moe} \& {Di Stefano}(2015)}]{2015ApJ...810...61M}
---, 2015, \apj, 810, 61

\bibitem[{{Montgomery} {et~al}\mbox{.}(2010){Montgomery}, {Provencal},
  {Kanaan}, {Mukadam}, {Thompson}, {Dalessio}, {Shipman}, {Winget}, {Kepler},
  \& {Koester}}]{2010ApJ...716...84M}
{Montgomery} M.~H. {et~al.}, 2010, \apj, 716, 84

\bibitem[{{Moon} \& {Dworetsky}(1985)}]{1985MNRAS.217..305M}
{Moon} T.~T., {Dworetsky} M.~M., 1985, \mnras, 217, 305

\bibitem[{{Murphy} {et~al}\mbox{.}(2019){Murphy}, {Hey}, {Van Reeth}, \&
  {Bedding}}]{2019MNRAS.485.2380M}
{Murphy} S.~J., {Hey} D., {Van Reeth} T., {Bedding} T.~R., 2019, \mnras, 485,
  2380

\bibitem[{{Murphy} {et~al}\mbox{.}(2018){Murphy}, {Moe}, {Kurtz}, {Bedding},
  {Shibahashi}, \& {Boffin}}]{2018MNRAS.474.4322M}
{Murphy} S.~J., {Moe} M., {Kurtz} D.~W., {Bedding} T.~R., {Shibahashi} H.,
  {Boffin} H. M.~J., 2018, \mnras, 474, 4322

\bibitem[{{Paxton} {et~al}\mbox{.}(2011){Paxton}, {Bildsten}, {Dotter},
  {Herwig}, {Lesaffre}, \& {Timmes}}]{2011ApJS..192....3P}
{Paxton} B., {Bildsten} L., {Dotter} A., {Herwig} F., {Lesaffre} P., {Timmes}
  F., 2011, \apjs, 192, 3

\bibitem[{{Paxton} {et~al}\mbox{.}(2013){Paxton}, {Cantiello}, {Arras},
  {Bildsten}, {Brown}, {Dotter}, {Mankovich}, {Montgomery}, {Stello}, {Timmes},
  \& {Townsend}}]{paxton2013}
{Paxton} B. {et~al.}, 2013, \apjs, 208, 4

\bibitem[{{Paxton} {et~al}\mbox{.}(2015){Paxton}, {Marchant}, {Schwab},
  {Bauer}, {Bildsten}, {Cantiello}, {Dessart}, {Farmer}, {Hu}, {Langer},
  {Townsend}, {Townsley}, \& {Timmes}}]{2015ApJS..220...15P}
---, 2015, \apjs, 220, 15

\bibitem[{{Paxton} {et~al}\mbox{.}(2018){Paxton}, {Schwab}, {Bauer},
  {Bildsten}, {Blinnikov}, {Duffell}, {Farmer}, {Goldberg}, {Marchant},
  {Sorokina}, {Thoul}, {Townsend}, \& {Timmes}}]{paxton:18}
---, 2018, \apjs, 234, 34

\bibitem[{{Paxton} {et~al}\mbox{.}(2019){Paxton}, {Smolec}, {Gautschy},
  {Bildsten}, {Cantiello}, {Dotter}, {Farmer}, {Goldberg}, {Jermyn}, {Kanbur},
  {Marchant}, {Schwab}, {Thoul}, {Townsend}, {Wolf}, {Zhang}, \&
  {Timmes}}]{paxton:19}
---, 2019, arXiv e-prints

\bibitem[{{Pr{\v{s}}a} {et~al}\mbox{.}(2016){Pr{\v{s}}a}, {Conroy}, {Horvat},
  {Pablo}, {Kochoska}, {Bloemen}, {Giammarco}, {Hambleton}, \&
  {Degroote}}]{2016ApJS..227...29P}
{Pr{\v{s}}a} A. {et~al.}, 2016, \apjs, 227, 29

\bibitem[{{Rappaport} {et~al}\mbox{.}(2019){Rappaport}, {Vanderburg},
  {Kristiansen}, {Omohundro}, {Schwengeler}, {Terentev}, {Dai}, {Masuda},
  {Jacobs}, {LaCourse}, {Latham}, {Bieryla}, {Hedges}, {Dittmann}, {Barentsen},
  {Cochran}, {Endl}, {Jenkins}, \& {Mann}}]{2019MNRAS.488.2455R}
{Rappaport} S. {et~al.}, 2019, \mnras, 488, 2455

\bibitem[{{Reese} {et~al}\mbox{.}(2009{\natexlab{a}}){Reese}, {MacGregor},
  {Jackson}, {Skumanich}, \& {Metcalfe}}]{2009A&A...506..189R}
{Reese} D.~R., {MacGregor} K.~B., {Jackson} S., {Skumanich} A., {Metcalfe}
  T.~S., 2009{\natexlab{a}}, \aap, 506, 189

\bibitem[{{Reese} {et~al}\mbox{.}(2009{\natexlab{b}}){Reese}, {MacGregor},
  {Jackson}, {Skumanich}, \& {Metcalfe}}]{2009ASPC..416..395R}
---, 2009{\natexlab{b}}, Astronomical Society of the Pacific Conference Series,
  Vol. 416, {Pulsation Modes with High Azimuthal Orders in Stellar Models Based
  on the Self-Consistent Field Method}, {Dikpati} M., {Arentoft} T.,
  {Gonz{\'a}lez Hern{\'a}ndez} I., {Lindsey} C., {Hill} F., eds., p. 395

\bibitem[{{Schmitt}, {Hartman} \& {Kipping}(2019){Schmitt}, {Hartman}, \&
  {Kipping}}]{2019arXiv191008034S}
{Schmitt} A.~R., {Hartman} J.~D., {Kipping} D.~M., 2019, arXiv e-prints,
  arXiv:1910.08034

\bibitem[{{Schr{\"o}der}, {Reiners} \& {Schmitt}(2009){Schr{\"o}der},
  {Reiners}, \& {Schmitt}}]{2009A&A...493.1099S}
{Schr{\"o}der} C., {Reiners} A., {Schmitt} J.~H.~M.~M., 2009, \aap, 493, 1099

\bibitem[{{Shibahashi}(2000)}]{2000ASPC..203..299S}
{Shibahashi} H., 2000, Astronomical Society of the Pacific Conference Series,
  Vol. 203, {The Oblique Pulsator Model for the Blazhko Effect in RR Lyrae
  Stars. Theory of Amplitude Modulation I.}, {Szabados} L., {Kurtz} D., eds.,
  pp. 299--306

\bibitem[{{Shibahashi} \& {Aerts}(2000)}]{2000ApJ...531L.143S}
{Shibahashi} H., {Aerts} C., 2000, \apjl, 531, L143

\bibitem[{{Shibahashi} \& {Kurtz}(2012)}]{2012MNRAS.422..738S}
{Shibahashi} H., {Kurtz} D.~W., 2012, \mnras, 422, 738

\bibitem[{{Shibahashi} \& {Saio}(1985)}]{1985PASJ...37..245S}
{Shibahashi} H., {Saio} H., 1985, \pasj, 37, 245

\bibitem[{{Shibahashi} \& {Takata}(1993)}]{1993PASJ...45..617S}
{Shibahashi} H., {Takata} M., 1993, \pasj, 45, 617

\bibitem[{{Skrutskie} {et~al}\mbox{.}(2006){Skrutskie}, {Cutri}, {Stiening},
  {Weinberg}, {Schneider}, {Carpenter}, {Beichman}, {Capps}, {Chester},
  {Elias}, {Huchra}, {Liebert}, {Lonsdale}, {Monet}, {Price}, {Seitzer},
  {Jarrett}, {Kirkpatrick}, {Gizis}, {Howard}, {Evans}, {Fowler}, {Fullmer},
  {Hurt}, {Light}, {Kopan}, {Marsh}, {McCallon}, {Tam}, {Van Dyk}, \&
  {Wheelock}}]{2006AJ....131.1163S}
{Skrutskie} M.~F. {et~al.}, 2006, \aj, 131, 1163

\bibitem[{{Smalley} {et~al}\mbox{.}(2017){Smalley}, {Antoci}, {Holdsworth},
  {Kurtz}, {Murphy}, {De Cat}, {Anderson}, {Catanzaro}, {Cameron}, {Hellier},
  {Maxted}, {Norton}, {Pollacco}, {Ripepi}, {West}, \&
  {Wheatley}}]{2017MNRAS.465.2662S}
{Smalley} B. {et~al.}, 2017, \mnras, 465, 2662

\bibitem[{{Takata} \& {Shibahashi}(1995)}]{1995PASJ...47..219T}
{Takata} M., {Shibahashi} H., 1995, \pasj, 47, 219

\bibitem[{{Tauris} \& {van den Heuvel}(2006)}]{2006csxs.book..623T}
{Tauris} T.~M., {van den Heuvel} E.~P.~J., 2006, {Formation and evolution of
  compact stellar X-ray sources}, Vol.~39, pp. 623--665

\bibitem[{{Thompson} {et~al}\mbox{.}(1978){Thompson}, {Nandy}, {Jamar},
  {Monfils}, {Houziaux}, {Carnochan}, \& {Wilson}}]{1978csuf.book.....T}
{Thompson} G.~I., {Nandy} K., {Jamar} C., {Monfils} A., {Houziaux} L.,
  {Carnochan} D.~J., {Wilson} R., 1978, {Catalogue of stellar ultraviolet
  fluxes : a compilation of absolute stellar fluxes measured by the Sky Survey
  Telescope (S2/68) aboard the ESRO satellite TD-1 /}

\bibitem[{{Trilling} {et~al}\mbox{.}(2007){Trilling}, {Stansberry},
  {Stapelfeldt}, {Rieke}, {Su}, {Gray}, {Corbally}, {Bryden}, {Chen}, {Boden},
  \& {Beichman}}]{2007ApJ...658.1289T}
{Trilling} D.~E. {et~al.}, 2007, \apj, 658, 1289

\bibitem[{{van Kerkwijk} {et~al}\mbox{.}(2010){van Kerkwijk}, {Rappaport},
  {Breton}, {Justham}, {Podsiadlowski}, \& {Han}}]{2010ApJ...715...51V}
{van Kerkwijk} M.~H., {Rappaport} S.~A., {Breton} R.~P., {Justham} S.,
  {Podsiadlowski} P., {Han} Z., 2010, \apj, 715, 51

\bibitem[{{Walczak} \& {Daszy{\'n}ska-Daszkiewicz}(2018)}]{2018CoSka..48...73W}
{Walczak} P., {Daszy{\'n}ska-Daszkiewicz} J., 2018, Contributions of the
  Astronomical Observatory Skalnate Pleso, 48, 73

\bibitem[{{Windemuth} {et~al}\mbox{.}(2019){Windemuth}, {Agol}, {Ali}, \&
  {Kiefer}}]{2019MNRAS.489.1644W}
{Windemuth} D., {Agol} E., {Ali} A., {Kiefer} F., 2019, \mnras, 489, 1644

\bibitem[{{Wojtkiewicz-Okulicz}(1925)}]{1925PulOB..10..19W}
{Wojtkiewicz-Okulicz} N., 1925, Pulkovo Observatory Bulletin, 10, 19

\end{thebibliography}

\end{document}